\documentclass{article}
\pdfoutput=1

\usepackage{arxiv}

\usepackage{amsmath}

\usepackage{algorithm}
\usepackage{algcompatible}
\usepackage{amsfonts}
\usepackage{amssymb}
\usepackage{amsbsy}
\usepackage[T1]{fontenc}
\usepackage{textcomp}
\usepackage{graphicx}
\usepackage{animate}
\usepackage{listings}
\usepackage{multicol}
\usepackage{multirow}
\usepackage[normalem]{ulem}
\usepackage{color}
\definecolor{ruby}{rgb}{0.6,0,0.3}
\definecolor{maroon}{rgb}{0.8,0,0.4}
\definecolor{rose}{rgb}{1.,0,0.4}
\definecolor{pinky}{rgb}{1.,0.5,0.5}
\definecolor{blue1}{rgb}{.75,.75,1}
\definecolor{blue2}{rgb}{0,.4,1}
\definecolor{blue3}{rgb}{0,0,.4}
\definecolor{fire0}{rgb}{0.000000,0.000000,0.000000}
\definecolor{fire1}{rgb}{0.134630,0.000508,0.000000}
\definecolor{fire2}{rgb}{0.204010,0.001525,0.000000}
\definecolor{fire3}{rgb}{0.251130,0.001904,0.000019}
\definecolor{fire4}{rgb}{0.298240,0.002775,0.000028}
\definecolor{fire5}{rgb}{0.346680,0.003885,0.000035}
\definecolor{fire6}{rgb}{0.402580,0.005471,0.000023}
\definecolor{fire7}{rgb}{0.453440,0.007244,0.000021}
\definecolor{fire8}{rgb}{0.505310,0.009437,0.000028}
\definecolor{fire9}{rgb}{0.558180,0.012159,0.000025}
\definecolor{fire10}{rgb}{0.618680,0.015920,0.000022}
\definecolor{fire11}{rgb}{0.673320,0.020187,0.000020}
\definecolor{fire12}{rgb}{0.728700,0.025608,0.000002}
\definecolor{fire13}{rgb}{0.784740,0.032732,0.000000}
\definecolor{fire14}{rgb}{0.848380,0.044235,0.000000}
\definecolor{fire15}{rgb}{0.904510,0.061777,0.000000}
\definecolor{fire16}{rgb}{0.952200,0.116600,0.000000}
\definecolor{fire17}{rgb}{0.979340,0.208460,0.000000}
\definecolor{fire18}{rgb}{0.992630,0.307640,0.000000}
\definecolor{fire19}{rgb}{0.997650,0.384710,0.000000}
\definecolor{fire20}{rgb}{0.999660,0.453950,0.000000}
\definecolor{fire21}{rgb}{1.000000,0.517610,0.000045}
\definecolor{fire22}{rgb}{1.000000,0.584020,0.001194}
\definecolor{fire23}{rgb}{1.000000,0.639720,0.003574}
\definecolor{fire24}{rgb}{1.000000,0.693190,0.007441}
\definecolor{fire25}{rgb}{1.000000,0.744990,0.013160}
\definecolor{fire26}{rgb}{1.000000,0.801740,0.022680}
\definecolor{fire27}{rgb}{1.000000,0.851160,0.035488}
\definecolor{fire28}{rgb}{1.000000,0.899800,0.056209}
\definecolor{fire29}{rgb}{1.000000,0.947520,0.120510}
\definecolor{fire30}{rgb}{1.000000,0.991500,0.516530}

\usepackage{graphicx}
\graphicspath{{./}}

\usepackage[hidelinks]{hyperref} 
\usepackage{fixltx2e}
\usepackage{soul}

\usepackage{tabularx}
\usepackage{longtable}
\usepackage{caption}
\usepackage{afterpage} 
\usepackage{colortbl}
\newcolumntype{Y}{>{\centering\arraybackslash}X}

\usepackage{scalerel}

\def\bsquare{\mathord{\scalerel*{\blacksquare}{gX}}}
\def\wdot{\mathord{\scalerel*{\cdot}{gX}}}

\newcommand{\ignore}[1]{}
\newcommand{\revisit}[1]{}

\usepackage[utf8]{inputenc} 

\usepackage{booktabs}       
\usepackage{microtype}      

\title{Modelling Orebody Structures:\\ Block Merging Algorithms and Block Model Spatial Restructuring Strategies Given Mesh Surfaces of Geological Boundaries}
\shorttitle{Block Merging Algorithms \& Spatial Restructuring Strategies}

\author{
  The final version published in the \textit{Journal of Spatial Information Science} will be available at\\ \url{doi:10.5311/JOSIS.2020.21.582}\\ \\
  \textbf{Raymond~Leung}\vspace{2mm} \\
  Australian Centre for Field Robotics (ACFR)\\
  Faculty of Engineering, The University of Sydney, Sydney, NSW 2006 Australia\\
  \texttt{raymond.leung@sydney.edu.au}
}

\renewcommand{\headerleft}{}
\renewcommand{\headeright}{Leung}

\begin{document}
\maketitle

\begin{abstract}
\vspace{-2mm}This paper describes a framework for capturing geological structures in a 3D block model and improving its spatial fidelity, including the correction of stratigraphic, mineralisation and other types of boundaries, given new mesh surfaces. Using surfaces that represent geological boundaries, the objectives are to identify areas where refinement is needed, increase spatial resolution to minimise surface approximation error, reduce redundancy to increase the compactness of the model and identify the geological domain on a block-by-block basis. These objectives are fulfilled by four system components which perform block-surface overlap detection, spatial structure decomposition, sub-blocks consolidation and block tagging, respectively. The main contributions are a coordinate-ascent merging algorithm and a flexible architecture for updating the spatial structure of a block model when given multiple surfaces, which emphasises the ability to selectively retain or modify previously assigned block labels. The techniques employed include block-surface intersection analysis based on the separable axis theorem and ray-tracing for establishing the location of blocks relative to surfaces. To demonstrate the robustness and applicability of the proposed block merging strategy in a more narrow setting, it is used to reduce block fragmentation in an existing model where surfaces are not given and the minimum block size is fixed. To obtain further insight, a systematic comparison with octree subblocking subsequently illustrates the inherent constraints of dyadic hierarchical decomposition and the importance of inter-scale merging. The results show the proposed method produces merged blocks with less extreme aspect ratios and is highly amenable to parallel processing. The overall framework is applicable to orebody modelling given geological boundaries, and 3D segmentation more generally, where there is a need to delineate spatial regions using mesh surfaces within a block model. 
\end{abstract}

\vspace{-2mm}\keywords{\\ Block merging algorithms\and Block model structure\and Spatial restructuring\and Mesh surfaces\and \\Subsurface modelling\and Geological structures\and Sub-blocking\and Boundary correction\and Domain identification\and \\Iterative Refinement\and Geospatial information system.\newline\newline
\textbf{CCS Concepts}:\newline \hspace{8mm}$\bullet$ Applied computing\,$\rightarrow$\, Earth and atmospheric sciences;\newline $\bullet$ Computing methodologies\,$\rightarrow$\,Volumetric models;\newline$\bullet$ Computing methodologies\,$\rightarrow$\, Mesh geometry models.
}

\newpage\section{Introduction}\label{sec:bsu-intro}
This article considers the spatial interaction of triangle-mesh surfaces with a block model as defined by Poniewierski \cite{poniewierski-19}. For this study, a block model may be conceptualised as a collection of rectangular prisms that span a modelled region in 3D space. Using a two-tier description, the block models of interest are formed initially by a set of non-overlapping `parent' blocks; these blocks have identical dimensions and are evenly spaced so they form a regular 3D lattice. Furthermore, a subset of the parent blocks --- particularly those that intersect with a surface --- are decomposed into smaller cuboids (often referred as `children' or sub-blocks) with the objective of preserving surface curvature subject to a minimum block size constraint. The sub-blocking problem represents a major theme in this paper. A distinguishing feature is that this problem is approached from a merging (bottom-up) perspective which offers opportunities for sub-block consolidation to minimise over-splitting, an issue often neglected in top-down approaches that focus exclusively on making block splitting decisions. The proposed framework allows new block models to be generated within a cell-based system using region partitioning surfaces. This comprises a spatial restructuring strategy that also allows iterative refinement of an existing block model given newer surfaces. At its core is a block merging algorithm that increases block model compaction and reduces spatial fragmentation due to subblocking. As motivation, we describe how this framework is deployed in a mine geology modelling system to illuminate key aspects of the proposal and illustrate what purpose they serve.

In mining, 3D geological models are used in resource assessment to characterise the spatial distribution of minerals in ore deposits \cite{luo-07}. A block model description of the geochemical composition is often created by fusing various sources of information from drilling campaigns, these include for instance: assay analysis, material or geophysical logging and alignment of stratigraphic units from geologic maps during the exploration phase. Due to the sparseness of these samples, the inherent resolution of these preliminary models are typically low. As the exploitation phase commences, denser samples may be taken strategically to develop a deeper understanding about the geology of viable ore deposits. This knowledge can assist miners with planning and various decision making processes \cite{leite-07}, for instance, to prioritise areas of excavation, to develop a mining schedule \cite{dimitrakopoulos-08}, to optimise the quality of an ore blend in a production plant. Of particular relevance to spatial modelling is that wireframe surfaces can be generated by geo-modelling software \cite{srk-18}\,\cite{mira-18}\, \cite{leapfrog-19}, or via kriging \cite{emery-05}, probabilistic boundary estimation \cite{ball-20}, boundary propagation (differential geometry) \cite{leung-19subsurface} and other inference techniques \cite{vasudevan-09} to minimise the uncertainty of interpolation at locations where data were previously unavailable.  For instance, triangle meshes may be created by applying the marching cubes algorithm \cite{newman-06} to Gaussian process implicit surfaces \cite{dragiev-11}. These boundary updates provide an opportunity to refine existing block models and remove discrepancies with respect to verified boundaries. The objective is to maximise the model's fidelity by increasing both accuracy and precision subject to some spatial constraints. The desired outcomes are improved localisation, reduced quantisation errors and less spatial fragmentation. In other words, the boundary blocks in the block model should accurately reflect the location of boundaries between geological domains; smaller blocks should be used to capture the curvature of regions near boundaries to minimise the surface approximation error; the model should provide a compact representation and have a low block count to limit spatial fragmentation.

\begin{figure}[h]
\centering
\includegraphics[width=80mm,trim={0mm 0mm 0mm 0mm},clip]{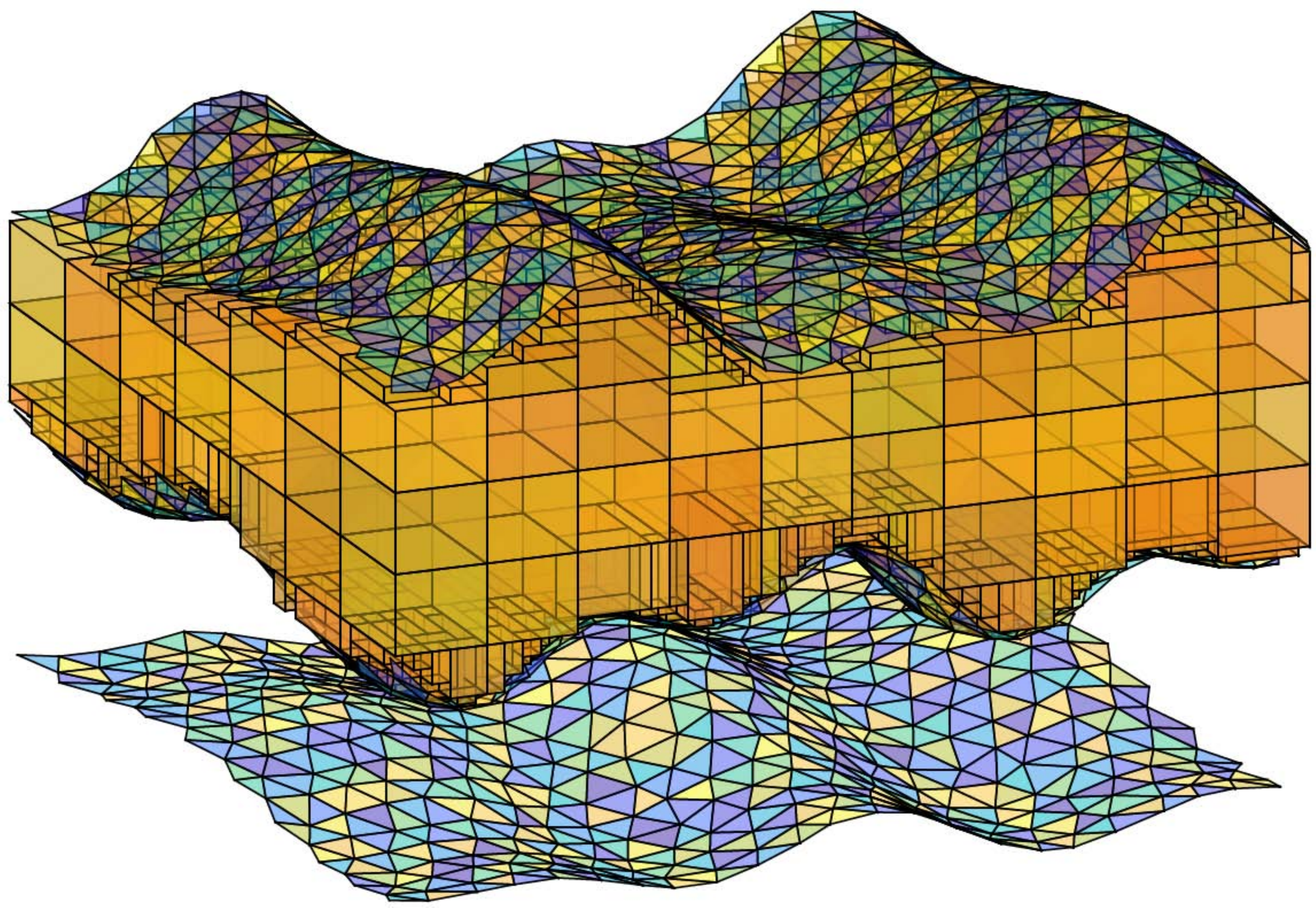}\\
\caption{Essence of block model spatial restructuring given mesh surfaces. A total of three surfaces are involved. For clarity, only the orange blocks situated between the top and middle surfaces are shown.}
\label{fig:bsu-intro-picture}
\end{figure}

Figure~\ref{fig:bsu-intro-picture} provides a visual summary of the primary objective. A key feature of spatial restructuring is that blocks are divided as necessary to adapt the block model to the curvature of the given surfaces. This process, known as \textit{sub-blocking}, is commonly performed in a top-down recursive manner which prioritizes splitting ahead of block consolidation. In some implementations, block consolidation is omitted altogether; this usually results in a highly fragmented and inefficient block representation. In this paper, surface-intersecting blocks are decomposed down to some minimum block size, then hierarchical block merging is performed in a bottom-up manner to consolidate the sub-blocks. In the ensuing sections, a framework for modifying the spatial structure of a block model using triangular mesh surfaces is first presented, the techniques underpinning each subsystem are described. Subsequently, we devote our attention to the block merging component, the algorithm is extended to support different forms of merging constraint. The proposed methods are applicable to orebody modelling given surfaces of mineralisation or stratigraphic boundaries --- see scenarios illustrated by \cite{wang-11}, \cite{feltrin-09}, \cite{mathers-11}, \cite{castagnac-11}; and general purpose 3D block-based modelling given other types of delineation.

\subsection{Definition of a surface}\label{sec:def-surfaces}
In this paper, the term `surface' encompasses both 2.5D and 3D surfaces. The former refers to open surfaces or warped 2D planes; these generally include mineralisation, hydration and stratigraphic surfaces in a mining context. The latter refers to closed surfaces that envelop a volume in 3D space. Examples include compact 2-manifolds that are topologically equivalent to a sphere. A simply-connected polyhedron surface would satisfy this requirement and have an Euler-Poincar\'{e} characteristic of 2. Closed surfaces may represent regions with local enrichment in an ore deposit or pockets with high level of contaminants.

\subsection{Justification for a block-based approach}\label{sec:justification}
In our envisaged application, the resultant block structure facilitates volumetric estimation of geochemical or geometallurgical  properties within a mine. A block-based representation provides spatial localisation while block labelling provides differentiation between geological domains. These properties are important for two reasons. First, it enables spatially varying signals and non-stationary geostatistics to be learned and captured at an appropriate granularity for mining. Individual blocks can be populated with local estimates of the geochemistry, material type composition or other physical attributes. Second, it prevents sample averaging from being applied across discontinuities (different domains) which can lead to incorrect probabilistic inferences and skewed predictions if applied unknowingly across a boundary with a steep attribute gradient. Although polyhedral, topological and surface-oriented models offer some interesting alternatives, block models have been well established and widely deployed within the resource industry. Introducing a radically different approach will likely cause significant disruption and require changes, acceptance and adaptation by various stakeholders. The proposed block model restructuring strategy is designed to work seamlessly with existing models. It makes changes which are transparent and compatible with subsequent processes such as mining exacavation which also operates on a block scale.

\section{Framework for Block Model Spatial Restructuring}\label{sec:bsu-framework}
\begin{figure}[h]
\centering
\includegraphics[width=87.5mm]{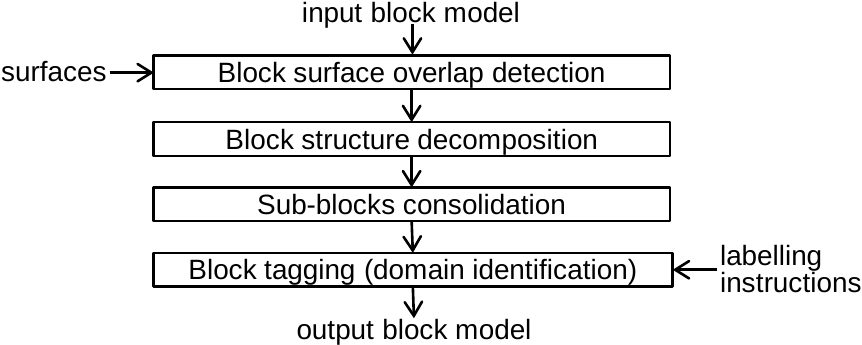}\\
\caption{Components in the block model spatial restructuring framework}
\label{fig:bsu-framework}
\end{figure}

This section develops a framework for altering the spatial structure of a block model to reconcile with the shape of the supplied surfaces as depicted in Fig.~\ref{fig:bsu-intro-picture}. The input block model consists of non-overlapping blocks of varying sizes (with uniform 3D space partitioning being a special case) and the surfaces, typically produced by boundary modelling techniques, represent the interface between different geological domains. The triangular mesh surfaces, together with the initial block model and block tagging instructions constitute the entire input. The tagging instructions simply assign to each block a label which classifies its location relative to each surface. The framework may be described in terms of four components: block surface overlap detection, block structure decomposition, sub-blocks consolidation and block tagging (domain identification) as shown in Fig.~\ref{fig:bsu-framework}.

\subsection{Block surface overlap detection}\label{sec:bsu-block-surface-overlap-detection}
The goal is to identify blocks in the input model which intersect with the given surface(s). These represent areas where model refinement is needed in order to minimise the surface approximation error. To establish a sense of scale, the input blocks (also called ``parent blocks'') typically measure $25\times 25\times 5$m in an axis-aligned local frame, ``axis-aligned'' means the edges of each block are parallel to one of the (x, y or z) axes.

\subsection{Block structure decomposition}\label{sec:bsu-block-structure-decomposition}
Block decomposition is performed on surface-intersecting parent blocks to improve spatial localisation. This process divides each block into smaller blocks of some minimum size (e.g., $6.25\times 6.25\times 1.25$m). These sub-blocks are disjoint and together, they span the whole parent block. Geometry tests are applied to determine which sub-block inside each surface-intersecting parent block actually intersects with a given surface. This process places blocks into one of three categories: (a) parent blocks that never intersect with any surface; or else sub-blocks inside surface-intersecting parent blocks that (b) intersect with a surface or (c) do not intersect with any surface.

\subsection{Sub-blocks consolidation}\label{sec:bsu-subblocks-consolidation}
This component coalesces sub-blocks to produce larger blocks in order to prevent over-segmentation. Recall that one of the objective is to minimise the total number of blocks in the output model. Therefore, if an array of $3\times 4\times 2$ sub-blocks could be merged into a perfect rectangular prism, it would be more efficient to represent them collectively by a single block with a new centroid and their combined dimensions.

In the event that the sub-blocks (within a surface-intersecting parent block) intersect with multiple different surfaces (say $s_1$ and
$s_2$) then merging considerations will be applied separately to the sub-blocks intersecting with $s_1$ and $s_2$. It is important to observe that block consolidation is applied to both category ``b'' and category ``c'' sub-blocks as defined above. Also, merging of sub-blocks across parent block boundaries is not permitted. In Sec.~\ref{sec:bsu-subblocks-consolidation-via-coordinate-ascent}, we describe in detail the proposed block merging approach which draws inspiration from the coordinate-ascent algorithm.

\subsection{Block tagging (domain identification)}\label{sec:bsu-block-tagging}
As a general principle, block tagging assigns abstract block labels to differentiate blocks located above and below a given surface. The objective is to support tagging with respect to multiple surfaces and two labelling policies. The first policy distinguishes surface-intersecting blocks from non-intersecting blocks. The second policy forces a binary decision and labels surface-intersecting blocks as strictly above or below a surface. The scheme also offers the flexibility of using a surface to limit the scope of an update, thereby leaving previously assigned labels intact above\,/\,below a surface. The processes described throughout (Sec.~\ref{sec:bsu-block-surface-overlap-detection} -- Sec.~\ref{sec:bsu-block-tagging}) are deterministic and applicable in an iterative setting. This completes our brief overview of the system.

\section{Techniques}\label{sec:bsu-techniques}
The techniques used in each component will now be further described. For maximum efficiency, the general set up assumes the modelled region aligns with the xyz axes. If the modelled deposit follows the inclination of a slope, it is assumed that an appropriate 3D rotation is applied to all spatial coordinates before subsequent techniques are applied. This has the effect of producing axis-aligned blocks in the modelling space that ultimately align with the principal orientation of the deposit in real space once the inverse transformation is applied at the conclusion of the process.

\subsection{Detecting block surface intersection}\label{sect:bsu-detect-block-surface-intersection}
The method we employed for determining if an axis-aligned rectangular prism intersects with a triangular patch from a surface is described by Akenine-M\"{o}ller in \cite{akenine-01}. This approach applies the Separating Axis Theorem (SAT) which states that two convex polyhedra, A and B, are disjoint if they can be separated along either an axis parallel to a normal of a face of either A or B, \textit{or} along an orthogonal axis computed from the cross product of an edge from A with an edge from B. The technical details can be found in Appendix~\ref{sec:appendix-block-triangle-intersect-detect-moller}.

In essence, block-surface intersection assessment consists of a series of ``block versus triangle'' comparisons where the triangles considered for each block are selected based on spatial proximity. The block-triangle overlap assessment operates on one basic principle: a ``no intersection'' decision is reached as soon as one of the tests returns \textsc{false}. A block-triangle intersection is found only when all 13 tests return \textsc{true}, when it failed to find any separation. These tests are used to identify the surface triangles (if any) that intersect with each parent block.

To maximise computation efficiency, each block is tested against a subset of triangles on each mesh surface rather than the entire mesh surface. A  kD-tree accelerator (a variation of binary space partitioning tree) \cite{pharr2016physically} is constructed using the 3D bounding boxes of each triangle. The subset includes only triangles whose bounding box overlaps with the block; only these candidates can intersect with the block. This pruning step limits the number of comparisons and speeds up computation considerably. The indices harvested here can subsequently reduce the test effort in the block structure decomposition stage.

\subsection{Block structure decomposition}\label{sec:bsu-block-structure-decomposition2}
The key premise is to decompose surface-intersecting blocks into smaller blocks to improve spatial localisation. The basic intuition is that the surface discretisation error decreases as spatial resolution increases. Precision increases when smaller blocks are used to approximate the surface curvature where the blocks meet the surface.

Block structure decomposition entails the following. For each block ($b$) that intersects with the surface, divide it volumetrically into multiple sub-blocks using the specified minimum block dimensions $(\Delta ^\text{block}_{x,\min},\Delta^\text{block}_{y,\min},\Delta^\text{block}_{z,\min})$. The main constraints are that sub-blocks cannot overlap and they must be completely contained by the parent block whose volume is the union of all associated sub-blocks.\footnote{Although we typically require parent blocks $(\Delta^{\text{block}}_{x}[b],\Delta^{\text{block}}_{y}[b],\Delta^{\text{block}}_{z}[b])$ to be divisible by the minimum block dimensions, the method works fine even if fractional blocks emerge during the division, i.e., the last block toward the end is smaller than the minimum size; the ratios in each dimension $n^{\text{block}}_x,n^{\text{block}}_y,n^{\text{block}}_z\notin \mathbb{Z}$ can be non-integers. This essentially means there is no fundamental limit on the minimum spatial resolution.} Within each surface-intersecting parent block, we also identify all sub-blocks that intersect with a surface and which surface they intersect with. This is accomplished using the associative mapping [obtained during block-surface overlap detection] which limits the relevant surface triangles to a small subset for each surface-intersecting parent block. The relevant attributes captured in the output include a list of surface-intersecting parent blocks, and attributes for each sub-block: viz., its centroid, dimensions, parent block index, intersecting surface and position within the parent block.

\subsection{Sub-blocks consolidation via coordinate-ascent}\label{sec:bsu-subblocks-consolidation-via-coordinate-ascent}
The consolidation component focuses on merging sub-blocks inside surface-intersecting parent blocks to reduce block  fragmentation. Its basic objective is to minimise the block count, although other mining or geologically relevant criteria such as block aspect ratio can be optimised for. These options are considered later in Sec.~\ref{sec:bsu-merging-conventions-optimisation-objectives} and Sec.~\ref{sec:bsu-results-statistical}. For now, these high resolution sub-blocks may themselves intersect or not intersect with any surface. This is indicated in the \textit{BlockStructureDecomposition} result which serves as input. This component returns the \textit{SubBlockConsolidation} result which describes the consolidated block structure. This encompasses all parent blocks processed, including those which do not intersect with any surface, as well as sub-blocks or super-blocks that constitute the surface-intersecting parent blocks.

The proposed merging algorithm is inspired by \textbf{coordinate-ascent optimisation} and may be summarised as follows.
\begin{itemize}
\item The algorithm is inspired by ``coordinate ascent'' where the search proceeds along successive coordinate directions in each iteration. The goal is to grow each block (a rectangular prism) from a single cell and find the maximum extent of spatial expansion, $\mathbf{k}=(k_x,k_y,k_z)$, without infringing other blocks or cells that belong to a different class.
\item For merging purpose, each parent block is partitioned uniformly down to the minimum block size. The smallest unit (with minimum block size) is referred as a \textit{cell}. Block dimensions are expressed in terms of the number of cells that span in the x, y and z directions. Accordingly, if a block $b$ with cell dimensions $\mathbf{k}=(k_x,k_y,k_z)$ is anchored at position $\mathbf{c}^{(b)}\in\mathbb{R}^3$, its bounding box would stretch from $\mathbf{c}^{(b)} - \frac{1}{2}\boldsymbol{\Delta}^\text{block}_{\min}$ to $\mathbf{c}^{(b)} + (\mathbf{k}-\frac{1}{2})\boldsymbol{\Delta}^\text{block}_{\min}$ for $k_x,k_y,k_z\in\mathbb{Z}\ge 1$.
\item Merging states are managed using a binary occupancy map (boolean 3D array) with cell dimensions identical to the parent block. Given a block anchor position $\mathbf{c}^{(b)}$ with $\mathbf{k}$ initialised to $(1,1,1)$, expansion steps are considered in each direction $\boldsymbol{\delta}\in\mathbb{Z}^3$ which must alternate through the sequence $\boldsymbol{\delta}_0=(1,0,0)$, $\boldsymbol{\delta}_1=(0,1,0)$ and $\boldsymbol{\delta}_2=(0,0,1)$.
\item A step in the direction $\boldsymbol{\delta}$ is feasible if all of the cells within the bounding box ($\mathbf{c}^{(b)} - \frac{1}{2}\boldsymbol{\Delta}^\text{block}_{\min}$, $\mathbf{c}^{(b)} + (\mathbf{k}+\boldsymbol{\delta}-\frac{1}{2})\boldsymbol{\Delta}^\text{block}_{\min}$) are 1 (active). In this case, the spatial extent is updated via $\mathbf{k}\leftarrow \mathbf{k}+\boldsymbol{\delta}$. Each iteration steps through $\boldsymbol{\delta}_0$, $\boldsymbol{\delta}_1$, $\boldsymbol{\delta}_2$ in turn. This continues until no expansion is possible in any direction, at which point the centroid and dimensions of the merged block are computed and the corresponding cells in the occupancy map are set to 0 (marked as inactive).
\item Sub-block merging terminates for a parent block when all cells in the occupancy map are set to 0.
\item The volume of a consolidated block is effectively the outer product $[0,k_x) \otimes [0,k_y) \otimes [0,k_z)$ where $k_x$, $k_y$, $k_z$ each represents some integer multiple of the minimum block size, $(\Delta ^\text{block}_{x,\min},\Delta^\text{block}_{y,\min},\Delta^\text{block}_{z,\min})$, with respect to x, y and z.
\item Different coordinate directions are used cyclically during the procedure. At all times, it must ensure the expansion does not include alien blocks in the encompassing cube, e.g., an L-shape block within a $2\times2\times1$ cube is not allowed.
\end{itemize}

\begin{figure}[!ht]
\centering
\includegraphics[width=135mm]{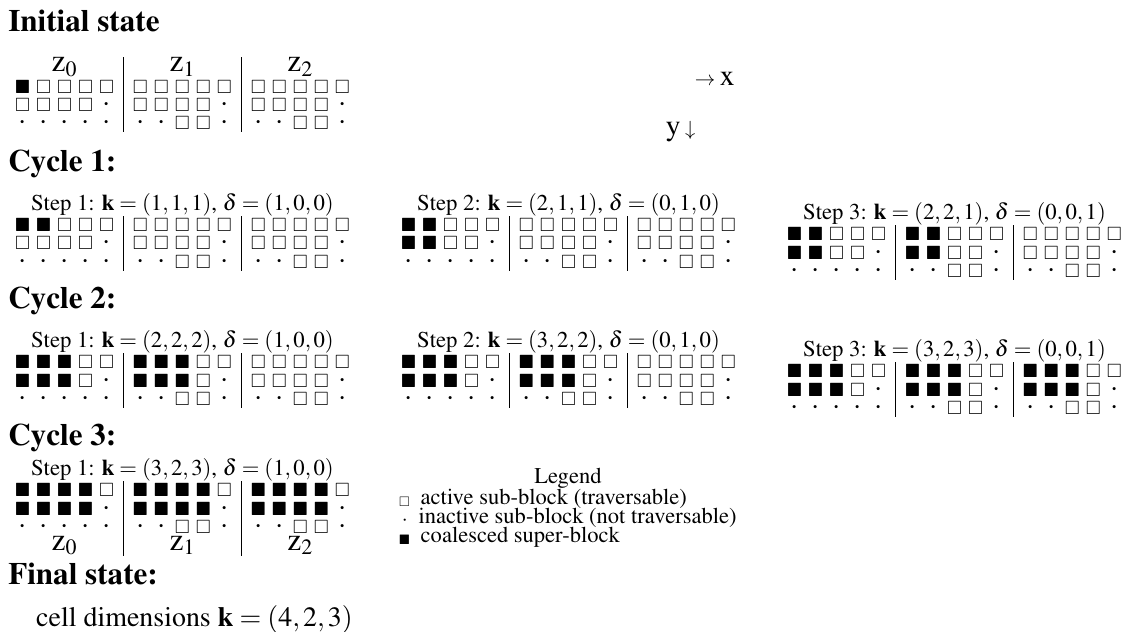}
\caption{Sub-blocks consolidation: illustration of the coordinate-ascent merging algorithm using occupancy map}
\label{fig:bsu-coalesce}
\end{figure}
The algorithm is explained by way of an example in Fig.~\ref{fig:bsu-coalesce}. To illustrate the algorithm, let us refer to blocks that belong to the class under consideration as the \textbf{active} cells. In the present context, this refers to either surface-intersecting sub-blocks, $\mathcal{S}_\text{intersect}$, or the non-intersecting sub-blocks, $\mathcal{S}_\text{non-intersect}$, inside a parent block. The \textbf{inactive} cells (labelled 0) refer to the complementary set to the active cells (labelled 1).

\vspace{3mm}In Fig.~\ref{fig:bsu-coalesce}, suppose the active sub-blocks (cells) refer to $\mathcal{S}_\text{intersect}$. The initial state shows the decomposition of a parent block in terms of three x/y cross-sections. There are $5\times3\times3$ sub-blocks (cells) within the parent block and 31 are considered ``active''. The white square cells all intersect with the same surface. It so happens the first cell encountered (with index $(i_Z\cdot n_Y + i_Y)\cdot n_X + i_X=0$) is the first active block. The algorithm considers expansion along each axis (x, y and z) in turn. In cycle 1, an expansion step in the x-direction is possible. The progressive expansion of the coalesced block is represented by black square cells. In cycle 2, further growth in the y-direction is not possible due to impediment by one or more ``inactive'' blocks (denoted by $\wdot$), however, the x and z dimensions each allow one step expansion. In cycle 3, expansion continues in the x direction, resulting in the formation of a $4\times2\times3$ super-block. At this point, a coalesced block is extracted as no further growth in any direction is now possible. Subsequently, the cells which have just been merged are marked out-of-bounds. The process repeats itself, starting with the next active cell it encounters in the raster-scan order.

\begin{figure}[!ht]
\centering
\includegraphics[width=85mm]{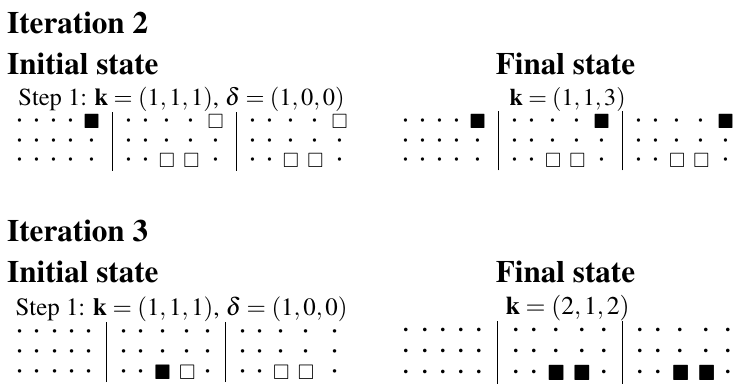}
\caption{Sub-blocks consolidation (continued) --- coalescing remaining active cells}
\label{fig:bsu-coalesce-continued1}
\end{figure}

\begin{figure}[!ht]
\centering
\includegraphics[width=85mm]{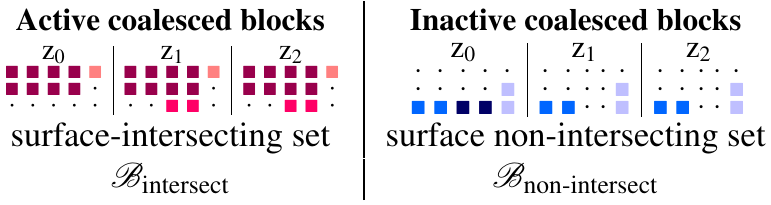}
\caption{Sub-blocks consolidation (continued) --- consolidated blocks in the surface-intersecting set $\mathcal{S}_\text{intersect}$ and non-intersecting set $\mathcal{S}_\text{non-intersect}$}
\label{fig:bsu-coalesce-continued2}
\end{figure}

Once all the sub-blocks (cells) are coalesced within the surface-intersecting parent block, we are left with 3 merged blocks (coloured in different shades of red in Fig.~\ref{fig:bsu-coalesce-continued2}). These have cell dimensions\footnote{These represent multiplying factors relative to the minimum block size.} ($k_x,k_y,k_z$) of $(4,2,3)$, $(1,1,3)$ and $(2,1,2)$ and relative centroids of $(\frac{4}{10},\frac{2}{6},\frac{3}{6})$, $(\frac{9}{10},\frac{1}{6},\frac{3}{6})$ and $(\frac{6}{10},\frac{5}{6},\frac{4}{6})$, respectively, with respect to the parent block's minimum vertex and parent block dimensions.

\begin{itemize}
\item When \textbf{multiple surfaces} are considered, the coalesce function will next be applied to other sub-blocks (within the same parent block) that intersect with other surfaces. This seldom happens but it is possible if the surfaces are close.
\item Finally, the coalesce function is \textbf{applied to inactive sub-blocks} (within the same parent block) which do not intersect with any surface. The result for inactive blocks (coloured in different shades of blue) are shown in Fig.~\ref{fig:bsu-coalesce-continued2}.
\end{itemize}

To summarise, coalescing is applied only to surface-intersecting parent blocks which are subject to decomposition. The sub-blocks contained within may intersect with one or more surfaces, or not intersect with any surface at all. After consolidation, the output contains both the original and merged blocks. Based on the sub-block and surface intersection status, the consolidated blocks may be arranged into two  separate sets, $\mathcal{B}_\text{intersect}$ and $\mathcal{B}_\text{non-intersect}$, as shown in Fig.~\ref{fig:bsu-coalesce-continued2}. Surface intersecting merged blocks ${\color{ruby}\blacksquare}{\color{maroon}\blacksquare}{\color{pinky}\blacksquare}$ (within surface-intersecting parent blocks) are placed in $\mathcal{B}_\text{intersect}$. Non surface-intersecting merged blocks ${\color{blue1}\blacksquare}{\color{blue2}\blacksquare}{\color{blue3}\blacksquare}$ (within surface-intersecting parent blocks) are placed in $\mathcal{B}_\text{non-intersect}$. Non surface-intersecting parent blocks are also appended to this. The coordinate ascent algorithm provides a method for merging cells (sub-blocks of minimum size) within the confines of a parent block. Coalesced sub-blocks must share the same label and form a rectangular prism.

\subsubsection{Practicalities}\label{sec:coalesce-practicalities}
Since the cell division lines are identical for all parent blocks of the same size, to avoid having to compute the sub-block centroids and dimensions repeatedly, the cell dimensions and local coordinates of each sub-block's centroid (relative to the parent block's minimum vertex) are stored in a look-up table and indexed by parent block dimensions to speed up computation.

\subsection{Block tagging: domain identification using surfaces}\label{sec:bsu-block-tagging2}
\begin{figure}[h]
\centering
\includegraphics[width=87.5mm]{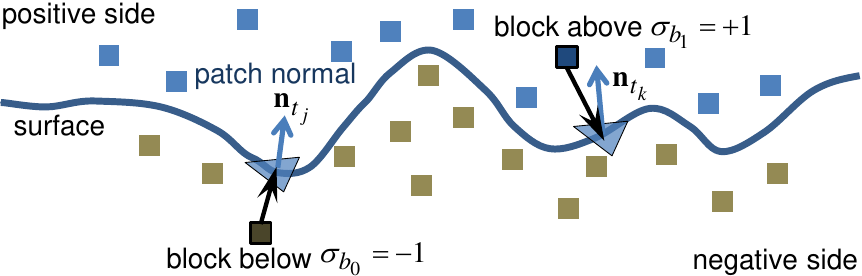}
\caption{Concepts relating to surfaces: surface normal (cross product) and block projected distance (inner product) associated with a triangular patch}
\label{fig:bsu-which-side-determination}
\end{figure}

The primary objective is to label whether a block is located above or below a surface. A picture of this is shown in Fig.~\ref{fig:bsu-which-side-determination}. More generally, the surface is not always horizontal, so it makes more sense referring to the space on one side of the surface as \textit{positive}, the other as \textit{negative}, however this is defined. 

The main points to appreciate in Fig.~\ref{fig:bsu-which-side-determination} are that
\begin{itemize}
\item Each triangular patch of the surface has a normal vector associated with it. For example, the normal $\mathbf{n}_{t_k}$ associated with triangle $t_k$ points in the upward direction. This normal is computed by taking the cross product between two of its edges, for instance, $(\mathbf{v}_{k,1}-\mathbf{v}_{k,0})\times(\mathbf{v}_{k,2}-\mathbf{v}_{k,0})$. This arrow would point in the opposite direction (rotate by 180\textsuperscript{o}) if the edge traversal direction is reversed; for instance, by swapping two vertices $\mathbf{v}_{k,1}$ and $\mathbf{v}_{k,2}$ in the triangle.
\item To determine ``which side of the surface'' a block is on, it suffices to consider the triangular patch located closest to the block. After establishing the positive side as the space `above' the surface, one can say block $b_1$ is located above the surface and has a positive sign ($\sigma_{b_1}=+1$) because the inner product $(\mathbf{c}_{t_k} - \mathbf{c}_{b_1})\cdot \mathbf{n}_{t_k}$ is negative. Here, $\mathbf{n}_{t_k}$ and $\mathbf{c}_{t_k}$ represent the normal and centroid of triangle $t_k$, likewise $\mathbf{c}_{b_1}$ represents the centroid of block $b_1$. Conversely, block $b_0$ is below the surface and has a negative sign ($\sigma_{b_0}=-1$) because the inner product $(\mathbf{c}_{t_j} - \mathbf{c}_{b_0})\cdot \mathbf{n}_{t_j}$ is positive. This `projection onto normal' approach provides the first method for block tagging.
\end{itemize}

\subsubsection{Discussion}\label{sec:bsu-tagging-projection-on-normal-discussion}
For this method to work, the edges for each triangle in the mesh surface must be ordered consistently (e.g., in the anti-clockwise direction) and any ambiguity in regard to surface orientation must be resolved to ensure the assigned labels ultimately conform to user's expectation --- e.g., the positive direction points upward with respect to an open surface (see below) or outward in the case of a closed surface.

\subsubsection{Labelling scheme: extension to multiple surfaces}\label{sec:bsu-labeling-scheme}
\begin{figure}[!h]
\centering
\includegraphics[width=146mm]{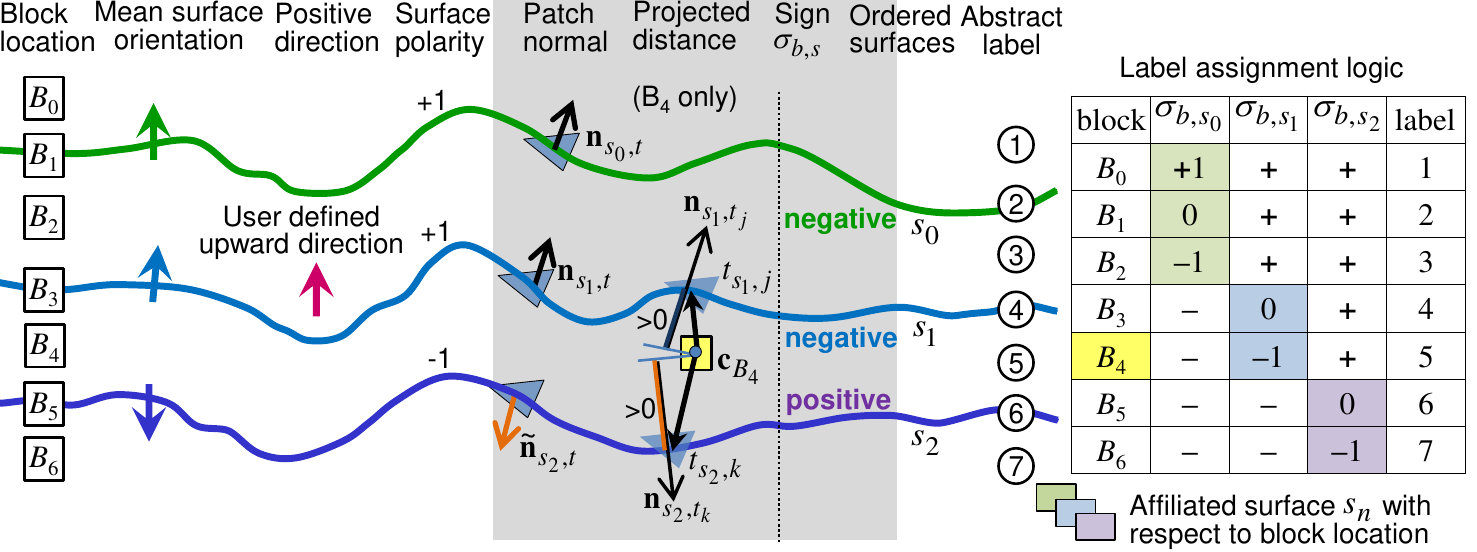}
\caption{Block tagging in a multi-surface scenario: terminologies and label assignment logic}
\label{fig:tagging-abstract-surface-layers}
\end{figure}

This tagging logic may be extended to multiple surfaces. Fig.~\ref{fig:tagging-abstract-surface-layers} depicts a multi-surface scenario and illustrates how abstract labels are assigned to multiple layers according to a logic table. Proceeding from left to right, blocks from seven distinct locations relative to the surfaces are shown. The \textit{mean surface orientation} shows the direction obtained by averaging over all triangular patch normals. This arrow can point up or down provided the ordering of triangle vertices is consistent. Consistency can be enforced by ensuring triangle edges are traversed only in the anti-clockwise (or clockwise) direction.

The \textit{positive direction} is a user-defined concept. By default, it points in the upward direction (+z axis). For a vertical surface (e.g., a geological feature such as a dyke) that runs across multiple layers, the positive direction may be defined as left (or right). \textit{Surface polarity} indicates if there is agreement between the mean surface orientation (a property conferred by the mesh) and the positive direction (intention of the user). If they oppose, as is the case for the bottom surface in Fig.~\ref{fig:tagging-abstract-surface-layers}, the polarity is set to negative. The significance is that the interpretation of the \textit{projected distance} between a block and relevant triangle, and subsequently what sign we assign to the block, depend on the surface polarity.

Focusing now on the shaded portion of Fig.~\ref{fig:tagging-abstract-surface-layers}, comparison of block $B_4$ with surface 1 yields $(\mathbf{c}_{t_{s_1,j}}\!-\mathbf{c}_{B_4})\cdot\mathbf{n}_{s_1,t_j}>0$, hence its sign $\sigma_{B_4,s_1}$ is \textit{negative} (it is below the $s_1$ surface). However, comparison with surface 2 yields a \textit{positive} sign $\sigma_{B_4,s_2}$ (it is above the $s_2$ surface) even though the projected distance $( \mathbf{c}_{t_{s_2,k}}\!-\mathbf{c}_{B_4})\cdot\mathbf{n}_{s_2,t_k}>0$; this is correct since $s_2$ has negative surface polarity which negates the logic. This method is adequate for simple surfaces such as those depicted in Fig.~\ref{fig:tagging-abstract-surface-layers}; however it can fail in more complex situations. Therefore, the \textit{projection-onto-patch-normal} approach is used here primarily as a vehicle for illustrating concepts and vulnerabilities. The pitfalls and robust solutions will be discussed in Sec.~\ref{sec:bsu-critical-reflection}.

\textit{Abstract labels} are assigned to merged blocks to distinguish between boundaries and embedded layers. In this example, layers are given odd-integer labels whereas boundaries (the interface between layers) are given even-integer labels, as indicated by the circles in Fig.~\ref{fig:tagging-abstract-surface-layers}. For the interleaved layers, given its sign $\sigma_{b,s_n}$ and affiliated surface $s_n$, the abstract label is given by $\lambda(s_n,\sigma_{b,s_n})=2\times(n+1)-\sigma_{b,s_n}$ for $n\ge 0$, where $\sigma_{b,s_n}\in\{-1,0,+1\}$ represent \{\text{below, across, above}\} the surface, respectively. The affiliated surface $s_n$ is deduced from the shaded $\sigma_{b,s_n}$ column of the logic table in Fig.~\ref{fig:tagging-abstract-surface-layers}. For instance, at block locations $B_3$ and $B_4$, the affiliated surface is $s_1$. With $\sigma_{b,s_0}=-1$ and $\sigma_{b,s_2}=+1$ for $b\in\{B_3,B_4\}$, the label only depends on $\sigma_{b,s_1}$ when $n=1$.

\subsubsection{Tagging instructions}\label{sec:tagging-instructions}
\begin{figure}[!ht]
\centering
\includegraphics[width=146mm]{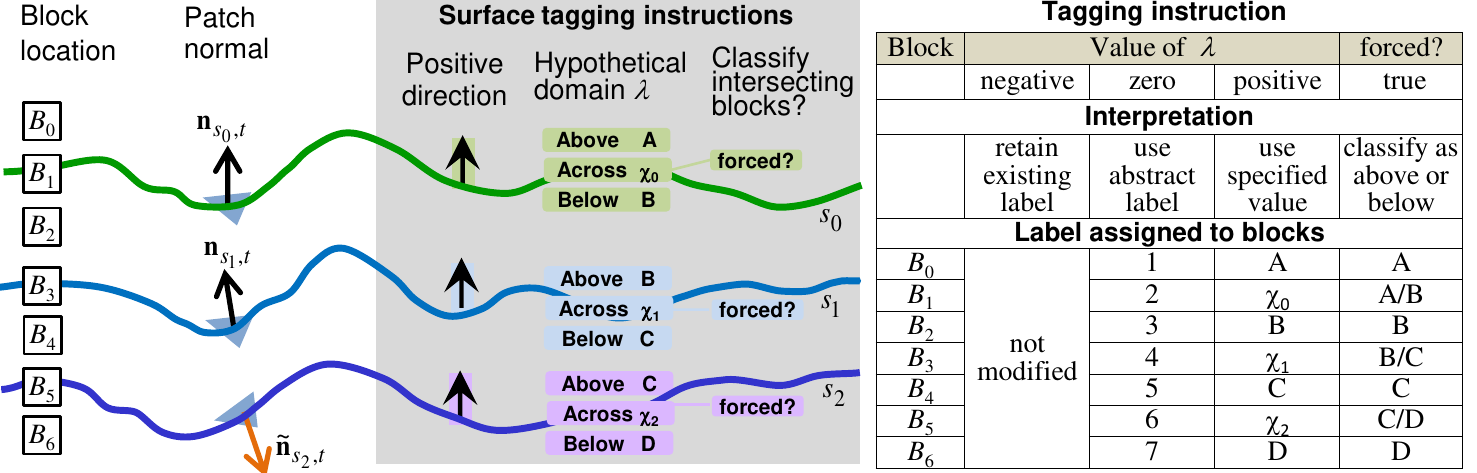}
\caption{Block tagging instructions: interpretation of the lambda value}
\label{fig:tagging-instructions}
\end{figure}

With block tagging, there is tremendous scope for creativity. The scheme described below provides a flexible framework for domain specification given arbitrary surfaces. The diagram shown in Fig.~\ref{fig:tagging-instructions} is similar to Fig.~\ref{fig:tagging-abstract-surface-layers} for the most part. What is different is the emphasis on \textit{surface tagging instructions}. Observe that there are three sets of tagging instructions: one for each surface. Each instruction specifies (1) a positive direction for each surface; (2) nominal labels for blocks located above, across and below each surface; (3) an override field (see ``forced?'') which communicates an intent to either \textbf{preserve} the surface intersecting blocks (by assigning a label different to any other layers) or \textbf{resolve} whether these blocks are strictly above or below the surface, i.e., force a binary decision.

A trivalent logic is built into the nominal labels specified in item 2. In Fig.~\ref{fig:tagging-instructions}, this is denoted by $\lambda$. User-specified values are assigned to blocks when $\lambda>0$. In addition, there are two special cases worth mentioning. First, an input value of $\lambda=0$ instructs the program to use abstract labels instead of specific domain values (see ``zero'' column in Fig.~\ref{fig:tagging-instructions}). Second, a negative input value $\lambda<0$ instructs the program to \textbf{leave current labels intact}. This retains any prior label which has been assigned to that block (see ``negative'' column in Fig.~\ref{fig:tagging-instructions}). This `retain existing labels' mode allows the spatial structure of a block model to be updated without invalidating previous domain assignments. For instance, a surface $s_0$ may be used as an upper bound. If $\lambda_{s_0}^\text{(above)}=-1$, then all blocks located above this surface will not have their labels modified. Similarly, $s_1$ may be used as a lower bound. If $\lambda_{s_1}^\text{(below)}=-1$, then all blocks located below this surface will not have their labels modified.

\subsubsection{Retaining existing domain labels}\label{sec:retain-existing-labels}
As a motivating example, suppose we rotate the picture in Fig.~\ref{fig:tagging-instructions} counter-clockwise by 45 degrees. Further, suppose the entire block model is currently labelled as domain A. Then, the space between the two tilted surfaces (below $s_0$ and above $s_1$) can be used to model a dolerite channel that runs diagonally across the layers. By using the following specification: $\lambda_{s_0}^\text{(above)}=-1$, $\lambda_{s_0}^\text{(below)}=\text{B}$, $\lambda_{s_1}^\text{(above)}=\text{B}$, $\lambda_{s_1}^\text{(below)}=-1$ and forced($s$)\,=\,\textsc{true} for both, blocks within the embedded layer will be tagged as domain B (dolerite) for some $\text{B}>$0.

\subsection{General interpretation}\label{sec:general-interpretation}
A volumetric block region (one comprising blocks of varying size) may be constructed in multiple passes and  labelled sequentially by a single or a combination of surfaces using logical \textsc{and}/\textsc{or} operators based on the plane partitioning property of open surfaces, or expressed as the intersection/union of multiple closed surfaces. As an example, a fault surface may partition a space into two disjoint sets, say, blocks to the left (respectively, right) of the divide. Input blocks in contact with the fault line may be partitioned (subblocked) to conform to the shape of the fault surface. Subsequently, block tagging is applied independently to these two sets. The bounding surfaces that correspond across the discontinuity will provide similar block labelling instructions despite their physical separation. This ensures blocks within the same layer will receive the same label  notwithstanding the vertical offset introduced by the fault.

\section{Visualisation}\label{sec:bsu-visualisation}
This section illustrates some of the results produced by the proposed framework. Fig.~\ref{fig:bsu-results-block-restructuring-multiple-surfaces}(a) shows a regular block structure which covers the region $x\in[1000,1250],y\in[750,925],z\in[590,670]$ with parent blocks measuring $25\times25\times5$m. Three horizontal surfaces are also given as input, these are shown in Fig.~\ref{fig:bsu-results-block-restructuring-multiple-surfaces}(b). Surface-intersecting parent blocks identified in (c) are subject to block decomposition. Using block-triangle overlap detection, the surface-intersecting sub-blocks, each measuring $5\times 5\times 1$m, are identified in Fig.~\ref{fig:bsu-results-block-restructuring-multiple-surfaces}(d). To reduce fragmentation, the decomposed sub-blocks in the surface-intersecting and non-intersecting sets are consolidated in Fig.~\ref{fig:bsu-results-block-restructuring-multiple-surfaces}(e)--(f). For the surface-intersecting set, the block-count decreases from 8497 to 2102 after sub-blocks are coalesced.

\begin{table}[!t]
\setlength\tabcolsep{0pt}
\begin{center}
\scriptsize
\begin{tabular}{p{50mm}p{50mm}p{50mm}}
(a) Regular block structure as input & (b) Multiple surfaces as input & (c) Surface intersecting parent blocks\\
\includegraphics[width=43.5mm,trim={0mm 0mm 0mm 0mm},clip]{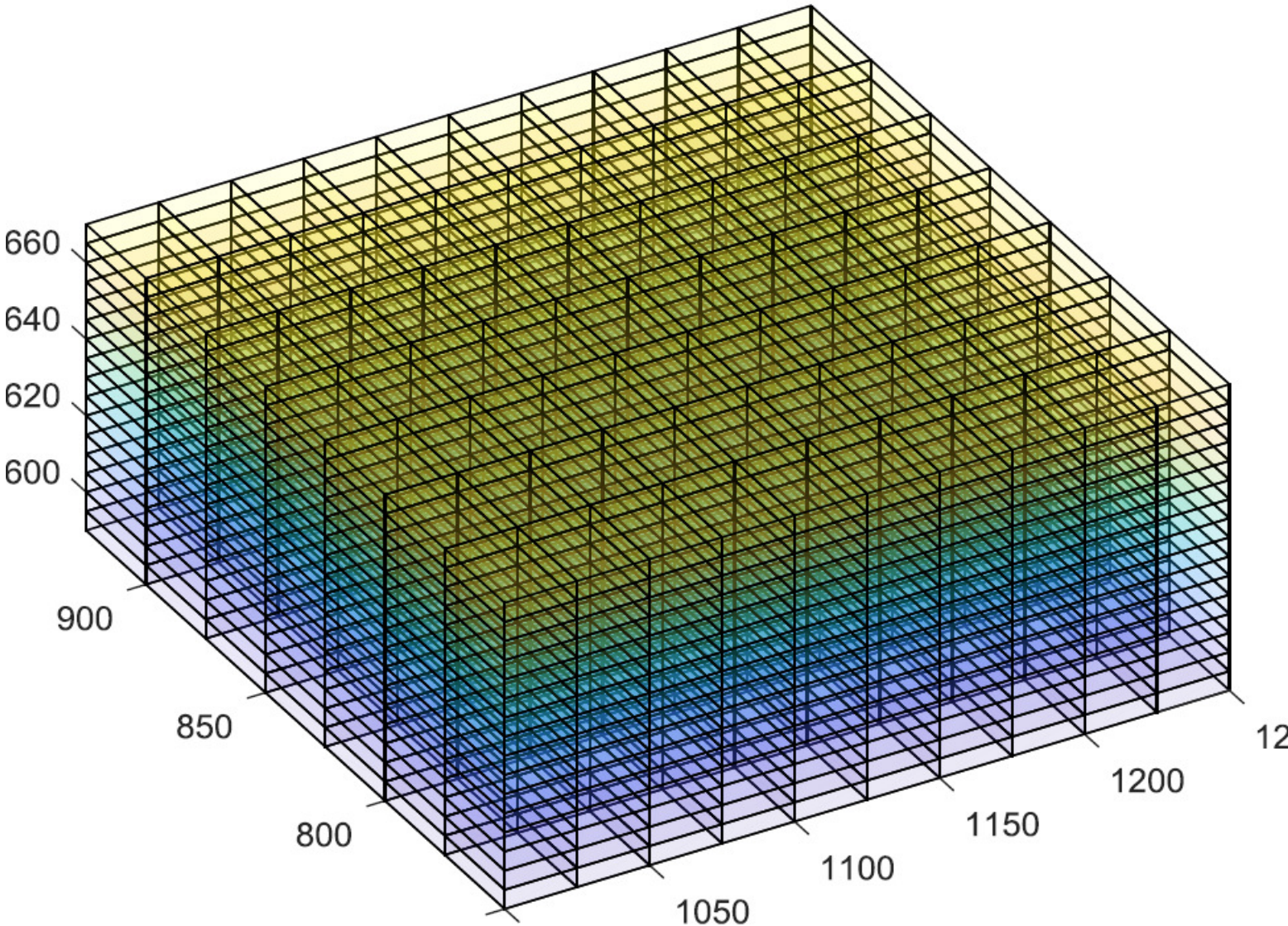} & \includegraphics[width=43.5mm]{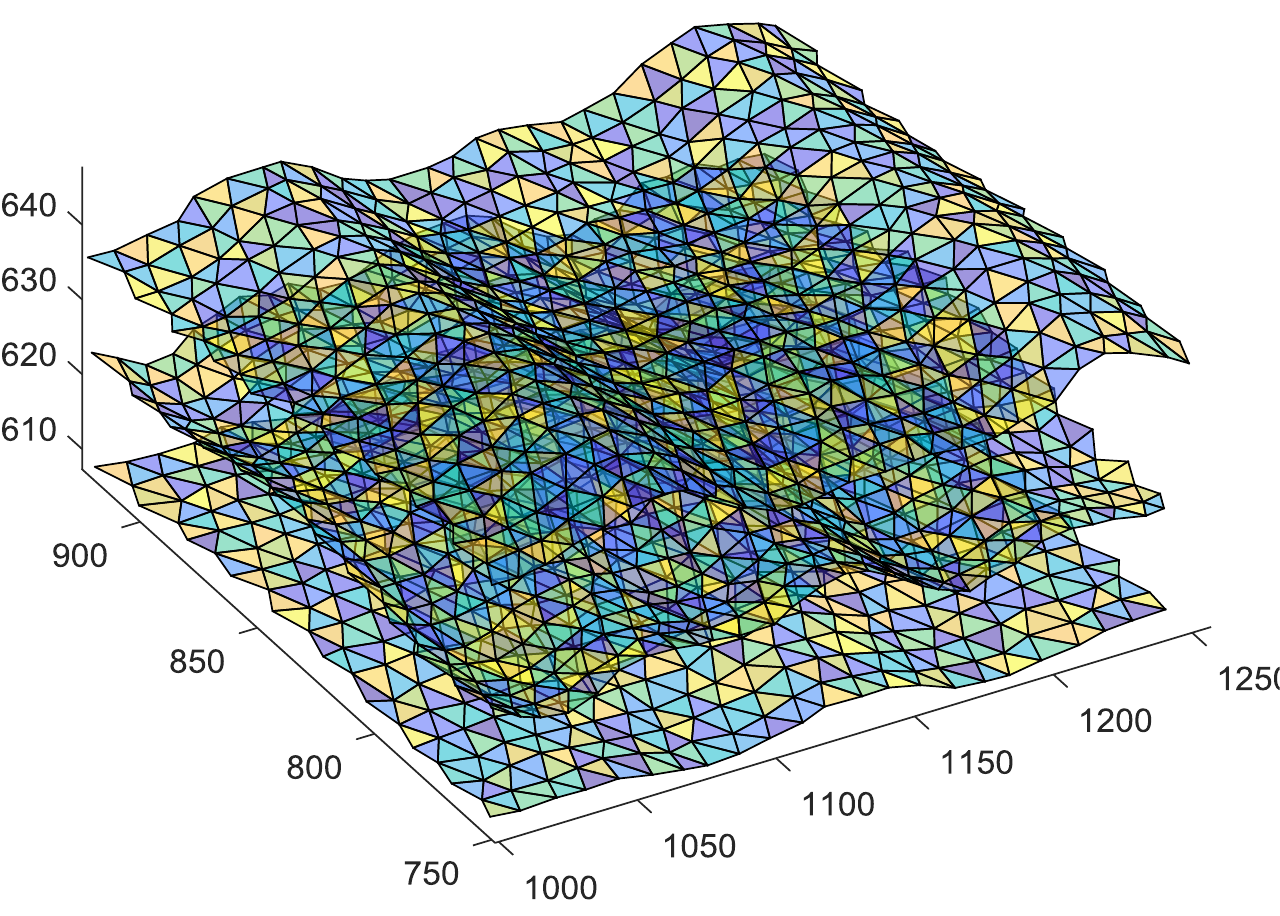} & \includegraphics[width=43.5mm,trim={0mm 0mm 0mm 0mm},clip]{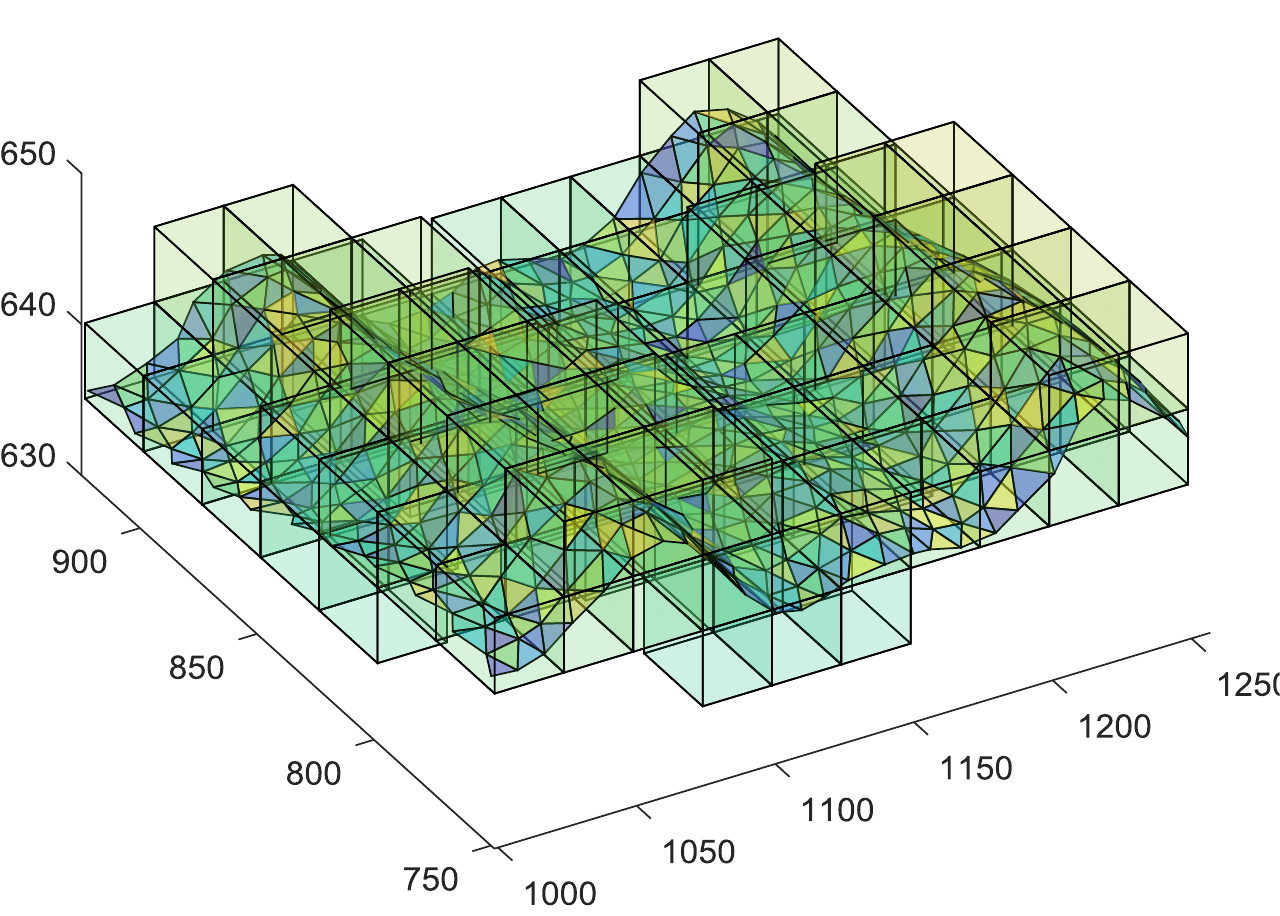}\\
(d) Surface intersecting sub-blocks & (e) Consolidated sub-blocks & (f) Consolidated sub-blocks from non-\\
& from surface-intersecting  set & intersecting (complementary) set\\
\includegraphics[width=43.5mm]{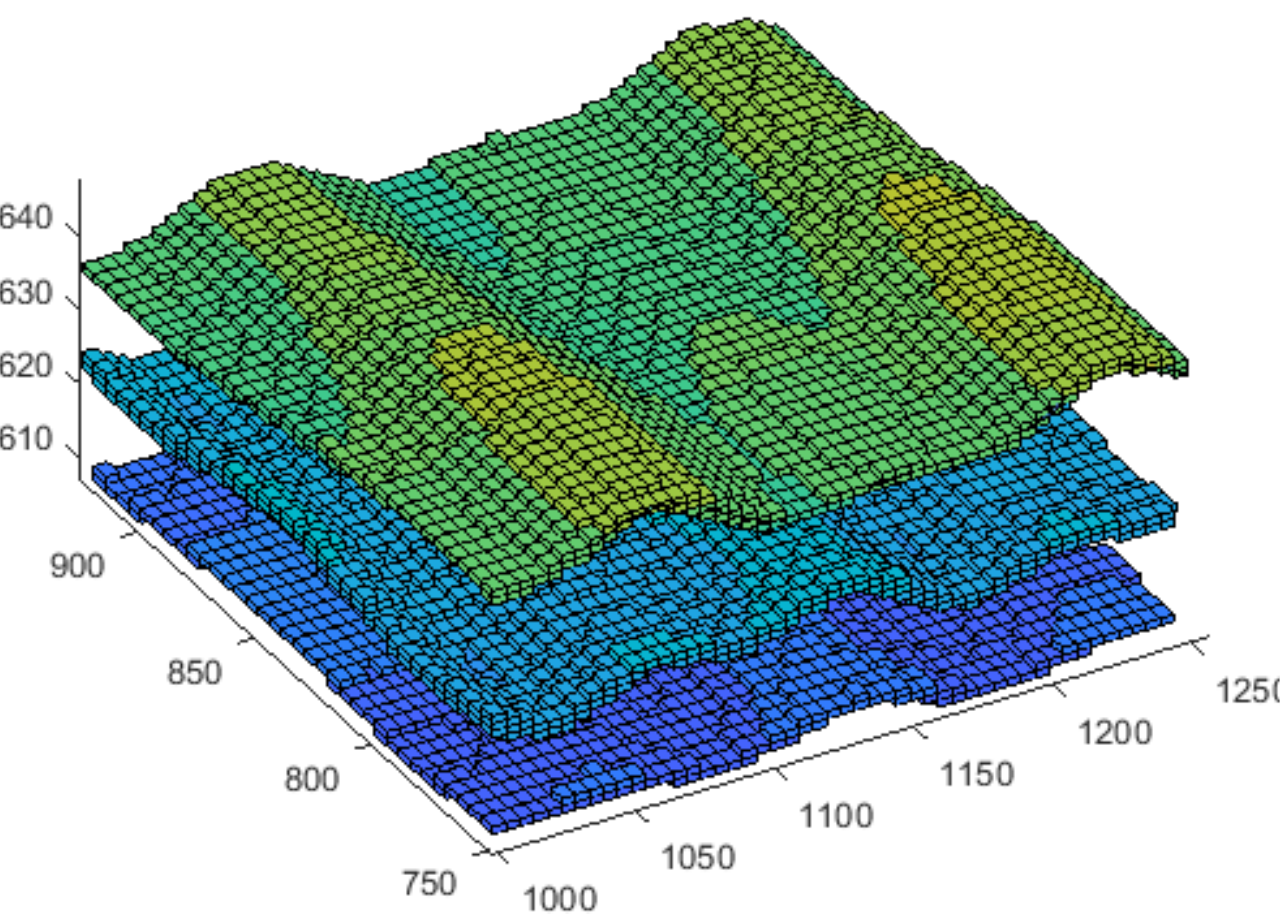} & \includegraphics[width=43.5mm,trim={0mm 0mm 0mm 0mm},clip]{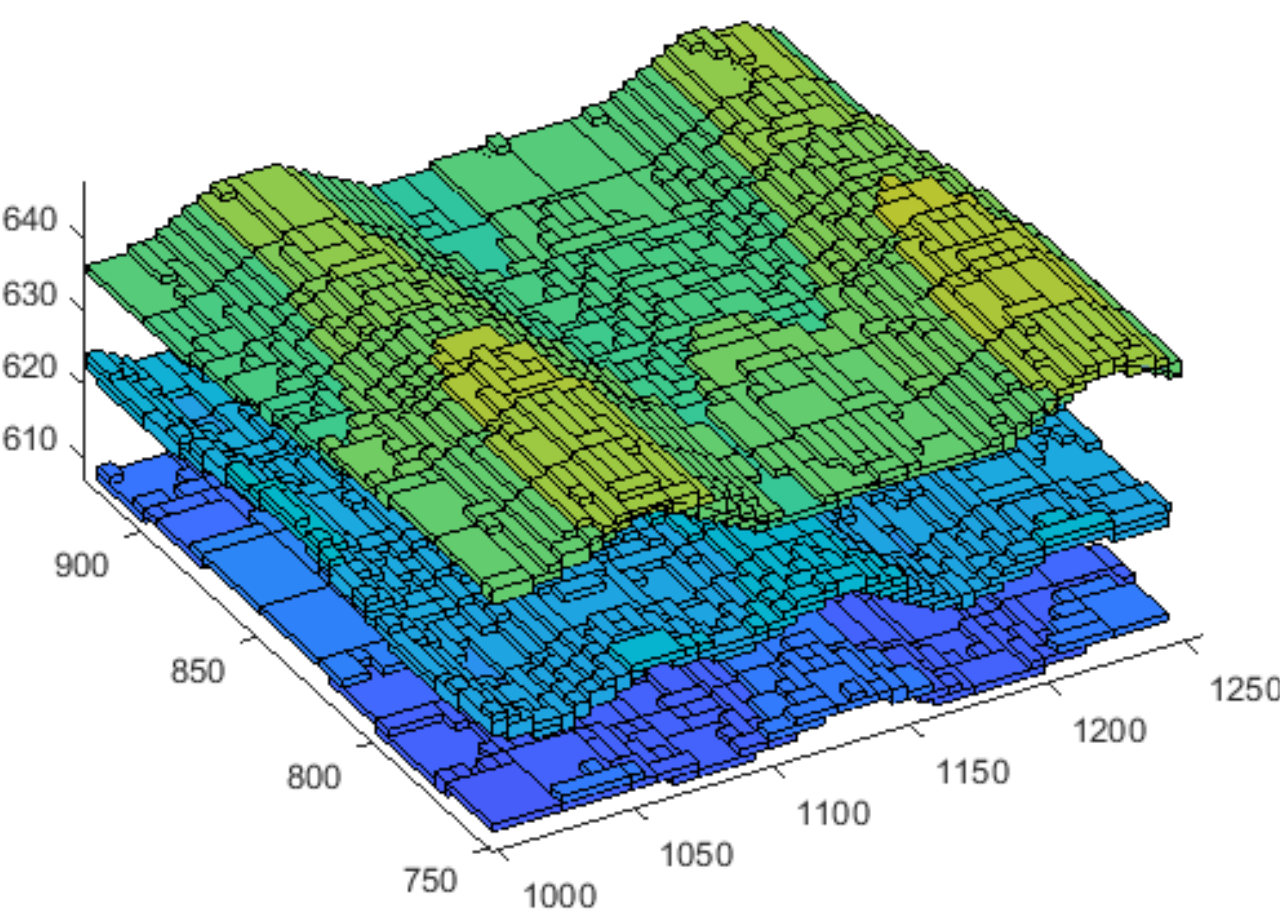} & \includegraphics[width=43.5mm]{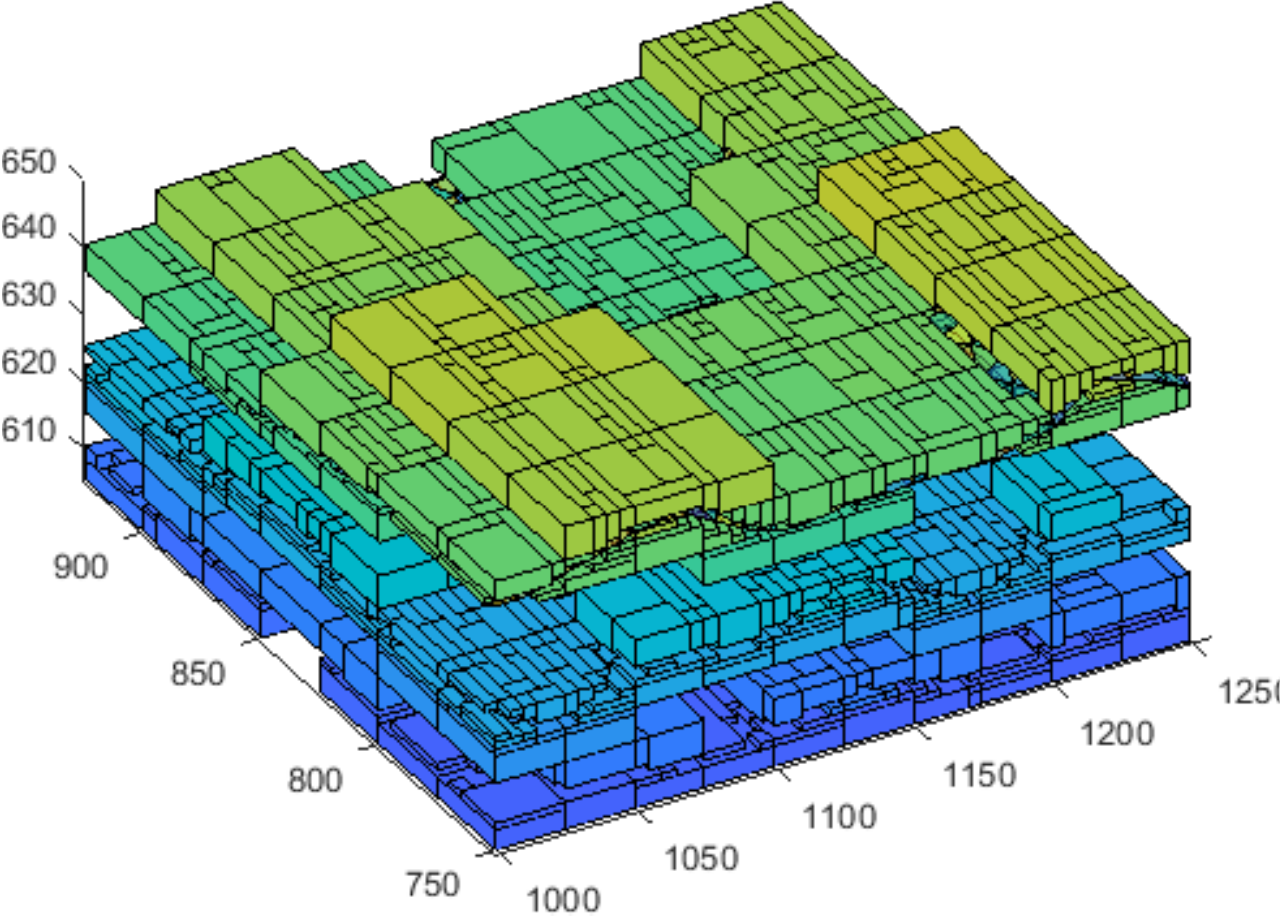}
\end{tabular}
\end{center}
\captionof{figure}{Block restructuring given multiple surfaces}
\label{fig:bsu-results-block-restructuring-multiple-surfaces}
\end{table}

In regard to the \textit{forced} option for block tagging, the differences between preserving the boundary and classifying the blocks at the interface as strictly above\,/\,below a surface are demonstrated in Fig.~\ref{fig:bsu-results-block-tagging-multiple-surfaces}(g)--(h). For clarity, Fig.~\ref{fig:bsu-results-block-tagging-multiple-surfaces}(i) shows only blocks labelled as A and C (belonging to two domains of interest) in isolation. Clearly, they conform to the shape of the relevant surfaces. In Fig.~\ref{fig:bsu-results-block-tagging-multiple-surfaces}(j), only blocks that intersect with the top surface (labelled 2) have been extracted.

\begin{table}[!t]
\setlength\tabcolsep{0pt}
\begin{center}
\scriptsize
\begin{tabular}{p{50mm}p{50mm}}
(g) Resolving boundary blocks & (h) Preserving boundary blocks\\
\textit{forced}=1 classifies as above/below & \textit{forced}=0 maintains unique identity\\
\includegraphics[width=43.5mm,trim={0mm 0mm 0mm 0mm},clip]{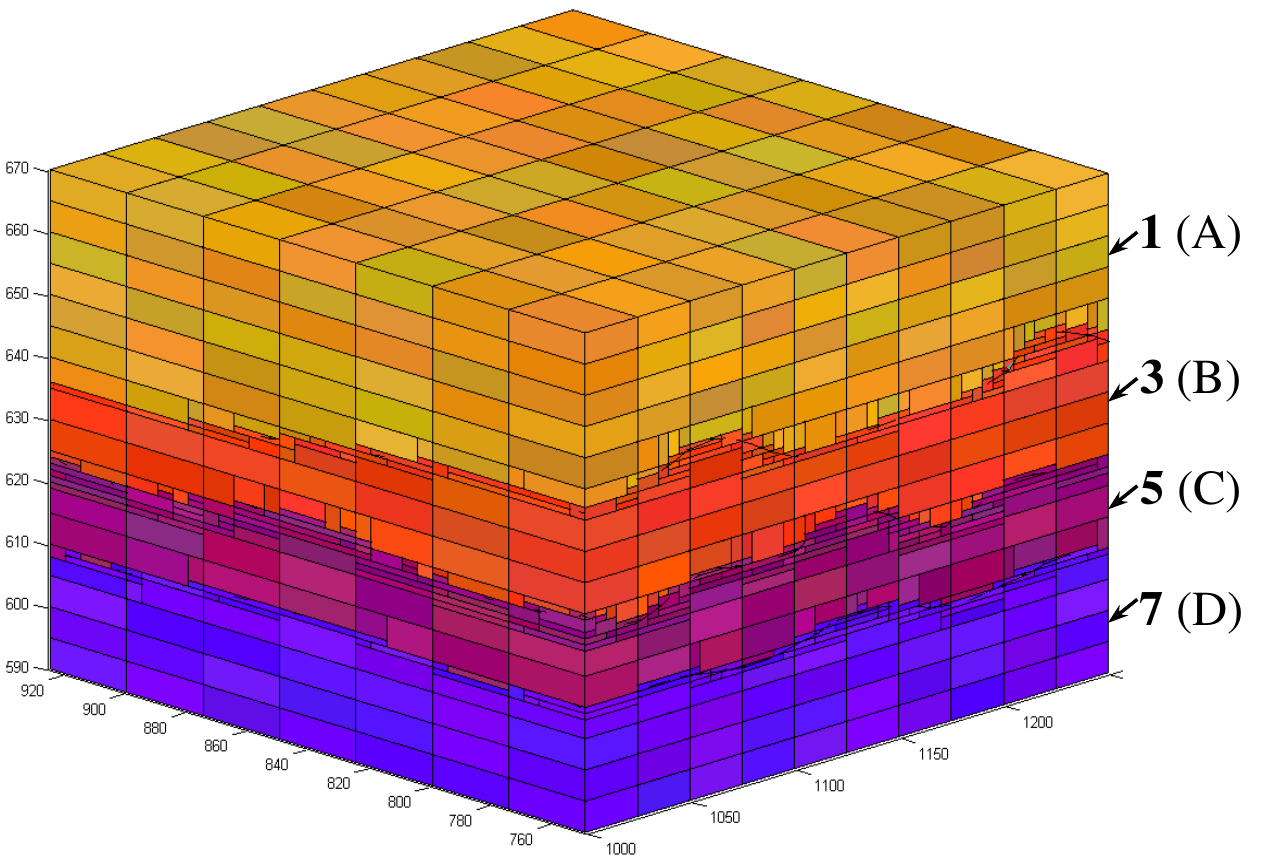} & \includegraphics[width=43.5mm]{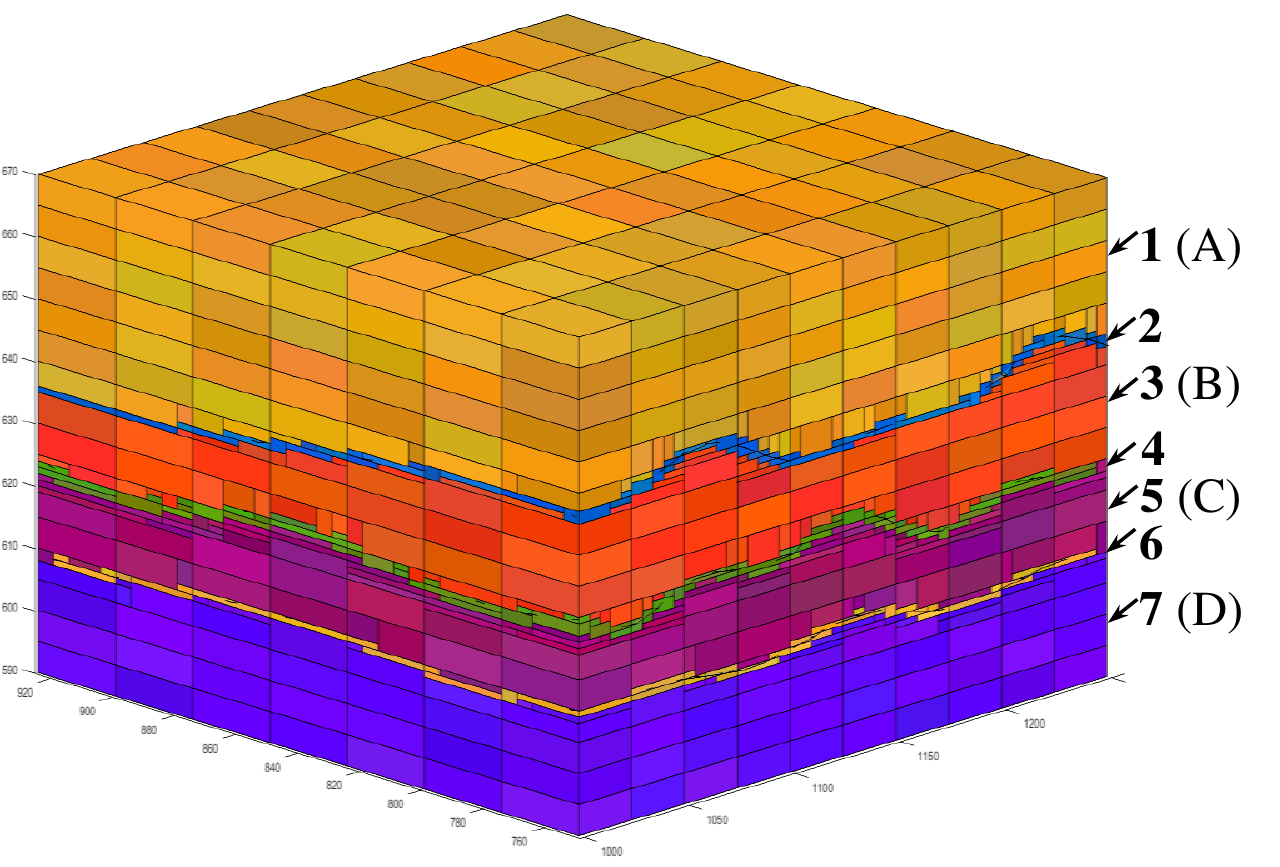}\\
(i) Isolated blocks from A and C & (j) Localised blocks extracted from\\
 --- two domains of interest  & the A/B interface\\
\includegraphics[width=43.5mm,trim={0mm 0mm 0mm 0mm},clip]{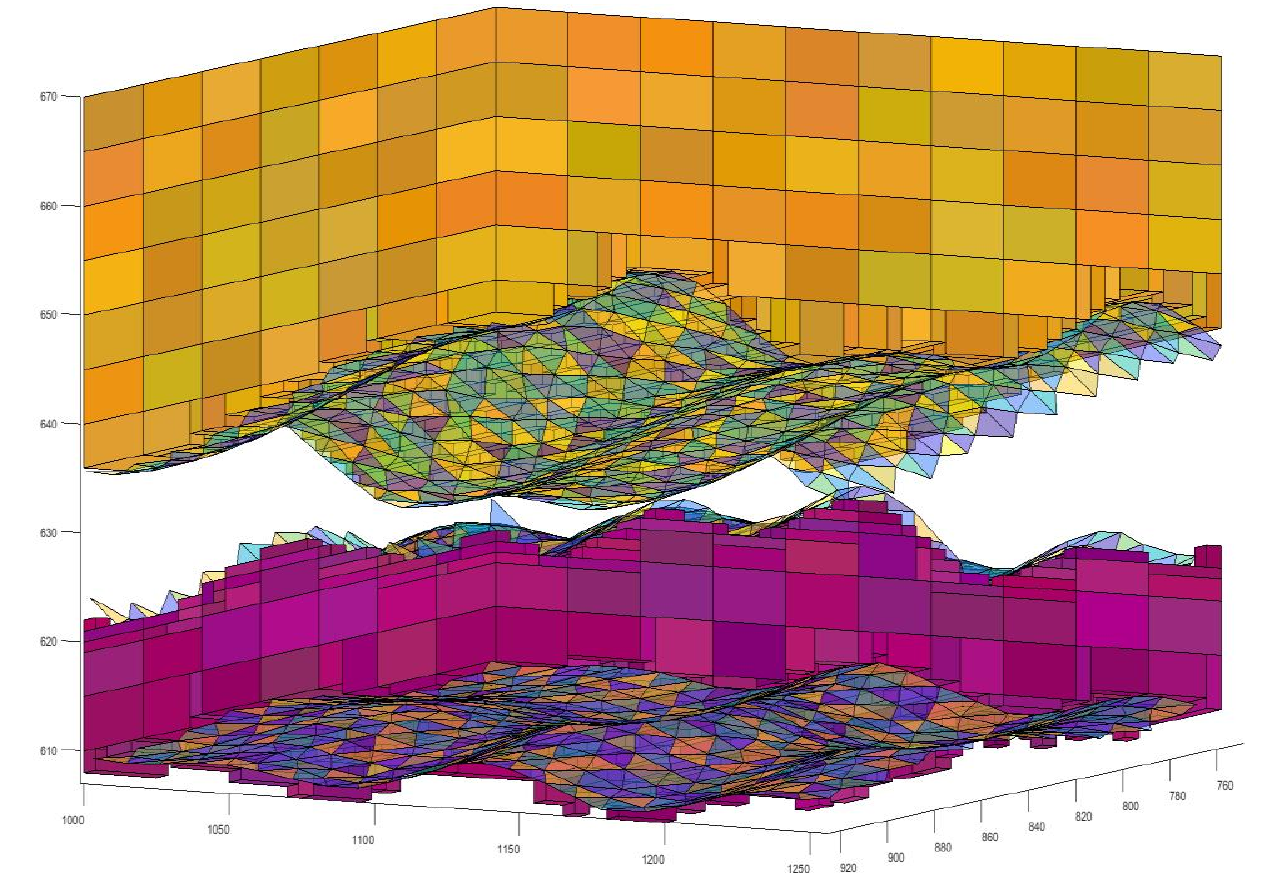} & \includegraphics[width=43.5mm]{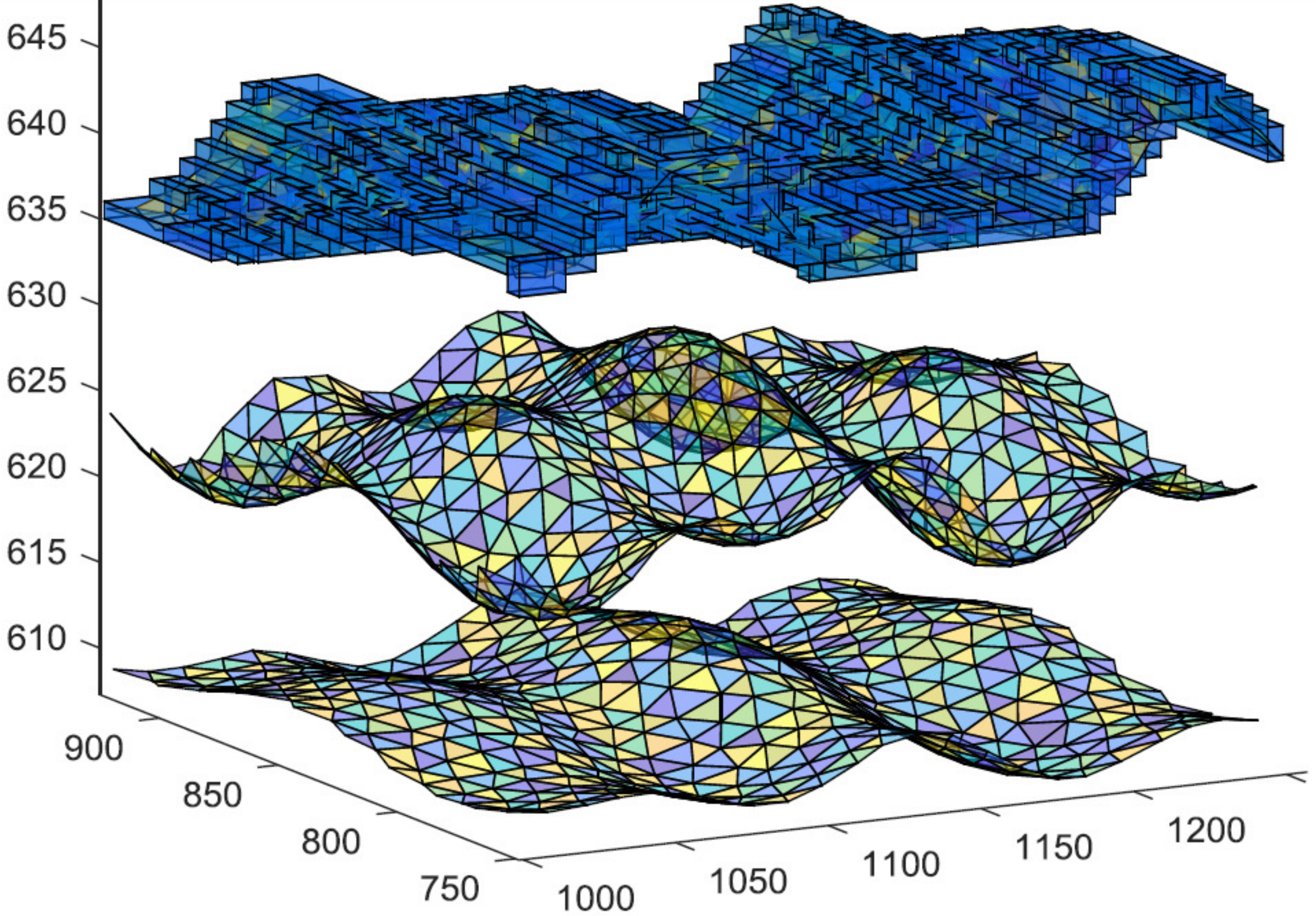}
\end{tabular}
\end{center}
\captionof{figure}{Block tagging given multiple surfaces}
\label{fig:bsu-results-block-tagging-multiple-surfaces}
\end{table}

\subsection{Iterative refinement: an application to tilted surfaces}\label{sec:bsu-iterative-application}
\begin{figure}[!ht]
\centering
\includegraphics[width=87.5mm]{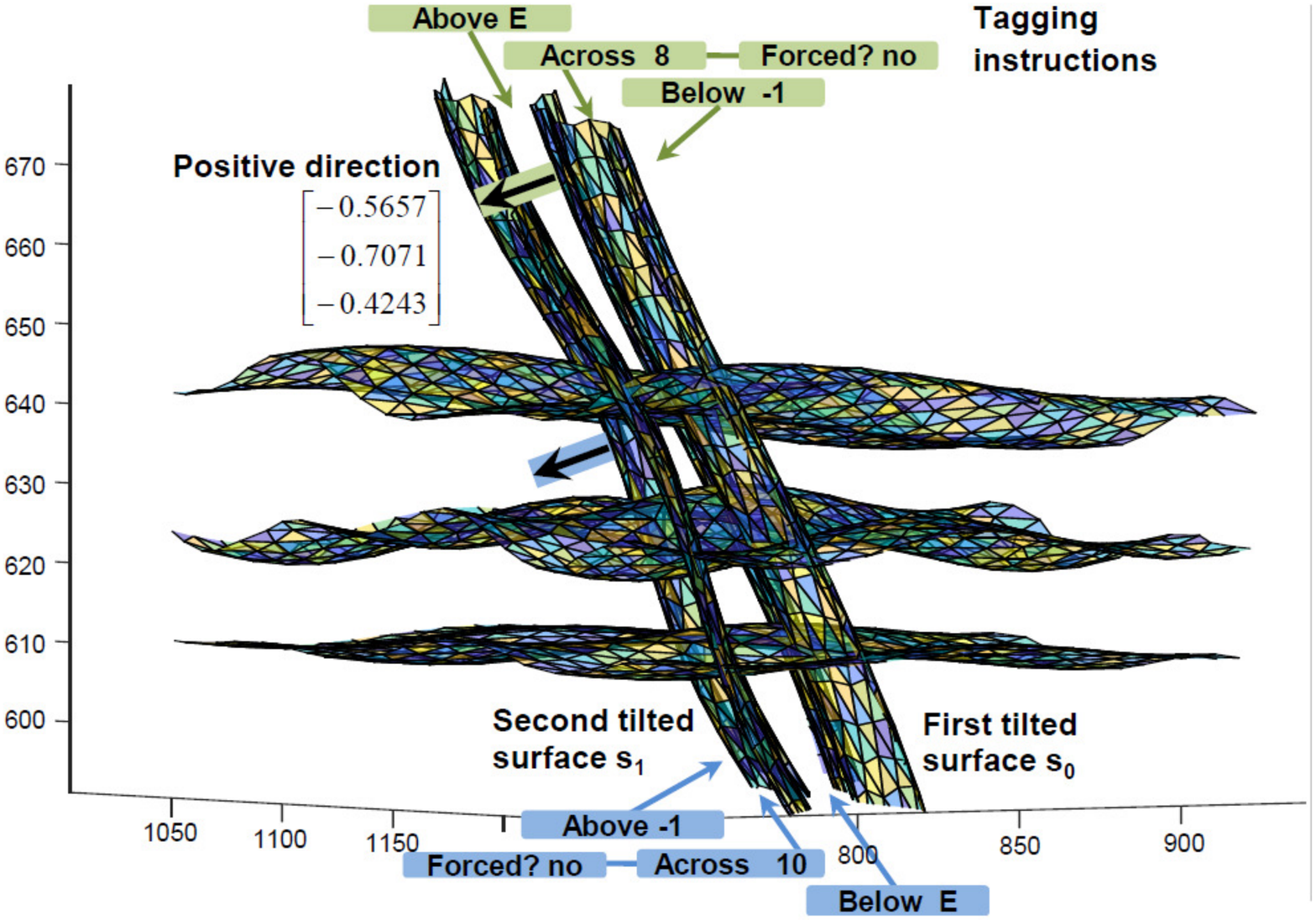}
\caption{Block tagging instructions for tilted surfaces. The labels E, 8 and 10 refer to entities shown in Fig.~\ref{fig:bsu-results-tilted-surfaces}.}
\label{fig:tagging-instructions-tilted-surfaces}
\end{figure}

The running example has thus far taken only a regular block structure as input. In this section, we demonstrate that the framework can also modify the spatial structure of a model with irregular (non-uniform) block dimensions. Of particular interest is that tilted surfaces are used to model a hypothetical dyke channel running through bedded layers. This highlights two significant features: (a) ability to iteratively improve the spatial structure of an existing block model whilst preserving the labels for horizontal strata which have been previously assigned; (b) ability to work with oblique surfaces and produce correct result when the surface orientation is ambiguous (positive may not point upward), thus user has to specify precisely what is meant by the positive direction in relation to the supplied tagging instructions. An example of this is shown in Fig.~\ref{fig:tagging-instructions-tilted-surfaces}.

Fig.~\ref{fig:bsu-results-tilted-surfaces} shows the recut block model follows the curvature of the tilted surfaces. In (k), blocks within the dyke are removed for clarity. Pre-existing labels (outside the dyke) remain intact. The spatial structure is only modified around the tilted surfaces. In Fig.~\ref{fig:bsu-results-tilted-surfaces}(l)--(m), blocks within the dyke and those located on the east\,/\,west interfaces are shown in isolation. Fig.~\ref{fig:bsu-results-tilted-surfaces}(n)--(p) show the existing labels for blocks in A, B, C and D have been perfectly preserved. As expected, new labels --- \{8, 10\} and E respectively --- have only been assigned to blocks that intersect with and sandwiched between the tilted surfaces.

\begin{table}[!t]
\setlength\tabcolsep{0pt}
\begin{center}
\scriptsize
\begin{tabular}{p{50mm}p{50mm}p{50mm}}
(k) Recut block model & (l) Blocks within the dyke (E) & (m) Blocks intersecting with\\
& & east/west tilted surfaces\\
\includegraphics[width=43.5mm]{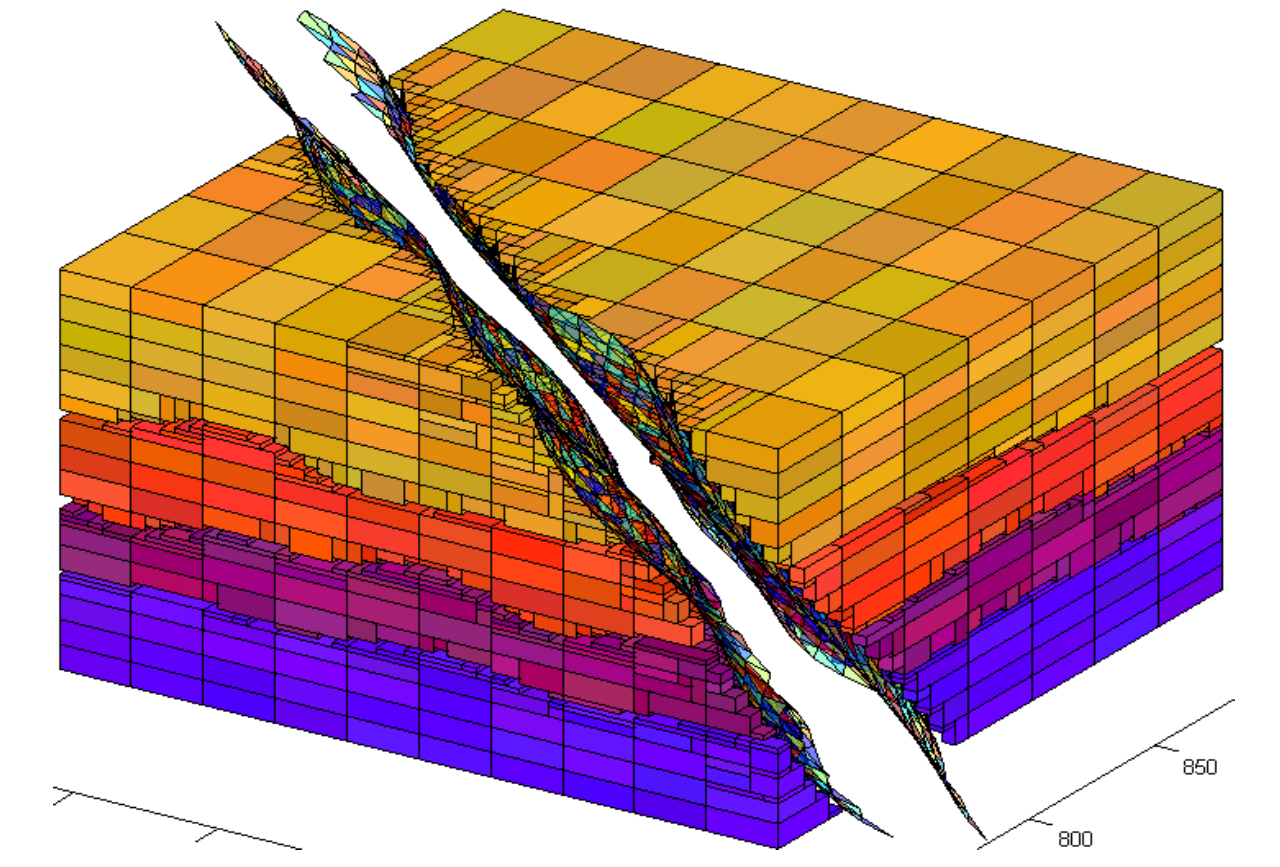} & \includegraphics[width=43.5mm]{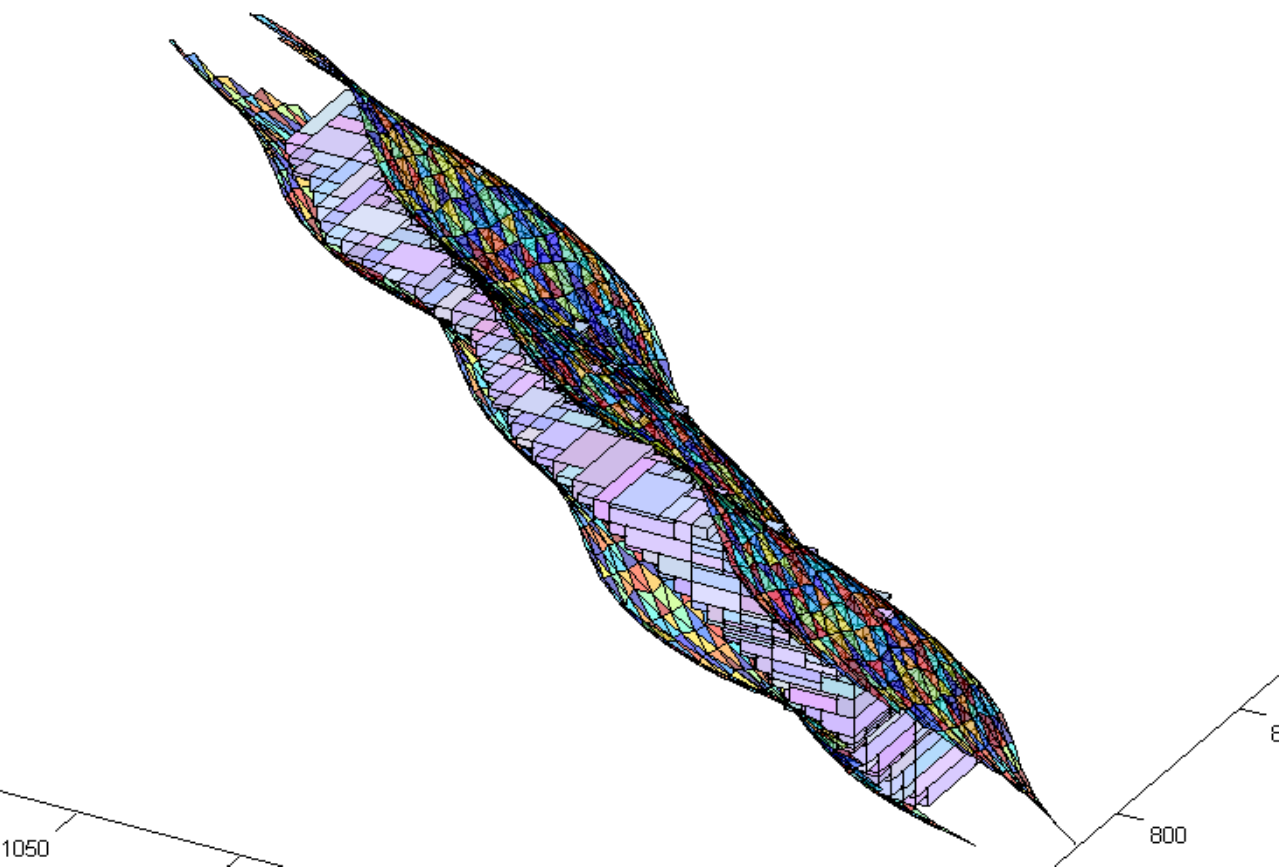} & \includegraphics[width=43.5mm]{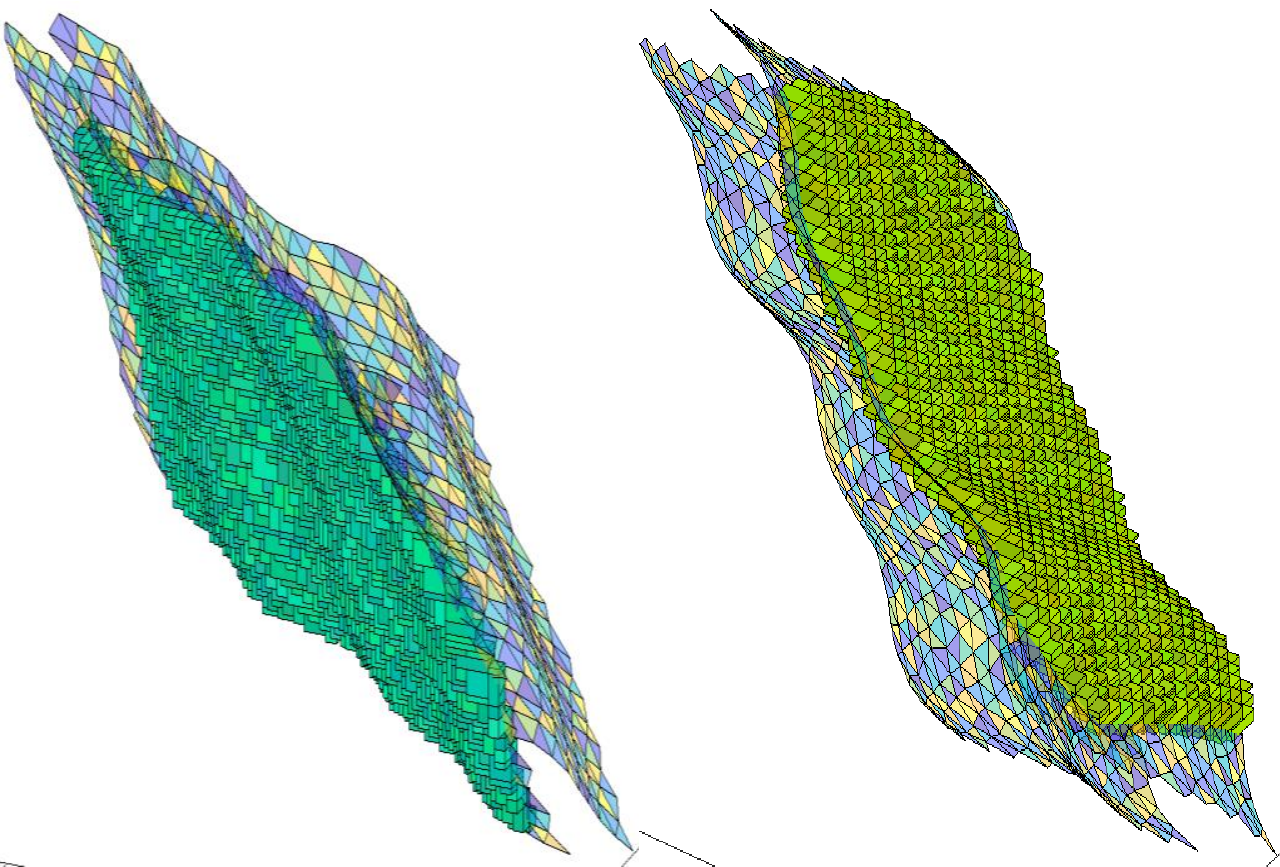}\\
(n) Blocks within the dyke & (o) Blocks in E, A and C & (p) Blocks in E, B and D\\
sandwiched in the middle. & & \\
\includegraphics[width=43.5mm]{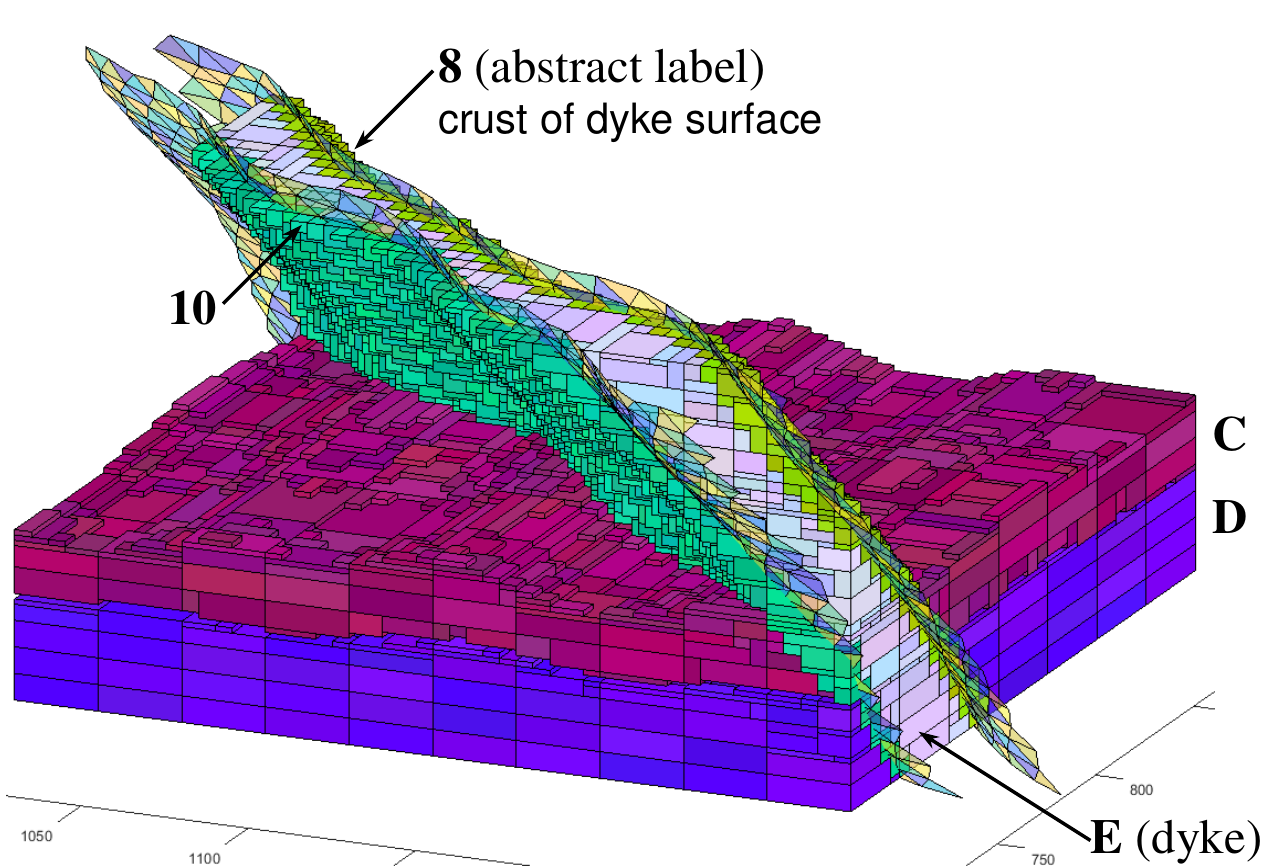} & \includegraphics[width=43.5mm]{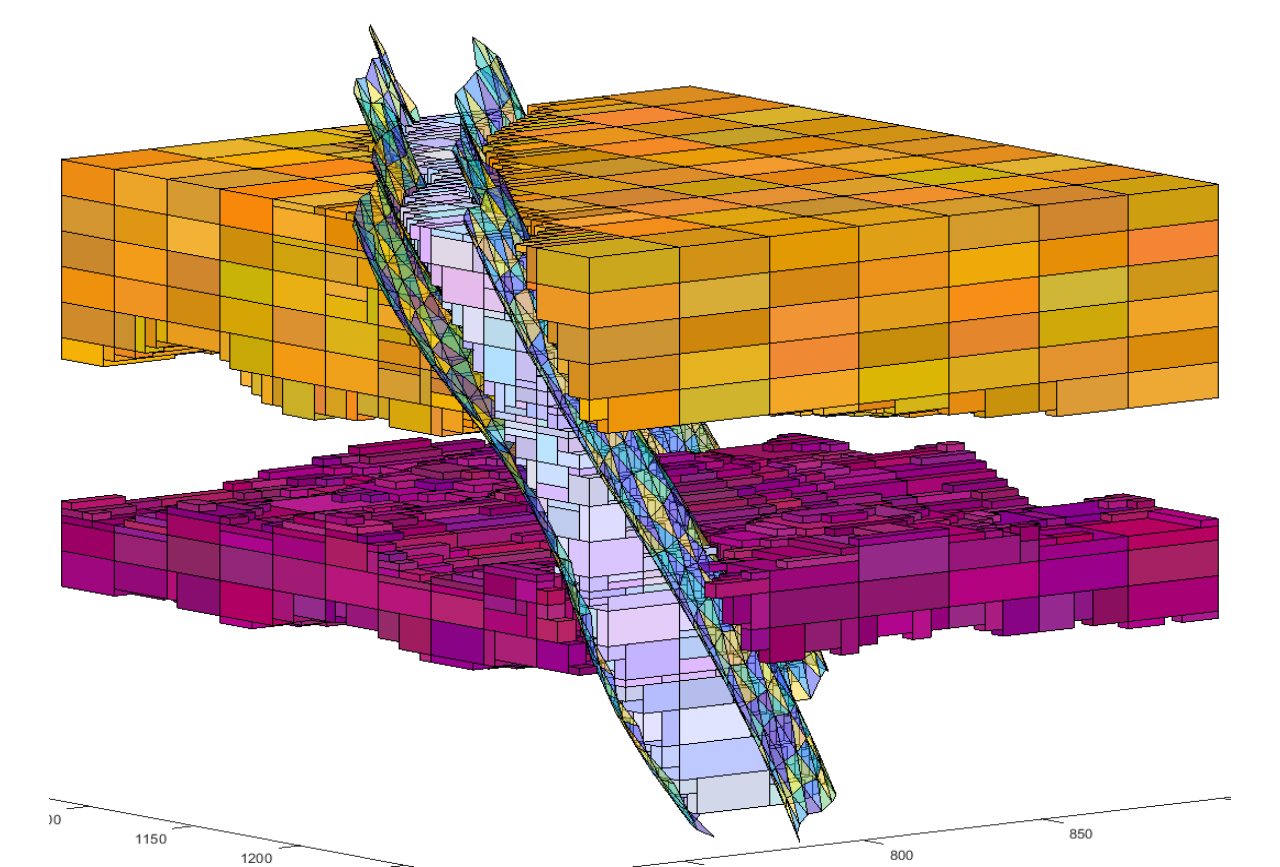} & \includegraphics[width=43.5mm]{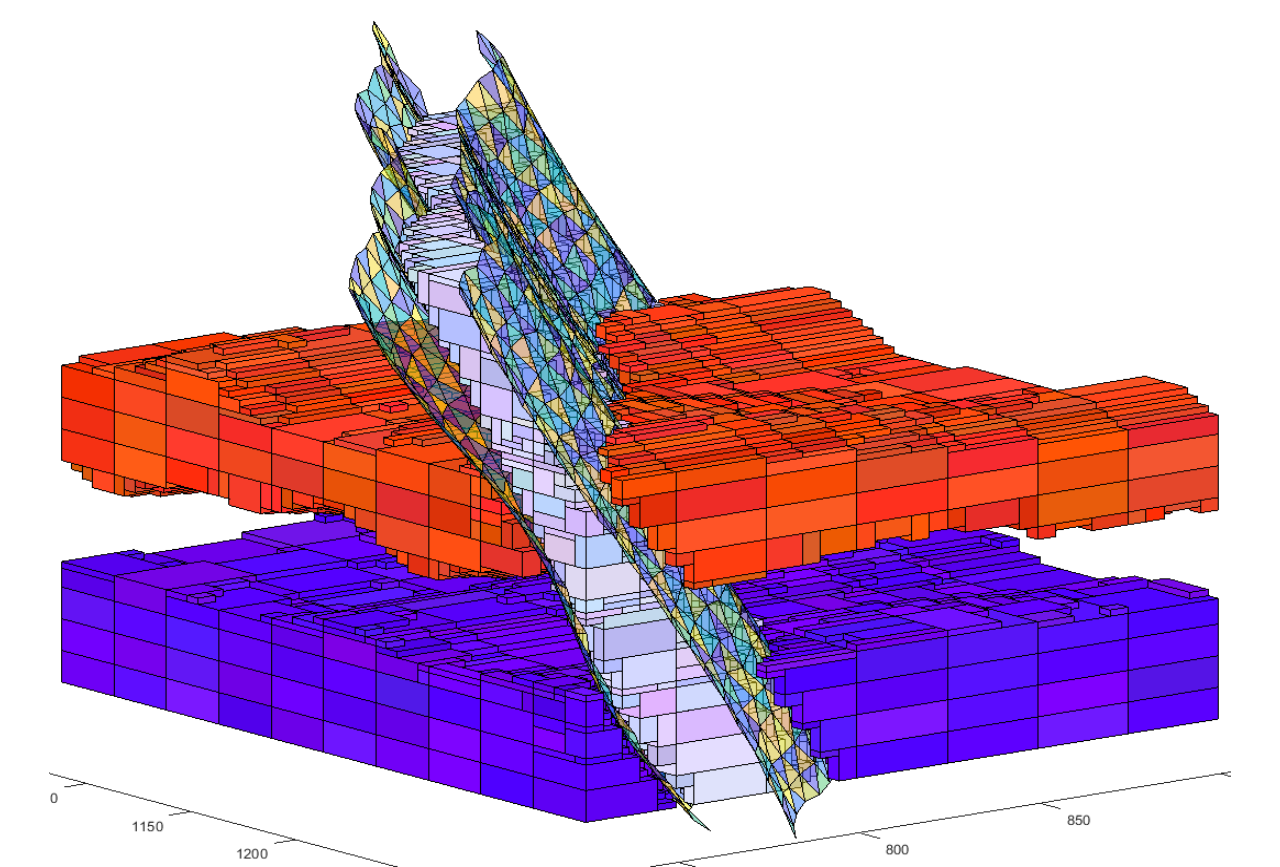}
\end{tabular}
\captionof{figure}{Iterative refinement incorporating tilted surfaces}
\label{fig:bsu-results-tilted-surfaces}
\end{center}
\end{table}

\section{Engineering perspectives and applications}\label{sec:bsu-critical-reflection}
Critical reflection and user feedback are both essential to designing robust and flexible systems \cite{arvidsson-08}. Guided by the principle of reflective and iterative design \cite{sedlmair-12}, there was an early and continued focus in our approach on real usage scenarios \cite{rosson-09}, as well as successive evaluation, modification and scenario-based testing. In this section, we describe two improvements made to eliminate flaws identified through this process which made the system more robust.

\subsection{Issue 1: Sign inversion due to sparse, jittery surface}\label{sec:issue1}
\begin{figure}[!ht]
\centering
\includegraphics[width=87.5mm]{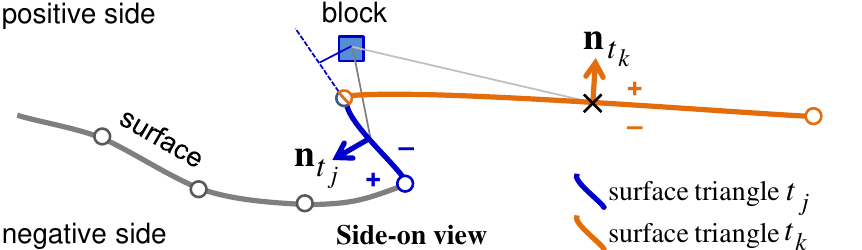}
\caption{Local sign inversion when triangle mesh surface is sparse and jittery}
\label{fig:issue-projection-onto-patch-normal-when-sparse}
\end{figure}
In regard to block tagging and `which-side-of-the-surface' determination, the `projection-onto-normal' method described in Section~\ref{sec:bsu-block-tagging} works well when the surface is smooth and triangle mesh is dense and uniform. Potential issues arise when the surface exhibits local jitters and the triangles are sparse. For instance, the mesh resolution is low relative to the parent block size when triangle patches stretch over distances of up to one kilometer in certain areas. As an illustration,  Fig.~\ref{fig:issue-projection-onto-patch-normal-when-sparse} shows a block associated with triangle $t_j$ (the nearest patch based on block-triangle centroid-to-centroid distance). This association yields the wrong result, a negative sign with respect to the normal $\mathbf{n}_{t_j}$ is obtained (according to the plane partitioning test) even though it lies above the surface. We observe the projection of the block lies outside the support interval of the referenced triangle $t_j$ and the same comparison with $t_k$ which is further away would produce the right result (a positive sign). This problem can be remedied by upsampling the mesh surface to increase its density. Fig.~\ref{fig:bsu-mesh-refinement} shows an example where triangles are recursively split along the longest edge until the maximum patch area and length of all edges fall below the thresholds of 1250m$^2$ and 100m. A better solution, however, is ray-tracing \cite{moller-05}. In general terms, ray-tracing determines whether a block is above or below an open surface (resp., inside or outside a closed surface) by counting the number of intersections between the surface and a ray casted from the block. This recommended approach is described in Appendix~\ref{sec:appendix-ray-tracing}. The key advantage is that ray tracing is not susceptible to variation in surface mesh density (the sign inversion problem due to folding), furthermore, it does not require dense surfaces or consistent (e.g. clockwise) ordering of triangle vertices. Therefore, ray-tracing will be the method of choice for block tagging moving forward. It replaces the projection-onto-normal approach in our proposal.

\begin{table}[!ht]
\setlength\tabcolsep{0pt}
\begin{center}
\scriptsize
\begin{tabular}{p{50mm}p{50mm}}
\qquad Before: non-uniform density & \qquad After: high density mesh\\
\includegraphics[width=43.5mm,trim={60mm 43mm 65mm 25mm},clip]{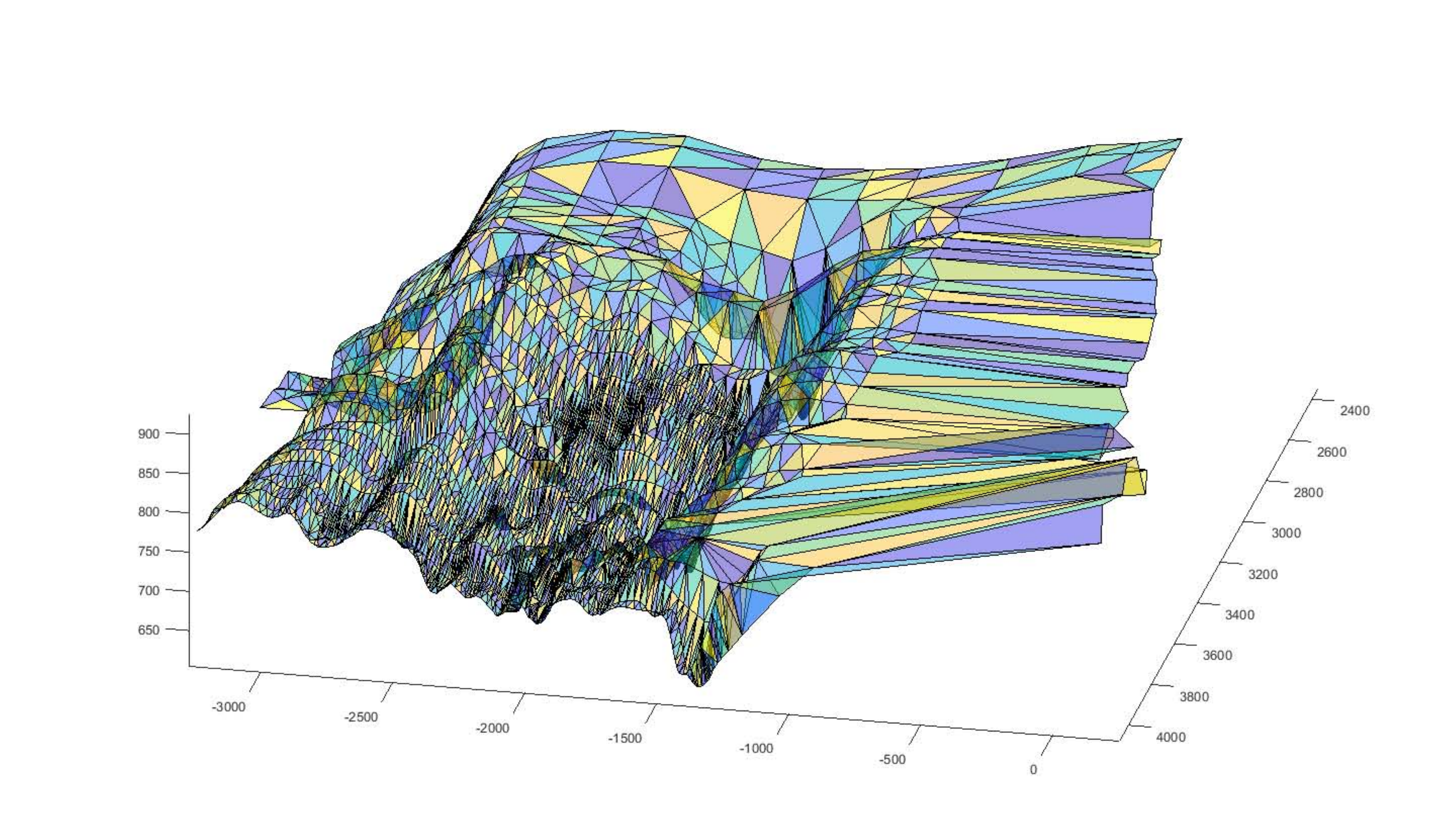} & \includegraphics[width=43.5mm,trim={60mm 43mm 65mm 25mm},clip]{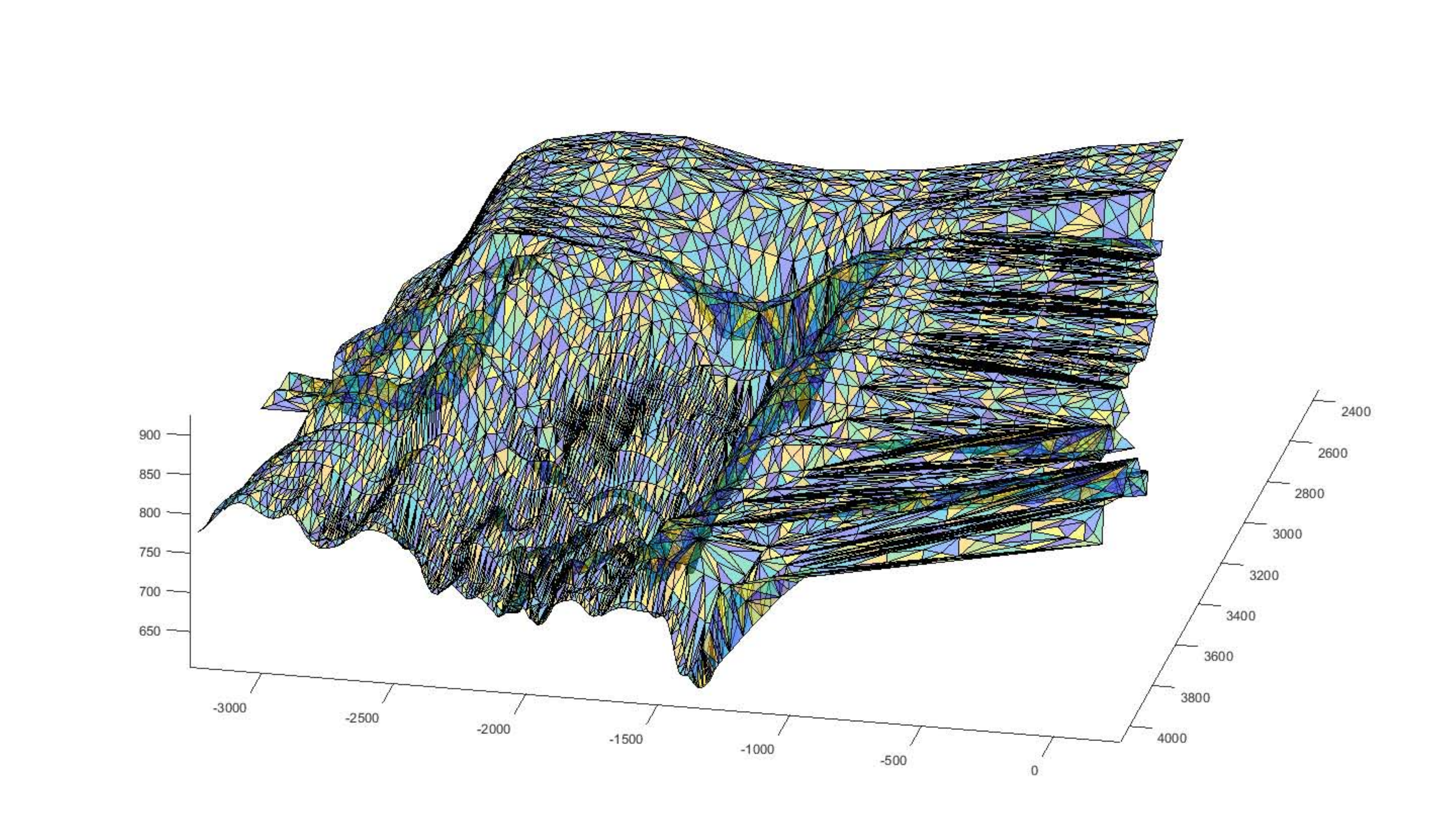}
\end{tabular}
\end{center}
\captionof{figure}{Increasing mesh density of a real surface by splitting recursively the longest edge of triangles whose area is above the threshold}
\label{fig:bsu-mesh-refinement}
\end{table}

\subsection{Issue 2: Boundary localisation accuracy}\label{sec:boundary-localisation-accuracy}
\begin{figure}[h]
\centering
\includegraphics[width=87.5mm]{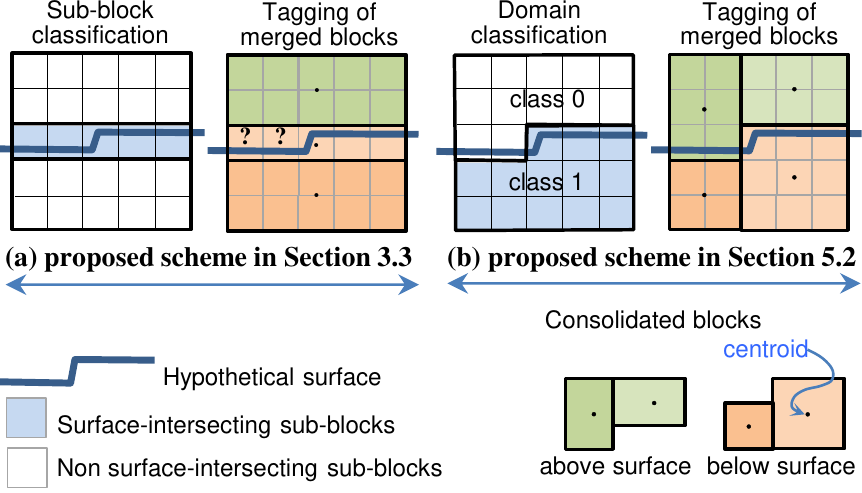}
\caption{Treating surface-intersecting and non-intersecting sub-blocks, $\mathcal{B}_\text{intersect}$ and $\mathcal{B}_\text{non-intersect}$, independently during block consolidation may reduce boundary localisation accuracy and limit merging potential. In the latest proposed scheme (Section~\ref{sec:boundary-localisation-accuracy}), sub-blocks are classified by their location relative to the surface before sub-blocks consolidation.}
\label{fig:issue-of-accuracy-compaction}
\end{figure}

The block consolidation component, as it currently stands, considers the sub-blocks that belong to the surface-intersecting set ($\mathcal{B}_\text{intersect}$) and non surface-intersecting set ($\mathcal{B}_\text{non-intersect}$) independently. As Fig.~\ref{fig:bsu-coalesce-continued2} has shown, the sub-blocks (cells) within each respective set are merged separately to form larger rectangular prisms within the confines of the parent blocks. While this split is useful for extracting surface-intersecting sub-blocks, it has two drawbacks. In terms of \textbf{boundary localisation accuracy}, Fig.~\ref{fig:issue-of-accuracy-compaction}(a) shows that surface-intersecting sub-blocks --- with centroids located on different sides of the surface --- are merged together irrespective of whether it is predominantly above or below the surface. In terms of compaction, \textbf{merging potential is limited} because adjacent cells from $\mathcal{B}_\text{intersect}$ and $\mathcal{B}_\text{non-intersect}$ cannot be coalesced even if they lie on the same side of the surface.

To reinforce the first point, the two cells marked with ``?'' in Fig.~\ref{fig:issue-of-accuracy-compaction}(a) ought to be labelled as above the surface. However, under the current regime, they are considered jointly with the three cells immediately to the right that also intersect with the surface, thus they are treated collectively as a 1-by-5 merged block. Since the centroid (black dot) lies marginally below the surface, the merged block will also be labelled as such. This distorts the boundary as it introduces a vertical bias of around $\Delta_\text{y,min}^\text{(block)}/2$ to the two left-most cells.

To accurately localise the boundary and achieve the result shown in Fig.~\ref{fig:issue-of-accuracy-compaction}(b), ray tracing is used to determine the location of cells with respect to each relevant surface that intersects the parent block. Given $S$ surfaces, the \{0=above (or no intersection), 1=below\} decisions naturally produce up to $2^S$ categories (or states) which would be treated separately during sub-blocks consolidation in lieu of $\mathcal{B}_\text{intersect}$ and $\mathcal{B}_\text{non-intersect}$. In practice, however, we suggest encoding the location with respect to each surface $s$ using 3 bits $b_{3s+2}b_{3s+1}b_{3s}$, where the mutually exclusive bits are set to 1 to denote the outcomes of \textit{above}, \textit{below} and \textit{untested}, respectively. The rationale is that when $b_{3s}=0$, domain identification needs not be attempted during block tagging with respect to surface $s$, since the decision has already been made here prior to sub-blocks consolidation (either $b_{3s+2}=1$ or $b_{3s+1}=1$) and this information is passed on. In fact, applying ray-casting at the cellular-level (highest resolution) yields more accurate results near the surface than applying to merged blocks, especially for undulating surfaces with high local curvature.

To summarise, administering ray-casting before sub-blocks consolidation helps divide cells along surface boundaries; this increases boundary localisation accuracy and ensures merging is performed within the right domains with maximum potential. Table~\ref{tbl:bsu-compare-techniques} provides a comparison of the techniques discussed.

\begin{table}[!ht]
\setlength\tabcolsep{0pt}
\begin{center}
\small
\begin{tabular}{l}
\includegraphics[width=92.5mm]{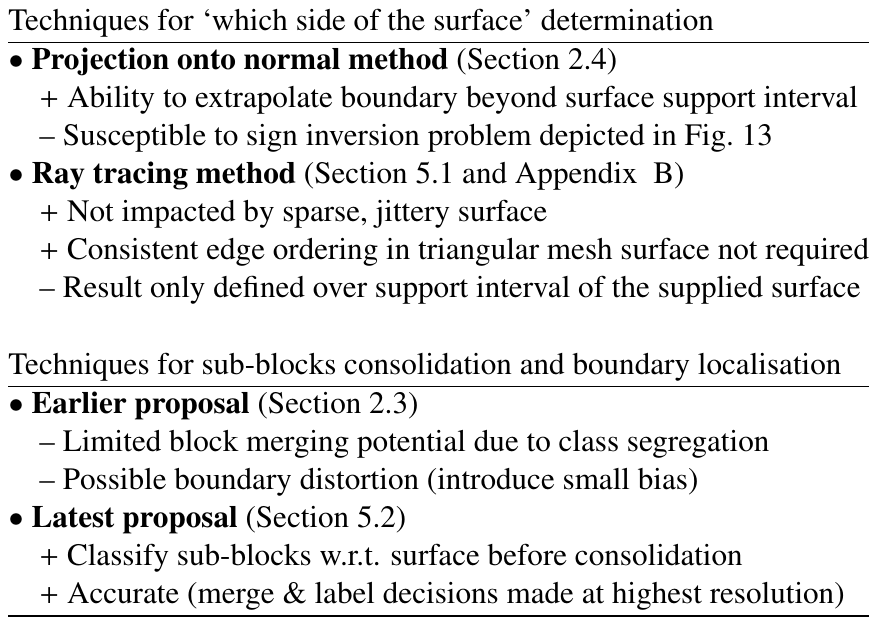}
\ignore{
Techniques for `which side of the surface' determination\\ \hline
\textbullet\ \textbf{Projection onto normal method} (Section~\ref{sec:bsu-block-tagging})\\
\quad + Ability to extrapolate boundary beyond surface support interval\\ 
\quad -- Susceptible to sign inversion problem depicted in Fig.~\ref{fig:issue-projection-onto-patch-normal-when-sparse}\\
\textbullet\ \textbf{Ray tracing method} (Section~\ref{sec:issue1} and Appendix~\ref{sec:appendix-ray-tracing})\\
\quad + Not impacted by sparse, jittery surface\\
\quad + Consistent edge ordering in triangular mesh surface not required\\
\quad -- Result only defined over support interval of the supplied surface\\ \\
Techniques for sub-blocks consolidation and boundary localisation\\ \hline
\textbullet\ \textbf{Earlier proposal} (Section~\ref{sec:bsu-subblocks-consolidation})\\
\quad -- Limited block merging potential due to class segregation\\
\quad -- Possible boundary distortion (introduce small bias)\\
\textbullet\ \textbf{Latest proposal} (Section~\ref{sec:boundary-localisation-accuracy})\\
\quad + Classify sub-blocks w.r.t. surface before consolidation\\
\quad + Accurate (merge \& label decisions made at highest resolution)\\ \hline
}
\end{tabular}
\end{center}
\captionof{table}{Comparison of techniques with emphasis on system robustness}
\label{tbl:bsu-compare-techniques}
\end{table}

\begin{table}[!ht]
\setlength\tabcolsep{0pt}
\begin{center}
\footnotesize
\begin{tabular}{c}
\includegraphics[width=123.5mm,height=56.5mm]{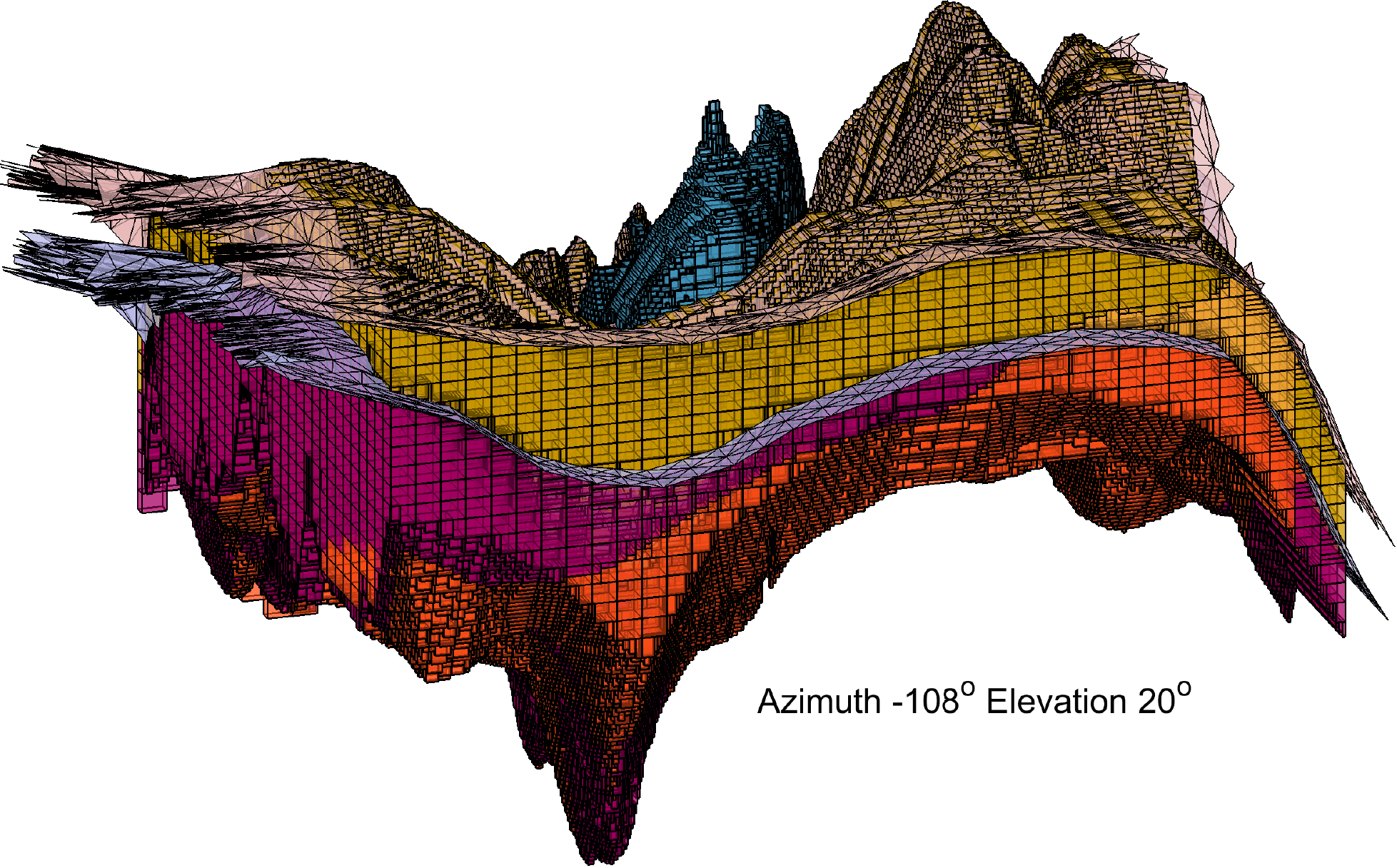}\\
\includegraphics[width=116.5mm,height=49.5mm]{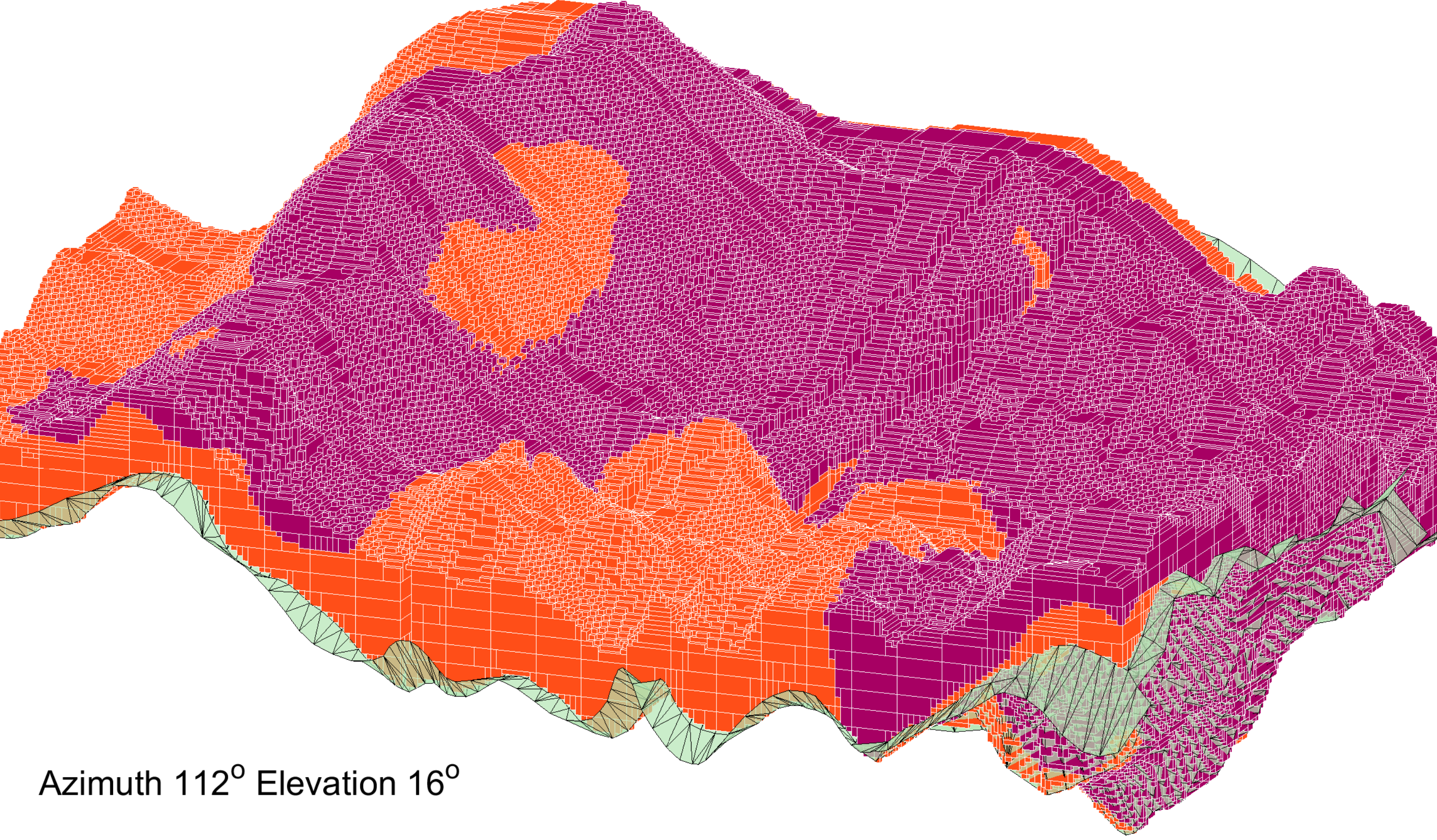}
\end{tabular}
\end{center}
\captionof{figure}{Typical block model spatial restructuring results. Top: blocks partitioned by surfaces into different domains (not all surfaces are shown). Bottom: reveals two block sets with different levels of mineralisation in an ore deposit. In the \textsc{pdf} article, zoom in to see individual blocks.}
\label{fig:bsu-typical-results}
\end{table}
\newpage\subsection{Applications}\label{sec:applications}
The first application we preview relates to block modelling for a typical mine site located in Western Australia. The block model spatial restructuring results shown in Fig.~\ref{fig:bsu-typical-results} illustrate a block-wise partitioning of the mine site into different domains. This is achieved using surfaces which were created to separate geological domains (mineralised, hydrated and waste) within the Brockman Iron Formation which contains members of interbedded BIF and shale bands in the Hamersley Basin Iron Province \cite{clout-06}. Although ore-genesis theories vary depending on the minerals or commodity, the ability to model formations and features such as igneous intrusions in ore deposits is of general interest in areas not limited to mining, but also in further understanding the structural geology of mineral deposits. Using open and closed surfaces to represent structures of varying complexity --- this may encompass volumes with exceptional geochemical or geophysical attributes --- it is possible to extract waste pockets with high concentration of trace elements, or regions with magnetic\,/\,gravity anomaly \cite{wang-15}. Equally, if the surfaces represent the boundary of aquifers separated by clay and lignite seams \cite{wycisk-09}, the process may serve as a basis for creating a structural hydrogeological model to study hydraulic and transport conditions in geotechnical or environmental risk assessment. The techniques developed for shaping a 3D block model can be used potentially in a variety of contexts, including surface buffer analysis (in GIS and structural modelling) for triangle mesh 3D boundary representation of localised objects \cite{li2017irregular}. In the next section, we focus on a specific application of block merging to reduce spatial fragmentation in a block model.

\section{Block merging to reduce spatial fragmentation}\label{sec:block-merge-reduce-fragmentation}
\begin{figure}[!ht]
\centering
\includegraphics[width=87.5mm]{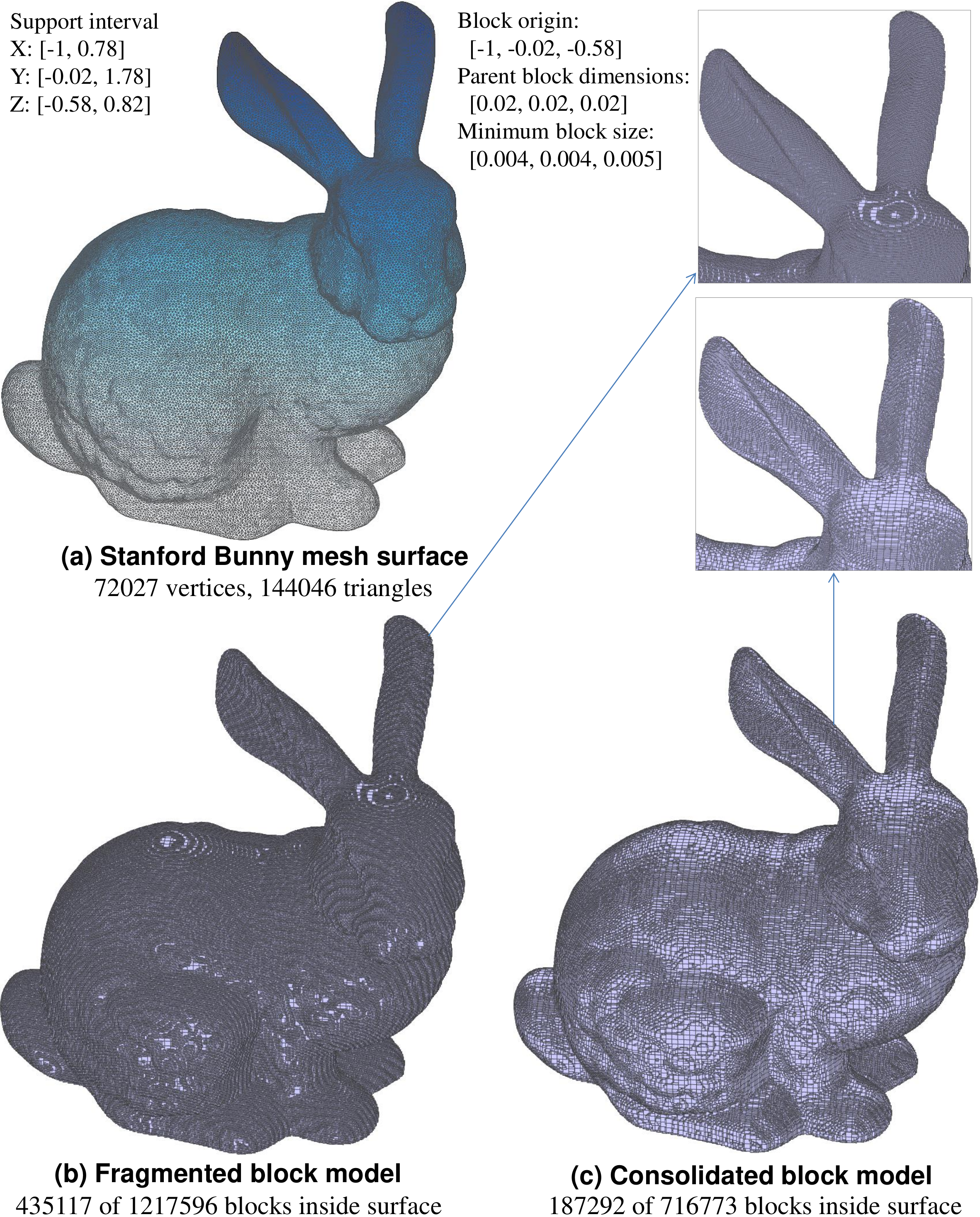}
\caption{Block merging applied to Stanford Bunny to reduce block fragmentation. Zoom in to see individual blocks.}
\label{fig:bsu-stanford-bunny}
\end{figure}

The proposed block merging algorithm can also be used to consolidate a fragmented block model that exists with or without reference to any mesh surface. Spatial fragmentation is used in this context to mean a highly redundant block model representation where blocks near the boundary are over-segmented or divided in an excessive manner to follow the curvature of a surface without regard for the compactness (total block count) of the model. As an illustration, Fig.~\ref{fig:bsu-stanford-bunny}(b) shows a highly fragmented block model for the \textit{Stanford Bunny} created by block decomposition without consolidation. In an effort to closely approximate the surface, numerous blocks at the minimum block size were produced near the surface. Fig.~\ref{fig:bsu-stanford-bunny}(c) shows a clear reduction in block density as blocks are appropriately merged. This results in a more compact block representation (3D segmentation) of the object.

The block merging algorithms are formally described in the Supplementary Material wherein a number of technical issues are discussed in depth. At its core, a notable feature is that `feasibility of cell expansion' is determined using a multi-valued (rather than boolean) 3D occupancy map, where the values correspond to the identity of the cells or sub-blocks. The spatial constraints governing cell expansion are somewhat different, in particular, the lateral dimensions orthogonal to the axis of expansion have to match for all the blocks involved in a merge. These, along with other relevant considerations and implementation details, are described in Appendix~\ref{sec:appendix-cell-expansion-feasibility}--\ref{sec:appendix-bsu-cell-expansion-feasibility-test}.

\subsection{Merging conventions and optimisation objectives}\label{sec:bsu-merging-conventions-optimisation-objectives}
\begin{figure}[!ht]
\centering
\includegraphics[width=146mm]{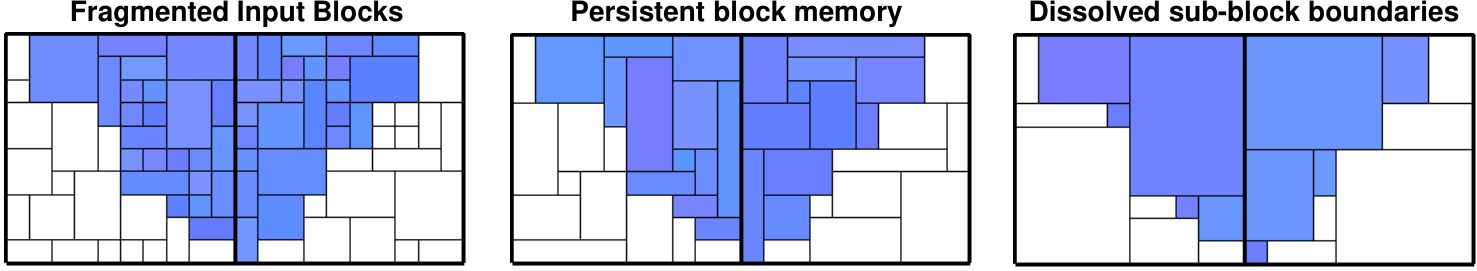}
\caption{Differences between two block merging conventions: persistent block memory vs dissolved sub-block boundaries in 2D}
\label{fig:persistent-vs-dissolved-raster}
\end{figure}

\begin{figure}[!ht]
\centering
\includegraphics[width=120mm]{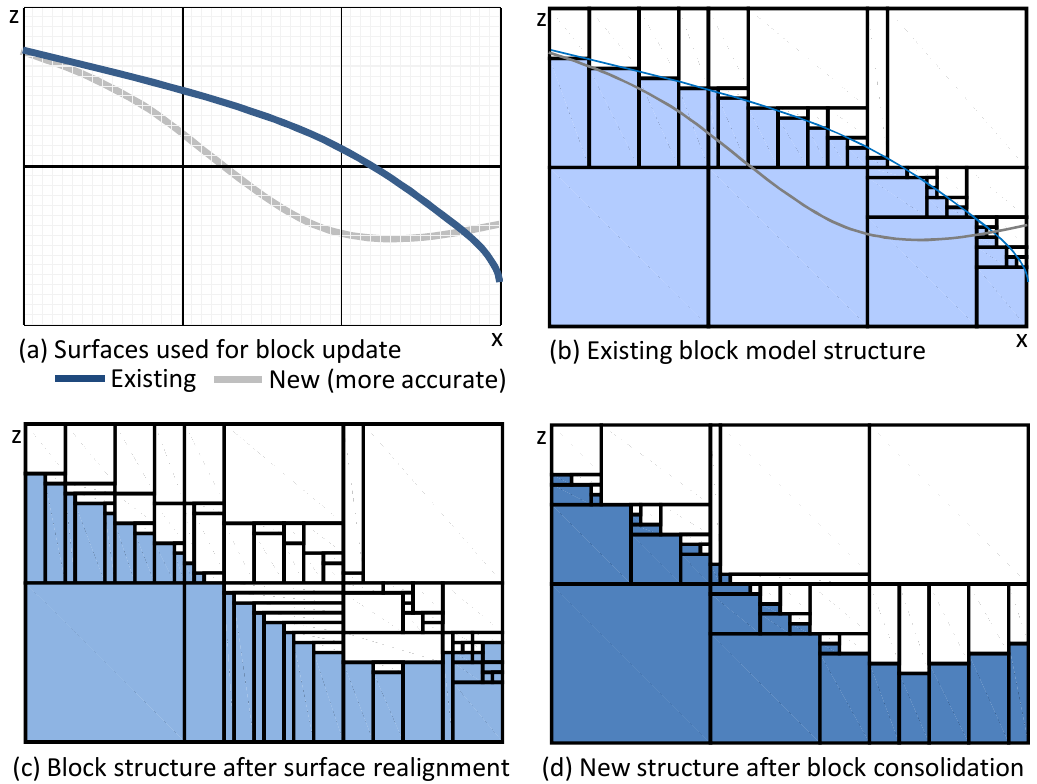}
\caption{Example of boundary healing through block consolidation --- refer to description in main text}
\label{fig:boundary-healing}
\end{figure}

The block merging algorithms also recognise that merging can be performed under different conventions. For example, in Algorithm~\ref{algo:bsu-coordinate-ascent2}, the procedure preserves the input block boundaries, it does not introduce new partitions (sub-divisions) that are not already present in a parent block. This merging convention is referred as \textbf{persistent block memory}. It has the property that each input block is mapped uniquely to a single block in the merged model. In contrast, Algorithm~\ref{algo:bsu-coordinate-ascent} implicitly erases the sub-block boundaries before block consolidation begins. This merging convention is referred as \textbf{dissolved sub-block boundaries}, it generally achieves higher compaction because it makes no distinction between input blocks from the same class and parent. It is able to grow blocks more freely and produce fewer merged blocks since the size compatibility constraints between individual blocks no longer apply when internal sub-block boundaries are ignored. These differences are illustrated in Fig.~\ref{fig:persistent-vs-dissolved-raster}.
The \textit{dissolved sub-block boundaries} convention can be useful for healing a fractured block model, for instance, over a region where a false geological boundary was given in a previous surface update. Under the \textit{dissolved sub-block boundaries} convention, coordinate-ascent can start from a clean slate and merge sub-blocks in a fragmented area back to the fullest extent in cases where individual sub-block dimensions or internal boundary alignments are otherwise incompatible. It is not bound by the consequences of prior model restructuring decisions. The technical details are given in  Appendix~\ref{sec:appendix-merging-convention}. For readers simply looking for a basic explanation, Fig.~\ref{fig:boundary-healing} reinforces these points and illustrates what \textit{healing} really means in practice.

Fig.~\ref{fig:boundary-healing}(a) shows an existing (poorly estimated) boundary created using limited data at an earlier point in time. Fig.~\ref{fig:boundary-healing}(b) shows the existing block model structure created using this surface which unbeknown to the user is not faithful to the actual boundary shown in gray. At a later point in time (perhaps after several months have elapsed), more assay measurements have been taken at previously unsampled locations. These denser observations help improve the estimated boundary contact points and a new (more accurate) surface is produced as a result; see bottom curve in (a). A second iteration of the block model spatial restructuring process is applied using the new surface. This workflow significantly improves the boundary localisation property of the new block model in (c), however it leaves behind remnant sub-blocks from the previous iteration. By relaxing the internal constraints essentially by dissolving the sub-block boundaries, block merging under the \textit{dissolved sub-block boundaries} convention helps promote healing in fractured areas. As evident in (d), fragmented blocks are coalesced into larger blocks in areas where the misplaced boundary (blue surface) once occupied.

The final block merging strategy (see Algorithm~\ref{algo:bsu-final-block-merging} in Appendix~\ref{sec:appendix-pseudocode}) can optimise results with respect to different objectives. For instance, it can minimise the block count or avoid extremely elongated blocks by optimising for the block aspect ratio. It also exploits symmetry by following different scan patterns to alleviate ordering effects associated with using a fixed starting point. These issues are further discussed in Appendixes~\ref{sec:appendix-scan-sequences} and \ref{sec:appendix-scan-implementation}. 

\subsection{Pseudocode}\label{sec:block-merge-pseudocode-pointer}
The block merging algorithms are formally described in Appendix~\ref{sec:appendix-pseudocode} which constitutes part of the Supplementary Material.

\subsection{Evaluation}\label{sec:experiments-evaluation}
Two experiments were performed targeting two different usage scenarios. The first experiment described in Sec.~\ref{sec:bsu-block-merge-results} uses only the block merging algorithms to reduce block fragmentation in an existing block model where mesh surfaces are not available and the minimum block size has been fixed by the supplier of the model a priori. This restricted setting requires only the use of the sub-blocks consolidation component depicted in the system diagram (see Fig.~\ref{fig:bsu-framework}). The main objective is to analyse its performance in terms of block compaction and multi-threaded execution.

The second experiment described in Sec.~\ref{sec:bsu-proposed-vs-octree} provides an end-to-end evaluation of the entire system (from block model creation to block tagging) which applies sub-block consolidation and ray-tracing in an iterative setting. This involves over 80 surfaces and creates approximately 30 different domains. The goal is to compare the proposed strategies with two hierarchical subblocking techniques based on octrees, to highlight the constraints of dyadic decomposition and the importance of inter-scale block merging.

\section{Experiment 1: Block merging efficacy on a mining resource block model}\label{sec:bsu-block-merge-results}
The proposed block merge strategy (Algorithm~\ref{algo:bsu-final-block-merging}) was implemented in C++ (with boost python bindings) and evaluated using a real model developed for a Pilbara iron ore mine in Western Australia. The model contains 697,097 input blocks of varying sizes\footnote{There were 200 unique input block dimensions, these range from $(5,5,1)$ to $(50,50,2)$ with x and y varying in increments of 5.} spanning 1342 parent blocks with dimensions $(200,200,20)$. The input block model is given as is without associated surfaces. The minimum block dimensions are also fixed by the user at $(5,5,1)$. A key requirement is that blocks smaller than $(5,5,1)$, or being some fractional multiples of it, must not be introduced in the output. This is guaranteed by the proposed system.

The metrics of interest are the block aspect ratio, merged block count and execution times with multi-threading. The \nobreak{\textit{AspectRatio}}($m$) obtained using method $m$ is defined by $f_{\pi^*}(\{\boldsymbol{\Delta}^{(b,\pi^*)}\}_{b\in\mathcal{S}_{\mathbf{p},\lambda}})$ in (\ref{eq:min-aspect-ratio-objective}). For comparison, a python implementation that uses a greedy merging strategy based on block edges connectivity is chosen to establish a baseline. Its sub-blocking approach is based on prime number factorisation. Given parent blocks with cell dimensions $(k_x,k_y,k_z)=(40,40,20)=2^2\cdot 5\times (2,2,1)$, it seeks to grow cells into subblocks with dimensions $2^n$ for $n\in\mathbb{Z}$ or 5 and merge iteratively.\footnote{This heuristic is clearly suboptimal since 3 does not appear in the prime factorisation of the cell dimensions, it is not possible for sub-blocks with $k_x,k_y,k_z$ of 3 to emerge in the first pass.}\ignore{The software for this is commercially sensitive and thus cannot be released. However, a formal evaluation against the well established octree hierarchical sub-blocking approach will be presented in Sec.~\ref{sec:bsu-proposed-vs-octree}.}

\newpage\subsection{Statistical perspective}\label{sec:bsu-results-statistical}
\begin{figure}[!ht]
\centering \includegraphics[width=87.5mm]{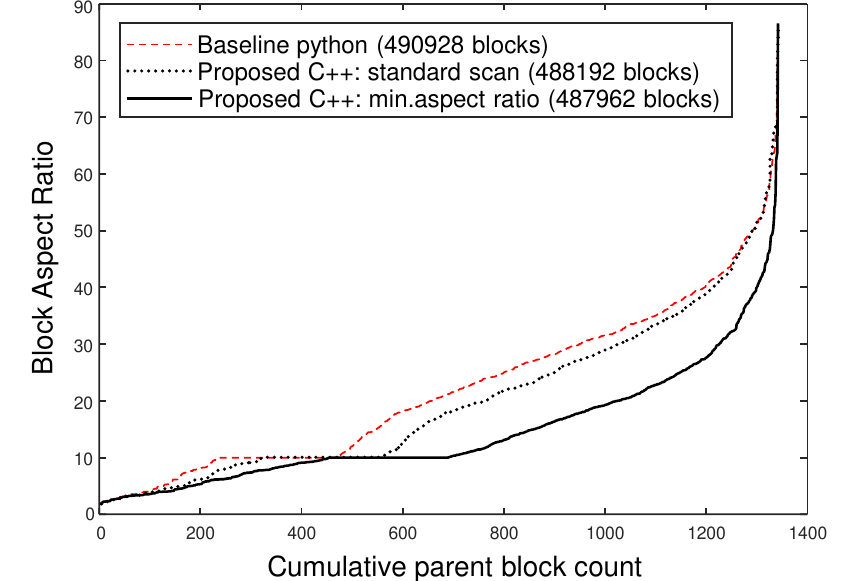}
\caption{Inverse cumulative distribution of volume-weighted block aspect ratio vs parent block count}
\label{fig:bsu-icdf-aspect-ratio-vs-block-count}
\end{figure}

Fig.~\ref{fig:bsu-icdf-aspect-ratio-vs-block-count} illustrates the block merging performance of Algorithm~\ref{algo:bsu-coordinate-ascent2} vs the baseline with emphasis on block aspect ratios. In this plot, performance is measured in terms of block aspect ratio. The volume-weighted aspect ratio is computed for each parent block and sorted in ascending order. Large aspect ratios correspond to long narrow blocks whereas small aspect ratios indicate more balanced, less extreme block dimensions which are more preferred in a mining context.  In this plot, the lower the curve, the more desirable it is. The main observation is that the proposed block merge strategy is superior to the baseline. Even when the coordinate-ascent procedure follows only the standard scan pattern, it produces higher quality merged blocks (dotted line is below the dash line). The performance margin increases significantly when multiple scans are deployed in Algorithm~\ref{algo:bsu-final-block-merging} to minimise the aspect ratio explicitly (note: solid line is consistently below the dash line).

\subsection{Spatial perspective}\label{sec:bsu-results-spatial}
\begin{figure}[!ht]
\centering
\includegraphics[width=87.5mm]{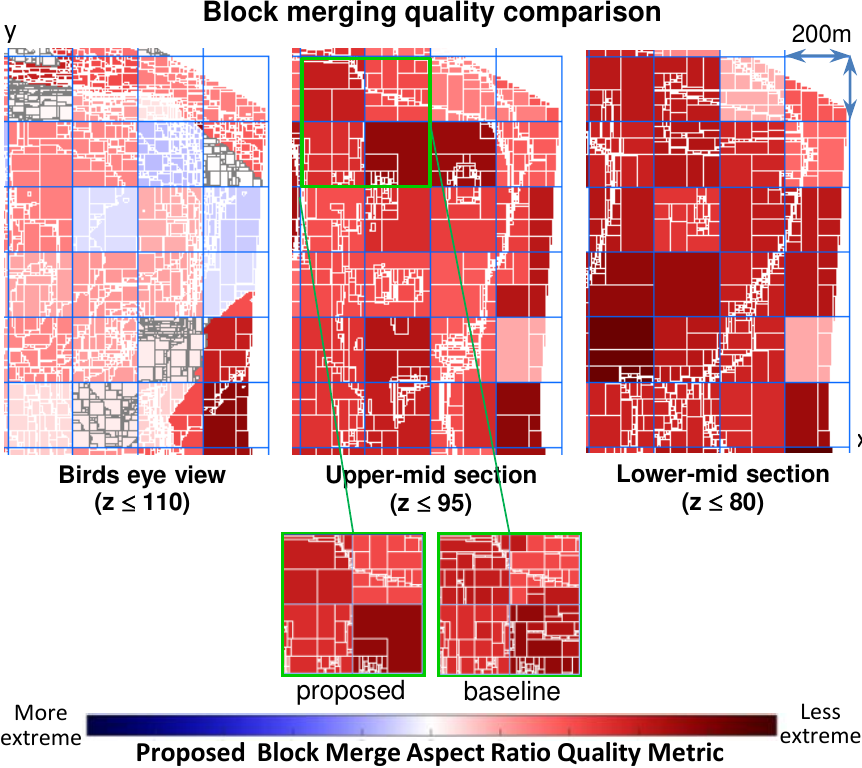}
\caption{Cross-sections of merged block model coloured by contrast in block aspect ratio}
\label{fig:bsu-xsections-block-merge}
\end{figure}

\begin{figure}[!ht]
\centering
\includegraphics[width=145mm]{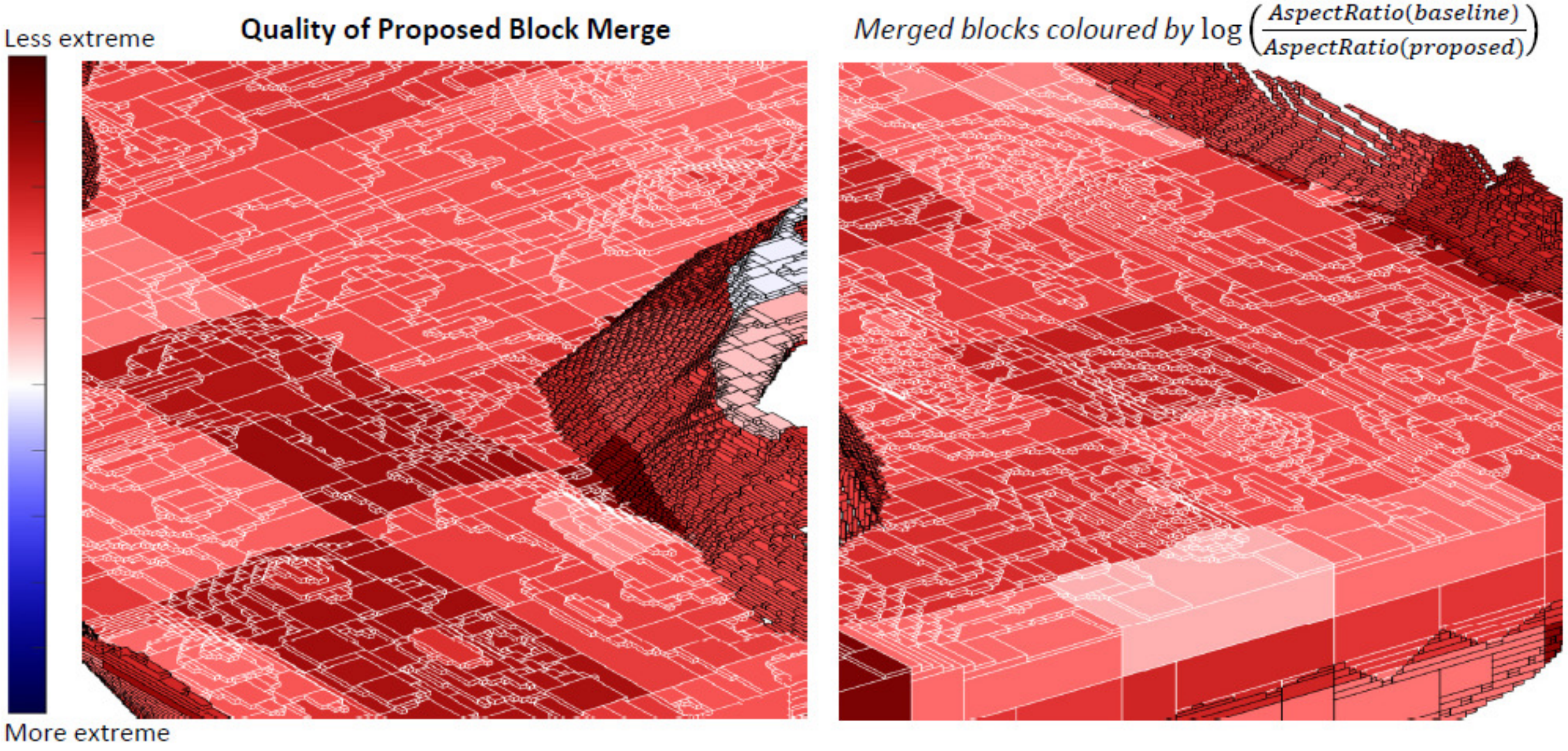}
\caption{Spatial perspective of merged block model in regions of interest}
\label{fig:bsu-perspective-block-merge}
\end{figure}

Fig.~\ref{fig:bsu-xsections-block-merge} provides an alternative view of the same result from a spatial perspective. A birds eye view and two cross-sections from the \textit{proposed} method are shown. Blocks are coloured to highlight differences in aspect ratio between the baseline and proposed method. The darker the red, the more superior is the proposed method. Conversely, blue blocks show the proposed method is inferior. The dominance of the red blocks show the proposed block merge method is able to consistently produce blocks with less extreme aspect ratios across the site. A magnified view for two regions of interest are shown in Fig.~\ref{fig:bsu-perspective-block-merge}.

\subsection{Local perspective}\label{sec:bsu-results-local}
\begin{figure}[!ht]
\centering
\includegraphics[width=87.5mm]{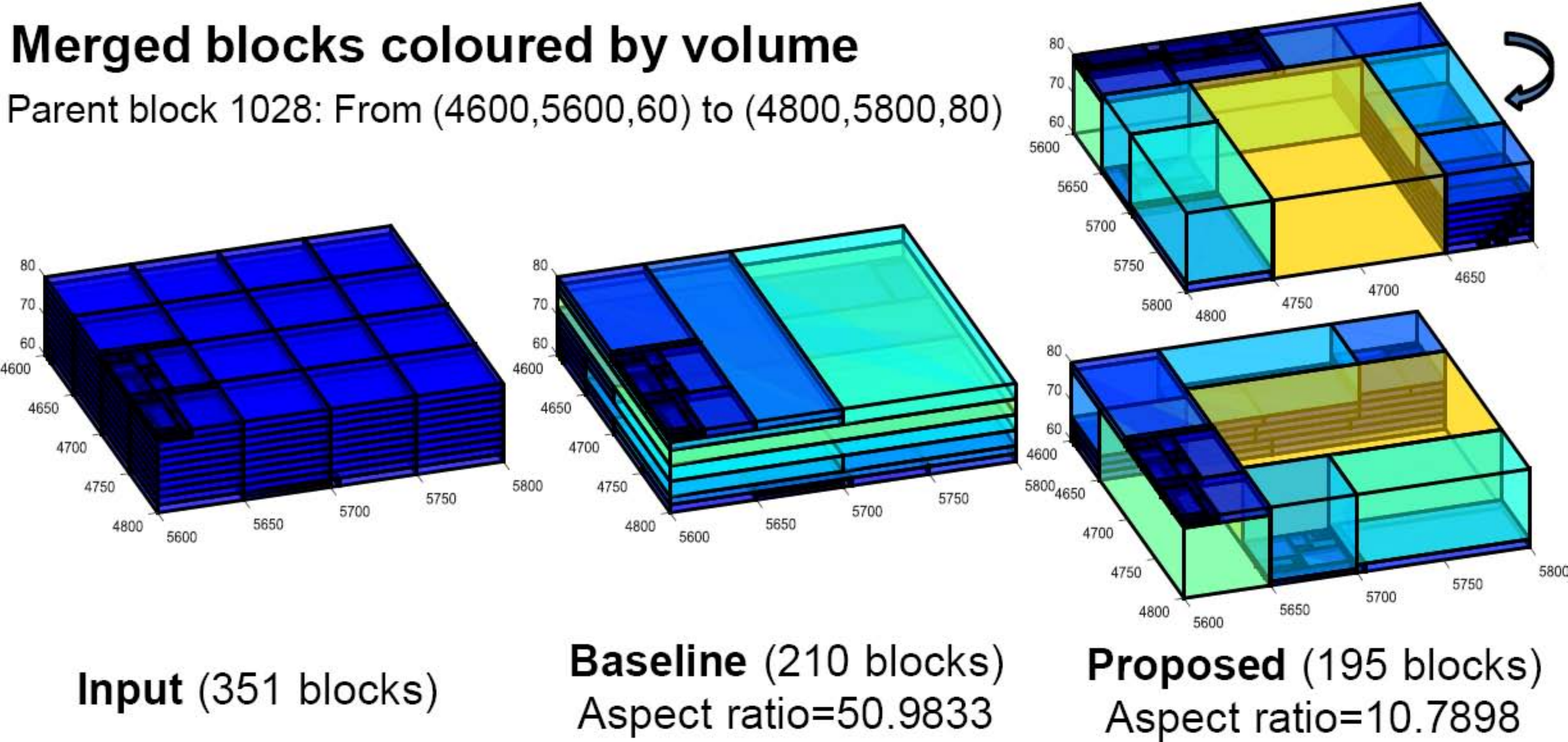}
\caption{Block merging local comparison 1 --- an instance where the log-contrast in block aspect ratio, $\log(AR_\text{baseline}/AR_\text{proposed})$, is highest amongst all parent blocks.}
\label{fig:bsu-result-local-comparison1}
\end{figure}

To verify these results, we compare the baseline and proposed method for parent blocks with a high contrast in $\log\left(\frac{AspectRatio(\text{baseline})}{AspectRatio(\text{proposed})}\right)$. Fig.~\ref{fig:bsu-result-local-comparison1} shows the sub-block structure within a parent block of interest, it shows from left to right: the input blocks, and merge results from the baseline and proposed methods. The key observation is the presence of thin narrow blocks in the baseline result; these disappeared under the proposed method. Algorithm~\ref{algo:bsu-final-block-merging} discovers block merging opportunities missed by the baseline method. It is able to minimise the block aspect ratio by conducting multiple scans during coordinate-ascent.

\begin{figure}[!ht]
\centering
\includegraphics[width=87.5mm]{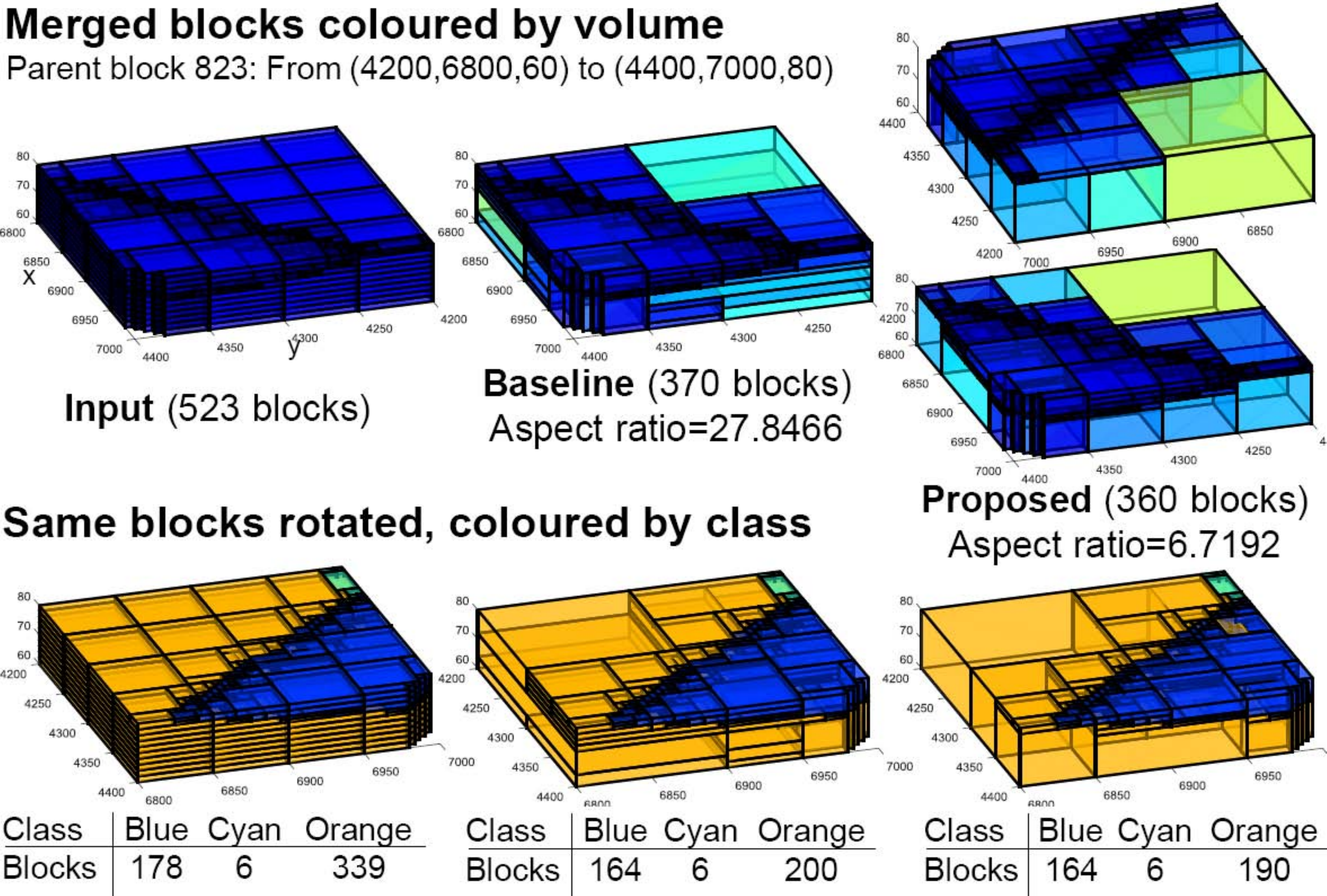}
\caption{Block merging local comparison 2 -- a second instance where the log-contrast in block aspect ratio is amongst the highest over all parent blocks.}
\label{fig:bsu-result-local-comparison2}
\end{figure}

Fig.~\ref{fig:bsu-result-local-comparison2} shows a second example. In the top panel, blocks are coloured by block volume such that bigger blocks are coloured in warm colours (yellow, for instance). This shows the proposed method is more effective at consolidating smaller blocks into larger blocks. The bottom panel shows the merge result for the same parent block from a different vantage point. Here, blocks are coloured by domain labels. This illustrates an interesting case where there is a limit to how much merging can be achieved because input blocks from different classes are processed independently even when they share the same parent.

\newpage\subsection{Block compaction}\label{sec:bsu-results-compaction}
Table~\ref{tab:bsu-block-merge-irregular-input-block-count-comparison} reports the number of output blocks produced by the baseline and proposed methods. The \textit{block count} column shows the proposed method is more efficient at coalescing blocks than the baseline method, as the block count under ``persistent block memory'' is lower. The block count reported for the proposed method under ``dissolved sub-block boundaries'' is lower still, this is consistent with our expectation and the reasoning given in Section~\ref{sec:appendix-merging-convention} based on constraint relaxation.

\begin{table}[!ht]
\captionof{table}{Block merging given irregular blocks: output block count comparison}
\label{tab:bsu-block-merge-irregular-input-block-count-comparison}
\small
\begin{center}
\begin{tabular}{|c|c|c|c|}\hline
merge method & convention & block count & relative \% \\ \hline
Input & -- & 697,097 & 100.0\%\\ \hline
Baseline & -- & 490,928 & 70.42\%\\ \hline
Proposed & persistent\,$\dag\star$ & 487,962 & 70.00\%\\
Proposed & dissolved\,$\ddag\star$ & 447,412 & 64.18\%\\ \hline
\multicolumn{4}{c}{$\dag$ = persistent block memory\,\,  $\ddag$ = dissolved sub-block boundaries}\\
\multicolumn{4}{c}{$\star$ both using multiple-scans, minimising block aspect ratio}\\
\end{tabular}
\end{center}
\end{table}

\subsection{Execution times and memory}\label{sec:bsu-results-times}
Table~\ref{tab:bsu-block-merge-time-comparison} reports the processing time for block merging and demonstrates the scalable nature of the proposed algorithm.\footnote{Experiments were conducted on a linux machine with 32 CPUs, 8 cores (Intel Xeon\textregistered\ CPU E5-2680 @ 2.70GHz), 64GB RAM and 2G swap memory\ignore{L1d/L1i/L2/L3 caches: 32K/32K/256K/20480K}} The parallel nature of the algorithm is a consequence of processing all input blocks that belong to the same parent block together independent of other parent blocks. This allows processing to be compartmentalised. From the description in Sec.~\ref{sec:bsu-subblocks-consolidation-via-coordinate-ascent}, it is clear that block consolidation and block tagging are applied locally to spatially disjoint regions; this is exploited in a multi-threaded implementation.

Merging under the ``dissolved sub-block boundaries'' convention given 697,097 blocks (of varying sizes) from the input model, is equivalent to merging 29,154,116 blocks (with single cell dimensions) directly under the ``persistent block memory'' convention. The problem size of ``\textit{dissolved}''$\ddag$ grows by a factor of 41.8, however, the computation time increased sub-linearly by a factor of only 9.5, this is partly due to the removal of sub-block alignment\,/\,compatibility constraints when internal sub-block boundaries are dissolved. Of course, the run time still increases relative to ``\textit{persistent}'', ultimately this may be viewed as a trade-off between speed and compaction when Table~\ref{tab:bsu-block-merge-time-comparison} is viewed along side Table~\ref{tab:bsu-block-merge-irregular-input-block-count-comparison}. The CPU utilisation and memory use columns are included for completeness, but these figures are otherwise unremarkable. Running 16 threads, the physical memory footprint amounts to $1\%$ of the RAM available.

\begin{table}[!ht]
\normalsize
\captionof{table}{Block merging processing times and memory footprints comparison}
\label{tab:bsu-block-merge-time-comparison}
\small
\begin{center}
\begin{tabular}{|c|c|c|c|c|c|c|}\hline
merge method & convention & threads & time (s) & relative & cpu\% & mem (kB)\\ \hline
Baseline & -- & -- & 254.181 & 1.0000 & 94.1 & 365,344\\ \hline
Proposed & persistent\,$\dag\ast$ & 1 & 4.870 & 0.0192 & -- & -- \\ \hline
Proposed & persistent\,$\dag\star$ & 1 & 32.852 & 0.1292 & 100.0 & 285,076\\
Proposed & persistent\,$\dag\star$ & 4 & 9.840 & 0.0387 & 400.0 & 380,884\\
Proposed & persistent\,$\dag\star$ & 16 & 4.162 & 0.0164 & 1553.0 & 614,540\\ \hline
Proposed & dissolved\,$\ddag\star$ & 1 & 469.462 & 1.8470 & 100.0 & 293,988\\
Proposed & dissolved\,$\ddag\star$ & 4 & 125.111 & 0.4922 & 400.0 & 370,108\\
Proposed & dissolved\,$\ddag\star$ & 16 & 39.675 & 0.1561 & 1547.0 & 609,064\\ \hline
\multicolumn{7}{c}{$\dag$ = persistent block memory\,\,  $\ddag$ = dissolved sub-block boundaries}\\
\multicolumn{7}{c}{$\ast$ = single scan, coordinate-ascent follows the standard pattern}\\
\multicolumn{7}{c}{$\star$ multiple scans, minimising block aspect ratio}\\
\end{tabular}
\end{center}
\end{table}

\section{Experiment 2: The proposed coordinate-ascent block merging technique vs octree subblocking}\label{sec:bsu-proposed-vs-octree}
\begin{figure}[!ht]
\centering
\includegraphics[width=78mm]{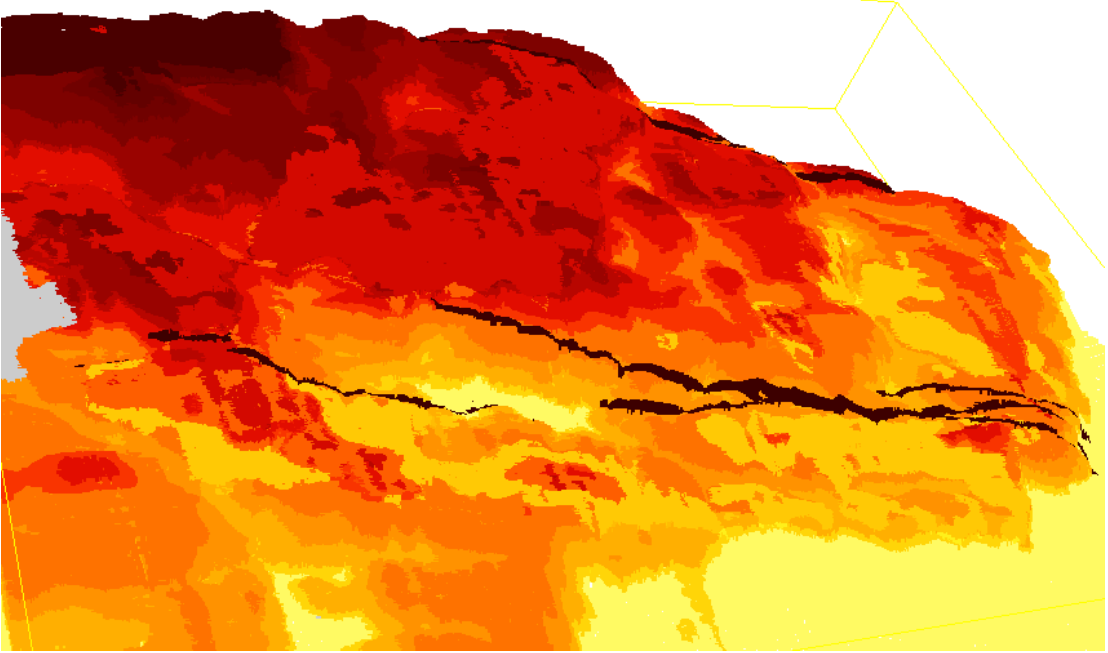}
\caption{Spatial structure of a mine where block model comparison between the proposed and octree representation are made. Around 30 geological domains are represented at this site, each coloured in a different shade. Black stripes represent igneous intrusions demarcated by a set of closed surfaces (not shown). The layers are peeled back one-by-one in an animated sequence in Appendix~\ref{sec:appendix-detailed-octree-subblocking-comparison}. Please refer to page~\pageref{fig:appendix-animated-site8-geological-domains}--\pageref{fig:appendix-animated-site8-geological-domains-dx} (the supplementary material).}
\label{fig:site8-geological-domain-structure}
\vspace{-3mm}
\end{figure}

Octree decomposition is a volumetric partitioning strategy that is well studied in the literature \cite{samet-88}. As this hierarchical structure has previously been considered as a potential representation for spatial information systems in geology \cite{jones-89}, it provides an interesting reference point to measure our proposed method against. To obtain a fair comparison, the parent blocks ($\boldsymbol{\Delta}_\text{parent}^\text{block}\in\mathbb{R}^3$) must be divisible by the minimum block size ($\boldsymbol{\Delta}_\text{min}^\text{block}$) by some vector  $\mathbf{k}=(k_\text{x},k_\text{y},k_\text{z})^T=(2^{K_\text{x}},2^{K_\text{y}},2^{K_\text{z}})^T\in\mathbb{Z}_{+}^3$ where each element is a power of two. This dyadic constraint is necessary for compatibility since the standard octree has to split each dimension into half along the x, y and z axes at each spatial decomposition step. The analysis presented earlier in Table~\ref{tab:bsu-block-merge-irregular-input-block-count-comparison} pertains to a block model where its parameters (e.g. $\boldsymbol{\Delta}_\text{min}^\text{block}$) are already fixed and the cell dimensions $\mathbf{k}=(40,40,20)^T$ are not divisible by 8; thus the maximum number of dyadic decomposition is limited to $D=2$. To facilitate a more wide-ranging comparison, we have chosen another site for which the block structure is not fixed a priori and that geological surfaces are available. This enables all system components in Fig.~\ref{fig:bsu-framework} including block model creation (not just block merging) to be applied. The chosen parameters $\boldsymbol{\Delta}_\text{parent}^\text{block}=(50,50,20)^T$ and $\boldsymbol{\Delta}_\text{min}^\text{block}=(1.5625,1.5625,0.625)^T$ satisfy the dyadic constraints and allow a maximum of $D\!=\!5$ levels of octree decomposition. Fig.~\ref{fig:site8-geological-domain-structure} illustrates the complexity of the domain structure. This surface-induced structure provides spatial separation between significant geological units. A detailed explanation in terms of stratigraphy, material type or geochemical  composition however is beyond the scope of this work. The main point is that deposits in the Hamersley Group typically contain interlayered Banded Iron Formations (BIF) and shale bands. Using multiple surfaces to express boundary constraints, it is possible to isolate, for instance, areas of localised iron enrichment within the BIFs where high grade iron ore deposits occur. Interested readers may refer to \cite{thorne2008banded} for geological context.

The schemes being compared are the two proposed coordinate-ascent block merging algorithms performed under the \textit{persistent block memory} and \textit{dissolved sub-block boundaries} conventions (denoted `Proposed-P' and `Proposed-D' respectively) versus octree decomposition and a variant that permits intra-scale merging (denoted `Octree' and `Octree+Merge' respectively). In a standard octree decomposition, a heterogenous node (viz.\,a block with dimensions $\boldsymbol{\Delta}_d=\boldsymbol{\Delta}_\text{parent}^\text{block}/2^d$) that contains smaller blocks with different labels is split into an octant with $2\times 2\times 2$ cells of size $\boldsymbol{\Delta}_d / 2$. This sub-division, similar to the approach implemented in the \textit{Surpac} software \cite{poniewierski-19}, is applied recursively to all heterogenous nodes at each decomposition level $d=1,2,\ldots,D$.  This process shatters a volume into many smaller blocks and thus vastly increases the block count in its pursuit of higher accuracy. The `Octree+Merge' variant allows 2 and 4 edge-connected cells to be consolidated into a single (rectangular or square-like) block thus achieves higher compaction. This octree merging step applies only to cells at the same scale within an octant in the spatial hierarchy. For completeness, a full description and pictorial examples of these concepts are given in Appendix~\ref{sec:appendix-octree-decomposition-merging}.

\subsection{Block count analysis}\label{sec:block-count-vs-octree-analysis}
In Table~\ref{tab:aggregate-block-counts-vs-octree}, it can be seen that performing octree decomposition without merging produces a large number of blocks (4241289 for $D\!=\!3$ levels). With octree merging, the block count is reduced to 43.167\%. The proposed strategies have a block count of about 25\% relative to standard octree decomposition. It is worth mentioning that despite this block count disparity, each block model represents the modelled region with the same fidelity and the total block volume for each domain classification remains the same irrespective of $D$ and the block count. In one sense, the `Octree+Merge' method provides an empirical bound on block model efficiency (as measured by block count) that other strategies can surpass. The impressive efficiency gains achieved by the proposed strategies may be attributed to inter-scale block merging without dyadic constraints. For instance, it is possible to combine two adjacent blocks $\mathcal{B}_i$ and $\mathcal{B}_j$ in neighbouring octants if their labels are the same and their cell-dimensions, say, $\mathbf{k}^{(i)}=(1,2,1)$ and $\mathbf{k}^{(j)}=(2,2,1)$ are compatible. This would not be possible using the na\"{i}ve `Octree+Merge' method as it involves blocks from different octants.

\begin{table*}[!th]
\begin{center}
\small
\setlength\tabcolsep{4pt}
\caption{Block model aggregate block counts: octree vs proposed methodology}\label{tab:aggregate-block-counts-vs-octree}
\begin{tabular}{|c|cccc|}\hline
&\multicolumn{4}{c|}{Total block count, $B_D$}\\ \hline
$D$ levels & Octree & Octree\,+\,Merge & Proposed-P & Proposed-D\\ \hline
3 & 4241289 & 1830826 & 1094917 & 1100585\\ \hline
4 & 17414755 & 7455284 & 4053310 & 4007018\\ \hline
5 & 74065301 & 31052736 & 15608162 & 15204844\\ \hline\hline
&\multicolumn{4}{c|}{Relative block count}\\ \hline
$D$ levels & Octree & Octree\,+\,Merge & Proposed-P & Proposed-D\\ \hline
3 & 100\% & 43.167\% & 25.816\% & 25.949\%\\ \hline
4 & 100\% & 42.810\% & 23.275\% & 23.009\%\\ \hline
5 & 100\% & 41.926\% & 21.074\% & 20.529\%\\ \hline
\multicolumn{5}{c}{\revisit{}The total block volume is approximately $3.693382\times 10^9$ m\textsuperscript{3}}\\
\multicolumn{5}{c}{A detailed breakdown of these figures is given in Appendix~\ref{sec:appendix-detailed-octree-subblocking-comparison}}\\
\end{tabular}
\end{center}
\end{table*}

\begin{table*}[!th]
\begin{center}
\small
\setlength\tabcolsep{4pt}
\caption{Block model block count growth factors: octree vs proposed methodology}\label{tab:aggregate-block-growth-factor-vs-octree}
\begin{tabular}{|c|cccc|}\hline
&\multicolumn{4}{c|}{Block count growth factor based on Table~\ref{tab:aggregate-block-counts-vs-octree}}\\ \hline
Growth rate & Octree & Octree\,+\,Merge & Proposed-P & Proposed-D\\ \hline
$B_4/B_3$ & 4.1060 & 4.0721 & 3.7019 & 3.6408\\ \hline
$B_5/B_4$ & 4.2530 & 4.1652 & 3.8507 & 3.7946\\ \hline
$B_5/B_3$ & 17.4629 & 16.9611 & 14.2551 & 13.8152\\ \hline
\end{tabular}
\end{center}
\end{table*}

From Table~\ref{tab:aggregate-block-counts-vs-octree}, it is evident the block count roughly grows by a factor of 4 with each additional decomposition level. This is made clear in Table~\ref{tab:aggregate-block-growth-factor-vs-octree} and expressed as a ratio $B_d/B_{d'}$ between levels $d$ and $d'$. 
Intuitively, the growth rate relates to the surface contact area with the blocks which scales with $O(n^2)$ where $n$ represents the number of cells measured in one dimension. Interestingly, the proposed coordinate-ascent block merging strategies have a lower growth rate compared to `Octree+Merge' --- moving from 3 to 4 levels, the block count increases by a factor of 3.6408 as opposed to 4.0721; from 3 to 5 levels, the growth rate is 13.8152 as opposed to 16.9611. In this respect, the proposed strategies scale better and they are more efficient at representing the same domain information. We caution however that block modelling is not ultimately about an endless pursuit for increasing precision. Block granularity (which may be specified in terms of the number of decompositions, $D$) is often limited by practical considerations. In our application, individual blocks in the model serve as containers for grade estimation. The blockwise spatial distribution of chemical attributes in turn informs what, where and how ore material are excavated. The heavy equipments used to extract and transport such material operate with physical or kinematic constraints (such as bucket size and turn radius), so the required spatial resolution in the case of mining is probably limited to 1.5m to 6m at best. This naturally imposes a lower bound on the spatial fidelity needed from a model. It informs the choice of $D$ in our evaluation, as $D\!=\!3$ and $D\!=\!5$ yield a resolution of 1.5625m and 6.25m, respectively, in a horizontal plane.

\subsection{Block aspect ratio analysis}\label{sec:block-aspect-vs-octree-analysis}
For grade estimation at a given level of spatial fidelity, a lower block count is desirable not so much for storage but from a computational perspective. A compact model contains fewer blocks that require the inference engine to estimate into, and by extension, it has fewer values to update when new information becomes available. For instance, the difference between `Octree+Merge' and `Proposed' can be as much as 50\% (based on $D\!=\!5$ in Table~\ref{tab:aggregate-block-counts-vs-octree}). A secondary consideration is an optimisation preference for more uniform blocks (as opposed to thin, narrow blocks) that have lower aspect ratios. In a pure octree decomposition scheme, the blocks always maintain the same aspect ratio (in our experiment, this value is 50/20 or 2.5). However, this is obtained at significant cost, resulting in a substantial increase in block count. In Table~\ref{tab:aggregate-block-aspects-vs-octree}, the main observation is that the weighted aspect ratios in the `Proposed-D' column are a marked improvement over `Proposed-P' when merging constraints are relaxed (when input block boundaries are no longer persistent in the sense defined in Sec.~\ref{sec:bsu-merging-conventions-optimisation-objectives}). It also increases slower than `Proposed-P' as the resolution (number of levels $D$) increases, and tends toward the values in the `Octree+Merge' column.

\begin{table*}[!th]
\begin{center}
\small
\setlength\tabcolsep{4pt}
\caption{Block model average block aspect ratios: octree vs proposed methodology}\label{tab:aggregate-block-aspects-vs-octree}
\begin{tabular}{|c|cccc|cccc|}\hline
& \multicolumn{4}{c|}{weighted by block volume}\\ \hline
$D$ levels & Octree & Octree\,+\,Merge & Proposed-P & Proposed-D\\ \hline
3 & 2.5 & 2.958828 & 3.668188 & 2.615373\\ \hline
4 & 2.5 & 2.968178 & 4.901418 & 2.941797\\ \hline
5 & 2.5 & 2.972378 & 7.395642 & 3.456381\\ \hline
\ignore{
& \multicolumn{4}{c|}{weighted by block count}\\ \hline
$D$ levels & Octree & Octree\,+\,Merge & Proposed-P & Proposed-D\\ \hline
3 & 2.5 & 3.535403 & 5.068056 & 2.839873\\ \hline
4 & 2.5 & 3.585874 & 7.861600 & 3.650470\\ \hline
5 & 2.5 & 3.585541 & 13.523983 & 5.078064\\ \hline
}
\end{tabular}
\end{center}
\end{table*}

\begin{figure}[h]
\centering
\includegraphics[width=120mm]{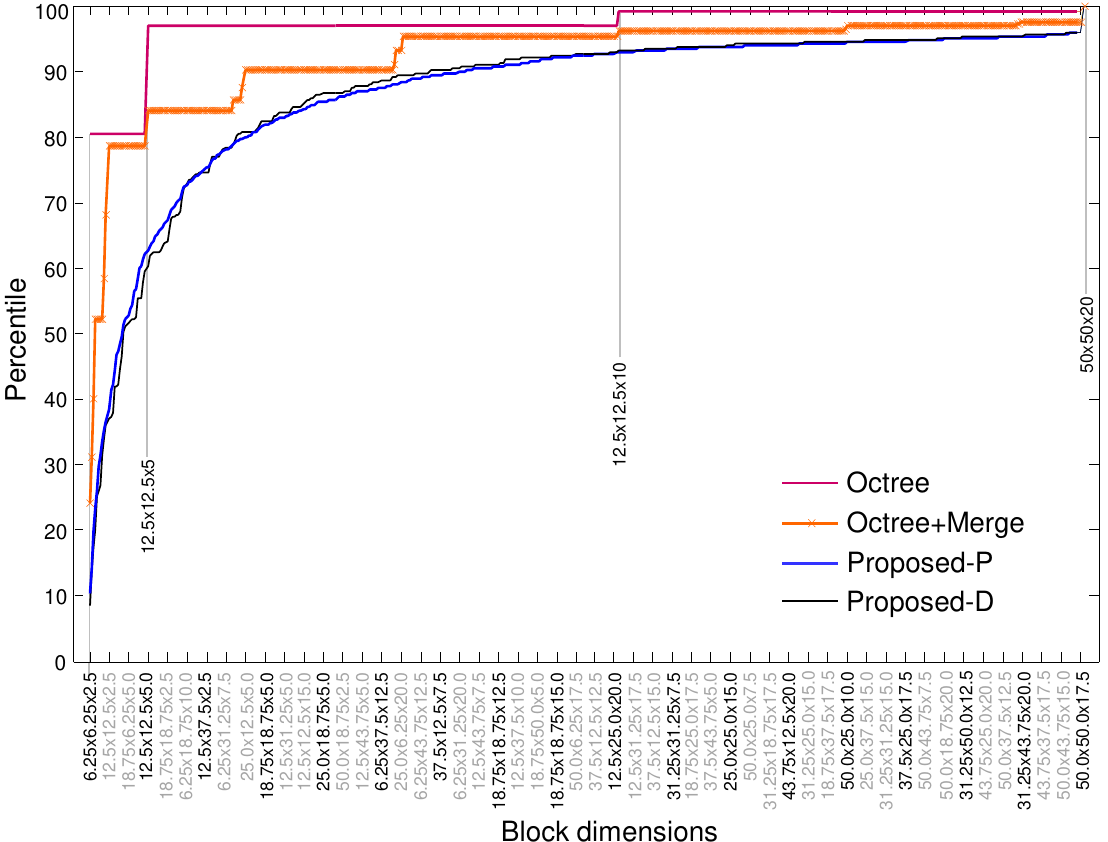}
\caption{Cumulative distribution of block dimensions for $D\!=\!3$: a comparison between the proposed and octree representations.}
\label{fig:octree-vs-proposed-blocksize-cdf}
\vspace{-3mm}
\end{figure}
\subsection{Block fragmentation analysis}\label{sec:block-fragmentation-vs-octree-analysis}
The diversity of blocks within each model is examined in Fig.~\ref{fig:octree-vs-proposed-blocksize-cdf}. The graph shows the cumulative distribution of block dimensions for $D\!=\!3$ decomposition levels, with block volumes arranged in ascending order (from left to right) and sorted by aspect ratio in the event of a tie. The staircase nature of the orange curve for `Octree+Merge' shows that there is a much smaller pool of possible block size permutations produced via octree decomposition with merging. To simplify the analysis, we measure blocks in multiples of  $\boldsymbol{\Delta}_\text{min}^\text{D\!=\!3}=(6.25,6.25,2.5)$, for instance, using $\mathbf{k}=(2,1,3)$ to denote a block with real dimensions $(12.5,6.25,7.5)$. Without merging, octree decomposition alone produces an overwhelming 80.44\% of blocks with cell dimensions $\mathbf{k}=(1,1,1)$, while 96.86\% and 98.93\% of blocks have dimensions at/below $(2,2,2)$ and $(4,4,4)$, respectively. The red curve demonstrates that the blocks in the `Octree' model are indeed highly fragmented. With intra-scale merging, the orange curve shows there is a larger spread of blocks and the proportion occupied by the smallest blocks has substantially reduced. Blocks with cell dimensions $(1,1,1)$, $(1,1,2)$, $(1,2,1)$ and $(2,1,1)$ now occupy the 24.09\textsuperscript{th}, 31.26\textsuperscript{st}, 39.98\textsuperscript{th} and 52.08\textsuperscript{nd} pecentiles, respectively. Because this strategy does not permit inter-scale merging (outside of an octant), it cannot produce blocks with cell dimensions $(1,1,3)$, $(1,3,1)$ or $(3,1,1)$ and indeed $(1,1,4)$. These lost opportunities are capitalised by the proposed strategies. However, the `Octree+Merge' strategy can still  produce blocks of size $(1,2,2)$, $(2,1,2)$ and $(2,2,1)$; the so called 4-connected cells occupy the 58.45\textsuperscript{th}, 68.14\textsuperscript{th} and 78.64\textsuperscript{th} percentiles, respectively.

Moving up the spatial hierarchy, the $(2,2,2)$, $(2,2,4)$, $(2,4,2)$ and $(4,2,2)$ blocks correspond to the 84.01\textsuperscript{th}, 85.50\textsuperscript{th}, 87.36\textsuperscript{th} and 90.13\textsuperscript{th} percentiles; as opposed to 96.86\% at the $(2,2,2)$ stage for octree decomposition without merging. With intra-scale merging, the $(2,4,4)$, $(4,2,4)$ and $(4,4,2)$  blocks occupy the 91.07\textsuperscript{st}, 92.99\textsuperscript{nd} and 95.25\textsuperscript{th} percentiles. Again, blocks with intermediate cell dimensions, such as $(2,3,4)$ and $(3,2,5)$, can only be synthesized by the proposed techniques through inter-scale merging and relaxation of the dyadic constraints. Similar observations hold for more levels.

\subsection{Discussion and insights}\label{sec:octree-vs-proposed-discussion}
The octree comparisons with the proposed methods highlight the constraints imposed by hierarchical decomposition and the importance of block merging. In cases where the block structure is subsequently consumed by a block estimation process, minimising the block count (while maintaining the same level of spatial fidelity and domain classification granularity) can substantially reduce the computational effort associated with probabilistic inference of block attributes. The analysis has shown, for instance, when a minimum block size of $(1.5625,1.5625,0.625)$ and 5 decomposition levels are applied to the test site, the block count of the proposed model and `Octree+Merge' model are 20.52\% and 41.92\% respectively relative to standard octree decomposition without block merging. The `Octree+Merge' approach is probably more indicative of the flexible subblocking approach used in the \textit{Datamine} software \cite{poniewierski-19} which adjusts the split depending on the angle of intersection of a particular block with the surface controlling the sub-division. It is however unclear if a split extends all the way down within a parent block, whether different permutations and aspect ratio optimisation are considered in \textit{Datamine}. Returning to the block count comparison, the factor of 2 reduction may be attributed to inter-scale merging and relaxation of dyadic constraints as illustrated in Sec.~\ref{sec:block-fragmentation-vs-octree-analysis}. The proposed coordinate-ascent block merging approach also scales better. For instance, the block count growth rate from 3 to 5 levels is kept at 13.8152 as opposed to 16.9611 (see Table~\ref{tab:aggregate-block-counts-vs-octree}). From the cumulative distribution in Fig.~\ref{fig:octree-vs-proposed-blocksize-cdf}, we observed the block models obtained using the proposed methods are spatially less fragmented.

Although the performance metrics all point in favour of the proposed methods, the real strength of the proposal is its flexibility. First, the block merging algorithm was used on its own to consolidate and heal a spatially fragmented block model given only the parent block size and minimum block dimensions (see Sec.~\ref{sec:bsu-block-merge-results}). Second, the same algorithm was used as part of a coherent block model spatial restructuring strategy to coalesce sub-blocks partitioned by intersecting surfaces (see Sec.~\ref{sec:bsu-proposed-vs-octree}). Finally, it provides a successive refinement framework for updating the spatial structure of a block model; this takes a dynamic (non-static) view and emphasizes the evolving nature of the model. In contrast, some commercial softwares \cite{poniewierski-19} produce block models that cannot be altered or easily manipulated. The iterative setting implies the input blocks can have different dimensions.
The standard octree paradigm is not equipped to deal with input blocks with varying sizes. As a case in point, given an input block with cell dimensions $(4,2,8)$, a three level dyadic decomposition would need to be applied asymmetrically to each axis, via $d_1(z),d_2(x,z),d_3(x,y,z)$, to not violate the minimum block size (lower bound and integer multiple) conditions. Furthermore, an input block with cell dimensions $(3,5,2)$ would leave the octree approach with no viable sub-blocking options unless the block is divided non-uniformly which would set it on a similar path as the proposed strategy notwithstanding the hierarchical embedding (rigid subblock alignment) constraints.

A crucial part of `iterative refinement' is that large portions of an existing model should remain intact, useful information in unaffected parts of the model are not invalidated through the update process, and changes affect only localised regions that interact with new surfaces. This allows creation of a new block model from scratch given a set of surfaces (representing geological boundaries or other types of delineation), as well as automatic correction for misplaced boundaries within an existing model. In \cite{leung-20}, the current proposal is portrayed as a system within a system, where new information captured by sampled assay data are used to manipulate the shape of relevant surfaces to maximise agreement with the latest observations. These are used in turn to update the block model to improve the accuracy of its domain structure by maximising the resolution and positional integrity of the blocks near those surfaces. This has been shown in \cite{leung-20} to have a positive effect on processes downstream. In particular, grade estimation performance improves as boundary (smearing) artefacts are reduced during inferencing.

\subsection{Summary of contributions}\label{sec:summary-of-contributions}
The overall contributions of this paper may be summarised as follows.
\begin{enumerate}
\item \textbf{Developing a flexible framework} capable of altering the spatial structure of a block model given boundary constraints in the form of triangle mesh surfaces and a set of tagging instructions.
\item \textbf{Describing computational techniques} for all system components
  \begin{itemize}
  \item From block-surface intersection detection (identifying areas requiring an update), block structure decomposition (improving boundary localisation), block consolidation (increasing compactness of representation) to block tagging (encoding the location of blocks relative to the given surfaces).
  \end{itemize}
\item \textbf{Devising a new algorithm} (coordinate-ascent inspired block merging) that is amenable to multi-threading.
\item \textbf{Demonstrating real applications}: The algorithm was used firstly to reduce block fragmentation in an existing mine resource estimation block model where parameters are fixed and no surfaces are given; secondly as an integrated component within the block model creation/augmentation workflow to coalesce subblocks partitioned by intersecting surfaces.
\item \textbf{Providing insights} through quantitative analysis, highlighting the constraints of dyadic hierarchical decomposition and importance of inter-scale merging for achieving a compact block model; discussing why this matters for subsequent block-based estimation processes (see Sec.~\ref{sec:block-count-vs-octree-analysis}--\ref{sec:octree-vs-proposed-discussion}).
\end{enumerate}

The first point includes correcting the position of previously misplaced boundaries in an existing model (see Fig.~\ref{fig:boundary-healing}). Philosophically, it challenges the perception that voxel-based models are hard to modify once they have been constructed. It demonstrates that local refinement (including consolidation of fragmented blocks from previous updates) is both possible and can be done efficiently in an iterative setting; with changes to the model confined to areas that interact with new surfaces. The second point is about greater transparency. In 3D geological modelling papers, models are often generated using commercial software where the fundamental techniques employed are unclear. This paper sheds light on this and discusses pit falls and robust solutions from an engineering perspective. The third point emphasizes flexibility. Depending on user requirement on whether internal subblock boundaries are to be respected, it can operate in two modes (see Fig.~\ref{fig:persistent-vs-dissolved-raster}): with \textit{persistent block memory} to ensure an input block always maps uniquely to a merged output block, or with \textit{dissolved subblock boundaries} to achieve higher block compaction (a lower block count). The specific objective in the fourth point was to capture geological structures of an ore body in a block model at a specified resolution with a minimal block count. However, the method can potentially be applied elsewhere.

\section{Conclusions}\label{sec:bsu-conclusion}
This paper described a framework for updating the spatial structure of a 3D geology block model given mesh surfaces. The system consists of four components which perform block-surface overlap detection, spatial structure decomposition, sub-blocks consolidation and block tagging, respectively. These processes are responsible for identifying areas where refinement is needed, increasing spatial resolution to minimise surface approximation error, reducing redundancy to increase the compactness of the model and establishing the domain of each block with respect to geological boundaries. A flexible architecture was presented which allows a model to be updated simultaneously, or iteratively, by multiple surfaces, to selectively retain or modify existing block domain labels. Robustness and accuracy of the system were considered during the design; one notable feature was using ray-tracing to establish the location of sub-blocks relative to surfaces, particularly those near boundaries, prior to block consolidation to minimise boundary distortion. Other techniques employed include block-surface intersection analysis based on the separable axis theorem.

The main contribution was a coordinate-ascent merging algorithm which is used during block consolidation in the proposed framework. This technique was extended to solve a related problem, viz., using block merging to reduce fragmentation in an existing block model where surfaces are not involved. Issues relating to scan pattern, merging conventions were discussed, differences between \textit{`persistent block memory'} and \textit{`dissolved sub-block boundaries'} were explained in terms of internal dimensions compatibility and sub-block alignment constraints imposed on the expansion feasibility test. Performance was evaluated with respect to block aspect ratio, output block count and processing time using a multi-threaded implementation. The results demonstrated the quality and scalability of the proposed technique. Systematic evaluation against octree subblocking futher highlights the limitations of hierarchical decomposition and the importance of inter-scale merging and relaxation of dyadic constraints. Algorithm~\ref{algo:bsu-final-block-merging} produced merged blocks with less extreme aspect ratios and the approach is well suited to parallel processing. The techniques described may apply to areas outside of geoscience (see Fig.~\ref{fig:bsu-stanford-bunny}) where 3D body localisation, also known as 3D region segmentation, is required inside a block model given some mesh surfaces.

\bibliographystyle{unsrt}  
\bibliography{ms}

\begin{thebibliography}{10}

\bibitem{poniewierski-19}
Julian Poniewierski.
\newblock Block model knowledge for mining engineers.
\newblock \url{https://www.deswik.com/wp-content/uploads/2019/07/Block-model-knowledge-for-mining-engineers-An-introduction.pdf}, 2019.
\newblock {A}ccessed: 2020-08-08.

\bibitem{luo-07}
Zhou~Quan Luo, Liu Xiao-ming, Su~Jia-hong, Wu~Ya-bin, and Liu Wang-ping.
\newblock Deposit 3{D} modeling and application.
\newblock {\em Journal of Central South University of Technology},
  14(2):225--229, 2007.

\bibitem{leite-07}
A~Leite and R~Dimitrakopoulos.
\newblock Stochastic optimisation model for open pit mine planning: application
  and risk analysis at copper deposit.
\newblock {\em Mining Technology}, 116(3):109--118, 2007.

\bibitem{dimitrakopoulos-08}
R~Dimitrakopoulos and S~Ramazan.
\newblock Stochastic integer programming for optimising long term production
  schedules of open pit mines: methods, application and value of stochastic
  solutions.
\newblock {\em Mining Technology}, 117(4):155--160, 2008.

\bibitem{srk-18}
{SRK Consulting}.
\newblock {B}lock {M}odelling (``3{D} {M}odeling {P}roject, {S}pout {L}ake,
  {B}ritish {C}olumbia, {C}anada'').
\newblock
  \url{http://webcache.googleusercontent.com/search?q=cache:https://www.srk.ru.com/en/block-modelling},
  2018.
\newblock {A}ccessed: 2018-12-25.

\bibitem{mira-18}
{Mira Geoscience}.
\newblock 3{D} {G}eological {M}odelling.
\newblock
  \url{http://webcache.googleusercontent.com/search?q=cache:http://www.mirageoscience.com/our-products/software-solutions/3d-geological-modelling},
  2018.
\newblock {A}ccessed: 2018-12-23.

\bibitem{leapfrog-19}
{Leapfrog}.
\newblock Leapfrog {G}eo: {T}rainagulated {M}eshes.
\newblock
  \url{https://help.leapfrog3d.com/Geo/4.3/en-GB/Content/meshes/triangulated-meshes.htm},
  2019.
\newblock {A}ccessed: 2019-10-25.

\bibitem{emery-05}
Xavier Emery.
\newblock Simple and ordinary multigaussian kriging for estimating recoverable
  reserves.
\newblock {\em Mathematical Geology}, 37(3):295--319, 2005.

\bibitem{ball-20}
Adrian Ball, Katherine Silversides, Anna Chlingaryan, and Arman Melkumyan.
\newblock Creating large scale probabilistic boundaries using {G}aussian
  {P}rocesses.
\newblock In {\em Proceedings of Geostats 2020}, Toronto, Canada, 2021.
  International Geostatistics Congress.

\bibitem{leung-19subsurface}
Raymond Leung.
\newblock Subsurface boundary geometry modeling: Applying computational
  physics, computer vision, and signal processing techniques to geoscience.
\newblock {\em IEEE Access}, 7:161680--161696, 2019.

\bibitem{vasudevan-09}
Shrihari Vasudevan, Fabio Ramos, Eric Nettleton, and Hugh Durrant-Whyte.
\newblock Gaussian process modeling of large-scale terrain.
\newblock {\em Journal of Field Robotics}, 26(10):812--840, 2009.

\bibitem{newman-06}
Timothy~S Newman and Hong Yi.
\newblock A survey of the marching cubes algorithm.
\newblock {\em Computers \& Graphics}, 30(5):854--879, 2006.

\bibitem{dragiev-11}
Stanimir Dragiev, Marc Toussaint, and Michael Gienger.
\newblock Gaussian process implicit surfaces for shape estimation and grasping.
\newblock In {\em Robotics and Automation (ICRA), 2011 IEEE International
  Conference on}, pages 2845--2850. IEEE, 2011.

\bibitem{wang-11}
Gongwen Wang, Shouting Zhang, Changhai Yan, Yaowu Song, Yue Sun, Dong Li, and
  Fengming Xu.
\newblock Mineral potential targeting and resource assessment based on 3{D}
  geological modeling in {L}uanchuan region, {C}hina.
\newblock {\em Computers \& Geosciences}, 37(12):1976--1988, 2011.

\bibitem{feltrin-09}
L~Feltrin, JG~McLellan, and NHS Oliver.
\newblock Modelling the giant, {Z}n--{P}b--{A}g {C}entury deposit,
  {Q}ueensland, {A}ustralia.
\newblock {\em Computers \& Geosciences}, 35(1):108--133, 2009.

\bibitem{mathers-11}
Stephen~J Mathers, Holger Kessler, and Bruce Napier.
\newblock British {G}eological {S}urvey: A nationwide commitment to 3-{D}
  geological modeling.
\newblock In Richard~C Berg~et al., editor, {\em Synopsis of Current
  Three-dimensional Geological Mapping and Modeling in Geological Survey
  Organizations}, volume 578, chapter~6, pages 25--30. 1Illinois State
  Geological Survey Circular, 2011.

\bibitem{castagnac-11}
Claire Castagnac, Catherine Truffert, Bernard Bourgine, and Gabriel Courrioux.
\newblock French {G}eological {S}urvey ({B}ureau de {R}echerches
  {G}\'eologiques et {M}ini\`eres): Multiple software packages for addressing
  geological complexities.
\newblock In Richard~C Berg~et al., editor, {\em Synopsis of Current
  Three-dimensional Geological Mapping and Modeling in Geological Survey
  Organizations}, volume 578, chapter~8, pages 42--47. 1Illinois State
  Geological Survey Circular, 2011.

\bibitem{akenine-01}
Tomas Akenine-M{\"o}ller.
\newblock Fast 3{D} triangle-box overlap testing.
\newblock {\em Journal of graphics tools}, 6(1):29--33, 2001.

\bibitem{pharr2016physically}
Matt Pharr, Wenzel Jakob, and Greg Humphreys.
\newblock {\em Physically based rendering: From theory to implementation}.
\newblock Morgan Kaufmann, 2016.

\bibitem{arvidsson-08}
Martin Arvidsson and Ida Gremyr.
\newblock Principles of robust design methodology.
\newblock {\em Quality and Reliability Engineering International},
  24(1):23--35, 2008.

\bibitem{sedlmair-12}
Michael Sedlmair, Miriah Meyer, and Tamara Munzner.
\newblock Design study methodology: Reflections from the trenches and the
  stacks.
\newblock {\em IEEE Transactions on Visualization and Computer Graphics},
  18(12):2431--2440, 2012.

\bibitem{rosson-09}
Mary~Beth Rosson and John~M Carroll.
\newblock Scenario based design.
\newblock In J~Jacko and A~Sears, editors, {\em The Human-Computer Interaction
  Handbook: Fundamentals, Evolving Technologies and Emerging Applications},
  chapter~53, pages 1032--1050. CRC Press, 2002.

\bibitem{moller-05}
Tomas M{\"o}ller and Ben Trumbore.
\newblock Fast, minimum storage ray/triangle intersection.
\newblock In {\em ACM SIGGRAPH 2005 Courses}, page~7. ACM, 2005.

\bibitem{clout-06}
JMF Clout.
\newblock Iron formation-hosted iron ores in the {H}amersley {P}rovince of
  {W}estern {A}ustralia.
\newblock {\em Applied Earth Science}, 115(4):115--125, 2006.

\bibitem{wang-15}
Gongwen Wang, Ruixi Li, Emmanuel John~M Carranza, Shouting Zhang, Changhai Yan,
  Yanyan Zhu, Jianan Qu, Dongming Hong, Yaowu Song, Jiangwei Han, et~al.
\newblock 3{D} geological modeling for prediction of subsurface {Mo} targets in
  the {L}uanchuan district, {C}hina.
\newblock {\em Ore Geology Reviews}, 71:592--610, 2015.

\bibitem{wycisk-09}
P~Wycisk, T~Hubert, W~Gossel, and Ch~Neumann.
\newblock High-resolution 3{D} spatial modelling of complex geological
  structures for an environmental risk assessment of abundant mining and
  industrial megasites.
\newblock {\em Computers \& Geosciences}, 35(1):165--182, 2009.

\bibitem{li2017irregular}
Nan Li, Leon Bagas, Mark Lindsay, Daniel Wedge, Lin Bai, and Xianglong Song.
\newblock An irregular triangle mesh buffer analysis method for boundary
  representation geological object in three-dimension.
\newblock {\em Earth Science Informatics}, 10(2):149--167, 2017.

\bibitem{samet-88}
Hanan Samet.
\newblock An overview of quadtrees, octrees, and related hierarchical data
  structures.
\newblock In {\em Theoretical Foundations of Computer Graphics and CAD}, pages
  51--68. Springer, 1988.

\bibitem{jones-89}
Christopher~B Jones.
\newblock Data structures for three-dimensional spatial information systems in
  geology.
\newblock {\em International Journal of Geographical Information System},
  3(1):15--31, 1989.

\bibitem{thorne2008banded}
W~Thorne, Steffen Hagemann, A~Webb, and J~Clout.
\newblock Banded iron formation-related iron ore deposits of the {H}amersley
  {P}rovince, {W}estern {A}ustralia.
\newblock In {\em Banded {I}ron {F}ormation-Related High-Grade Iron Ore}, pages
  197--221. Society of Economic Geologists, 2008.

\bibitem{leung-20}
Raymond Leung, Alexander Lowe, Anna Chlingaryan, Arman Melkumyan, and John
  Zigman.
\newblock Bayesian surface warping approach for rectifying geological
  boundaries using displacement likelihood and evidence from geochemical
  assays.
\newblock {\em arXiv e-prints}, page arXiv:2005.14427, May 2020.

\bibitem{leung-19a}
Raymond Leung.
\newblock Modelling orebody structures: Block merging algorithms and block
  model spatial restructuring strategies given mesh surfaces of geological
  boundaries.
\newblock {\em Journal of Spatial Information Science}, 2020.

\bibitem{dietrich-07}
Andreas Dietrich, Enrico Gobbetti, and Sung-Eui Yoon.
\newblock Massive-model rendering techniques: a tutorial.
\newblock {\em IEEE Computer Graphics and Applications}, 27(6):1--18, 2007.

\bibitem{mcguire-17}
Morgan Mc{G}uire.
\newblock Computer {G}raphics {A}rchive, July 2017.

\bibitem{stanford-bunny-14}
Greg Turk and Marc Levoy.
\newblock The {S}tanford 3{D} {S}canning {R}epository, August 2014.

\bibitem{tamminen1984efficient}
Markku Tamminen and Hanan Samet.
\newblock Efficient octree conversion by connectivity labeling.
\newblock {\em ACM SIGGRAPH Computer Graphics}, 18(3):43--51, 1984.

\end{thebibliography}

\newpage\section{Supplementary Material}\label{sec:supplementary}
This section elaborates on the computational aspects. In part~\ref{sec:appendix-block-triangle-intersect-detect-moller}, a method is described for finding blocks in the model that intersect with triangular patches on a given surface. This is used to identify areas where \textit{model refinement} is needed to accurately reflect the location of boundaries and more closely approximate the curvature of said surfaces. In part~\ref{sec:appendix-ray-tracing}, the concept of ray-tracing is described, this is used to establish the location of blocks relative to the surface(s) in the \textit{block tagging} system component. Part~\ref{sec:appendix-bsu-extension} deals with the technical aspects of block merging and discusses various considerations fundamental to its design. This in-depth discussion explains the differences between two block merging conventions, the constraints, the block merging optimisation objective, and how different scanning sequences are implemented in practice. It should be noted that the overall block merging technique can be applied to areas outside of geoscience as shown in part~\ref{sec:appendix-stanford_bunny}, to reduce redundancy\,/\,fragmentation in a parent-grid aligned block model, and in instances where 3D segmentation is desired given some triangle mesh surface for an object. Part~\ref{sec:appendix-pseudocode} provides the pseudocode for the coordinate-ascent inspired block merging algorithms which is the main contribution of \cite{leung-19a}. Finally, detailed commentary and results on octree subblocking are given in Part~\ref{sec:appendix-octree-decomposition-merging} and Part~\ref{sec:appendix-detailed-octree-subblocking-comparison}.

\appendix
\section{Akenine-M\"{o}ller method for block triangle overlap detection}\label{sec:appendix-block-triangle-intersect-detect-moller}
Assessment for ``block-triangle'' intersection involves at most 13 tests:
\begin{itemize}
\item 3 along the x, y, z axes, the orthonormal bases are denoted $\mathbf{e}_0=(0,0,1), \mathbf{e}_1=(0,1,0), \mathbf{e}_2=(0,0,1)$
\item 9 for cross-products between edges of A and B, viz., cross($\mathbf{e}_i,\mathbf{f}_j$) for $i,j\in\{0,1,2\}$
\item 1 for the normal of the triangle based on cross($\mathbf{f}_i,\mathbf{f}_j$) given vertices $\mathbf{v}_0,\mathbf{v}_1,\mathbf{v}_2$, edge vectors $\mathbf{f}_i=\mathbf{v}_{\text{mod}(i+1,3)}-\mathbf{v}_i$
\end{itemize}
Suppose a triangle has vertices $\mathbf{v}_0,\mathbf{v}_1,\mathbf{v}_2\in\mathbb{R}^3$, a block has centroid $\mathbf{b}_k=(b_x,b_y,b_z)$ and dimensions $\boldsymbol{\Delta}_k=(\Delta_x,\Delta_y,\Delta_z)$, the SAT test for axes x, y, z asserts ``no overlap'' if $\mathbf{v'}_{\min}[c]>\boldsymbol{\Delta}_k[c]/2$ or $\mathbf{v'}_{\max}[c]<-\boldsymbol{\Delta}_k[c]/2$ for any $c\in\{x,y,z\}$ where $\mathbf{v'}_{\min}$ and $\mathbf{v'}_{\max}$ represent the minimum and maximum coordinates of the translated vertices, $\mathbf{v}'_i=\mathbf{v}_i-\mathbf{b}_k$, after the block centroid is subtracted from the triangle vertices.

The SAT test for cross($\mathbf{e}_i,\mathbf{f}_j$) exploits the properties of axis-aligned blocks. Its efficiency derives from terms cancellation in the cross-product expansion when the geometry of interest is limited to axis-aligned prisms and triangles. This uses only simple algebra; the relevant formulas may be found in \cite{akenine-01}.

The last SAT test for plane-block overlap requires $p_{\min}=\text{dot}(\mathbf{\hat{n}},\boldsymbol{\delta}_{\min})+d$ and $p_{\max}=\text{dot}(\mathbf{\hat{n}},\boldsymbol{\delta}_{\max})+d$ to be computed, where $\mathbf{\hat{n}}=\mathbf{n}/\lVert\mathbf{n}\rVert$ is the unit length plane normal, $\mathbf{n}=(a,b,c)$, $d$ is the plane distance from origin, assuming the plane passing through the triangle is described by the equation $ax+by+cz+d=0$. The quantities  $\boldsymbol{\delta}_{\min}[c]=(1-2\times\mathcal{I}(\mathbf{n}[c]>0))\cdot\boldsymbol{\Delta}_k[c]/2$ and  $\boldsymbol{\delta}_{\max}[c]=(2\times\mathcal{I}(\mathbf{n}[c]>0)-1)\cdot\boldsymbol{\Delta}_k[c]/2$ evaluate to $\pm\boldsymbol{\Delta}_k[c]/2$. The test asserts ``no overlap'' if $p_{\min}>0$ or $p_{\max}<0$.

\section{Side-of-surface determination via ray tracing}\label{sec:appendix-ray-tracing}
Ray tracing is a well known technique in the computer graphics community \cite{dietrich-07}. In the affiliated paper \cite{leung-19a}, it is used to establish where a block is located with respect to one or more triangle mesh surfaces, rather than for rendering purpose. A ray emanating from a block (specifically, its centroid) is casted in some specified direction.\footnote{For an open surface, this direction might be the upward (positive) direction specified in the tagging instructions. For a closed surface, the direction matters little, it is generally taken as the outward normal for the surface.} The idea is to count the number of intersections between this ray and the relevant surface. An even number of intersections (including 0) result when the block is located above (respectively, outside) an open (respectively, closed) surface, and an odd number of interesections is interpreted as below (respectively, inside) the surface. The tests are based on the M{\"o}ller--Trumbore algorithm \cite{moller-05} which is explained below.

\subsection{Intersection between a ray and a plane}
\begin{figure}[!h]
\centering
\includegraphics[width=82.5mm]{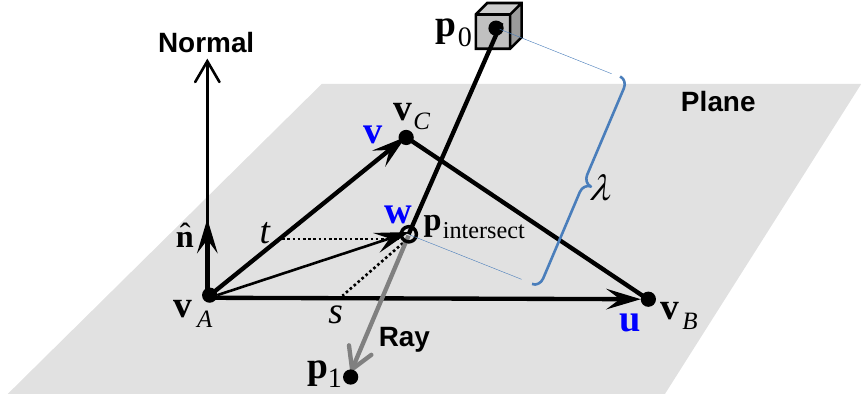}
\caption{Ray-triangle intersection analysis}
\label{fig:bsu-ray-triangle-intersection-analysis}
\end{figure}

A ray extending from $\mathbf{p}_0$ to $\mathbf{p}_1$ intersects with a plane $\pi(\mathbf{v}_A,\mathbf{n})$ that passes through $\mathbf{v}_A\in\mathbb{R}^3$, with normal $\mathbf{n}=\mathbf{v}_A\times\mathbf{v}_B$, at $\mathbf{p}_\text{intersect}=\mathbf{p}_0+\lambda(\mathbf{p}_1-\mathbf{p}_0)$ when $\lambda\in[0,1]$ where
\begin{align} \label{eq:vesk119-lambda}
\lambda=\dfrac{\mathbf{\hat{n}}\cdot(\mathbf{v}_A-\mathbf{p}_0)}{\mathbf{\hat{n}}\cdot(\mathbf{p}_1-\mathbf{p}_0)}
\end{align}
A picture of this is shown in Fig.~\ref{fig:bsu-ray-triangle-intersection-analysis}

\begin{itemize}
\item When $\lambda < 0$, the ray does not intersect with the triangle described by vertices $\mathbf{v}_A$, $\mathbf{v}_B$, $\mathbf{v}_C$ and plane $\pi(\mathbf{v}_A,\mathbf{n})$.
\item When the denominator in (\ref{eq:vesk119-lambda}) is zero, the ray is parallel to the triangle's plane. If the numerator is also zero, the ray intersects with the face of the triangle along a line. Otherwise, there is no intersection.
\end{itemize}

\subsection{Intersection between a ray and a triangle}
When $\lambda\in[0,1]$, the ray intersects with the triangle at $\mathbf{p}_\text{intersect}=\mathbf{v}_A+s\mathbf{u}+t\mathbf{v}$\,\ \textit{if} the barycentric coordinates $s$ and $t$ (see Fig.~\ref{fig:bsu-ray-triangle-intersection-analysis}) satisfy $s\ge 0, t\ge 0$ and $s+t\le 1$ where
\begin{align}\label{eq:vesk119-edge-vectors}
\mathbf{u}=\mathbf{v}_B-\mathbf{v}_A,\quad \mathbf{v}=\mathbf{v}_C-\mathbf{v}_A\\ \label{eq:vesk119-barycentric-coords}
s=\dfrac{(\mathbf{u}\cdot\mathbf{v})(\mathbf{w}\cdot\mathbf{v})-(\mathbf{v}\cdot\mathbf{v})(\mathbf{w}\cdot\mathbf{u})}{\Delta}\\
t=\dfrac{(\mathbf{u}\cdot\mathbf{v})(\mathbf{w}\cdot\mathbf{u})-(\mathbf{u}\cdot\mathbf{u})(\mathbf{w}\cdot\mathbf{v})}{\Delta}\\
\Delta=(\mathbf{u}\cdot\mathbf{v})^2-(\mathbf{u}\cdot\mathbf{u})(\mathbf{v}\cdot\mathbf{v})\\
\mathbf{w}=\mathbf{p}_\text{intersect}-\mathbf{v}_A
\end{align}
This involves only five distinct inner products, and the quantities ($\mathbf{u}\cdot\mathbf{u},\mathbf{v}\cdot\mathbf{v}$ and $\mathbf{u}\cdot\mathbf{v})$ may be precomputed as they are independent of $\mathbf{p}_\text{intersect}\in\mathbb{R}^3$, unlike $\mathbf{w}$ which is a function of the block centroid and ray direction.

\subsection{Practicalities}
Degenerate conditions must be handled to obtain proper results. First, when a ray intersects a surface at a common edge or vertex shared by multiple triangles, one needs to be careful that over-counting does not occur. In our implementation, a unique set of intersecting points is maintained for each ray to ensure the same intersecting point is not repeated. Second, when the denominator and numerator in (\ref{eq:vesk119-lambda}) are both zero, $\lambda$ is undefined, as the ray lies on the face of a triangle. This can be overcome by changing the direction slightly for the casted ray. Finally, triangles that collapse to a line segment or a single point need to be removed from the test surface since  $\Delta\rightarrow\infty$ when either edge vector $\mathbf{u}=\mathbf{0}$ or $\mathbf{v}=\mathbf{0}$ in (\ref{eq:vesk119-edge-vectors}) and $\Delta\rightarrow 0$ when $\mathbf{u}$ and $\mathbf{v}$ are parallel. Users may wish to perform surface integrity checks as a preprocessing step to eliminate these conditions.

\section{Demonstration on the Stanford Bunny}\label{sec:appendix-stanford_bunny}
\begin{figure}[!ht]
\centering
\includegraphics[width=87.5mm]{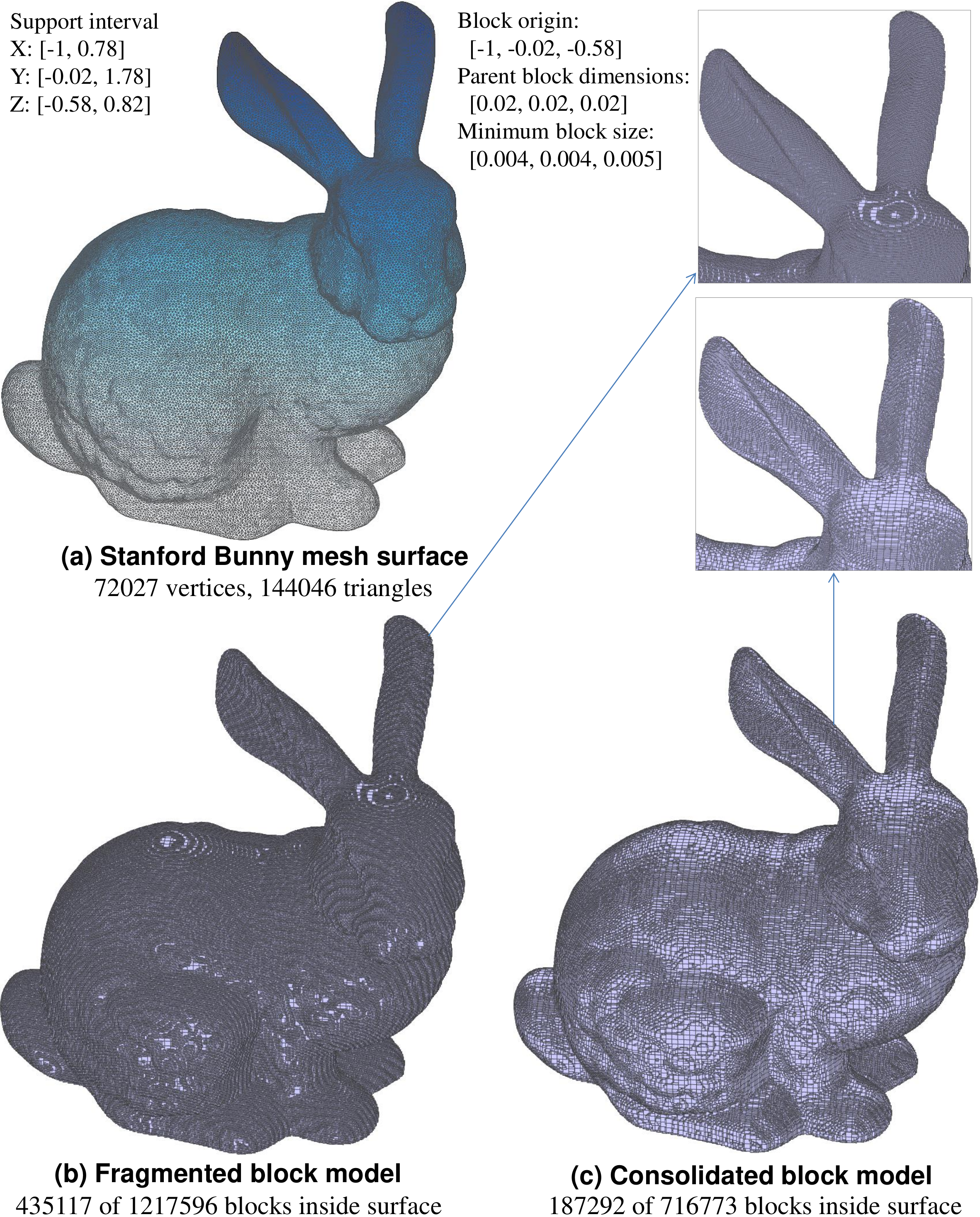}
\caption{Block merging applied to Stanford Bunny to reduce block fragmentation. Zoom in to see individual blocks.}
\label{fig:bsu-stanford-bunny-appendix}
\end{figure}

The block merging technique described in Algorithm~\ref{algo:bsu-coordinate-ascent2} is applicable to more complex surfaces outside the geoscience domain. Fig.~\ref{fig:bsu-stanford-bunny-appendix} (a) shows a triangular mesh surface \cite{mcguire-17} of the terracotta bunny obtained using multiple range scanners at the Stanford Computer Graphics Laboratory \cite{stanford-bunny-14}. Fig.~\ref{fig:bsu-stanford-bunny-appendix} (b) shows a highly fragmented block model created by block decomposition without consolidation. In an effort to closely approximate the surface, numerous blocks at the minimum block size were produced near the surface. Fig.~\ref{fig:bsu-stanford-bunny-appendix} (c) shows a reduction in block density and increase in clarity as blocks are merged under the ``dissolve sub-block boundaries'' convention (see \ref{sec:appendix-merging-convention}). This resulted in a more efficient block representation (3D segmentation) of the object.


\vspace{-3mm}
\section{Extended discussion about using block merging to reduce fragmentation}\label{sec:appendix-bsu-extension}
The adjustments foreshadowed in Section~\ref{sec:bsu-critical-reflection} of the paper \cite{leung-19a} improve both the robustness and accuracy of the block model spatial restructuring system which utilises at least one surface. This section considers how the block consolidation component can be extended to serve the needs of a \textit{block merging} application where the key objective is to coalesce blocks in a fragmented block model without any input surface. This extension builds upon the ideas described in Section~\ref{sec:bsu-subblocks-consolidation}. The characteristics and constraints of the problem will be described next. Henceforth, the established framework for block model \textit{spatial restructuring using surfaces} from Section~\ref{sec:bsu-critical-reflection} and new \textit{block merge} application will be abbreviated as SRUS and BM, respectively.

\subsection{Problem description}\label{sec:appendix-bsu-problem-description}
In block model spatial restructuring using surfaces (SRUS), merging follows block-surface intersection detection and block decomposition, so we know precisely which input block (parent) a sub-block (cell) comes from. These input blocks may have different dimensions, particularly if the pipeline is repeated when individual surfaces are processed in cascade (see example in Section~\ref{sec:bsu-iterative-application} where the output from the first iteration becomes the input in the second iteration). 
For the application envisaged in surface-free \textit{block merge} (BM), the input contains only the labels, locations and dimensions of sub-blocks which are integer multiples of the minimum block size. Whilst the parent block and origin continues to provide a uniform grid structure that covers the 3D space, these parent-blocks have constant dimensions and only exist on a conceptual level for the purpose of grouping together the sub-blocks. Furthermore, ray-casting needs not be performed to determine which side of a surface a block is located, since the input provides domain labels for each block. The goal of BM is to consolidate the input blocks into larger rectangular prisms to minimise fragmentation.

\subsection{Constraints}\label{sec:appendix-bsu-constraints}
\begin{figure}[h]
\centering
\includegraphics[width=87.5mm]{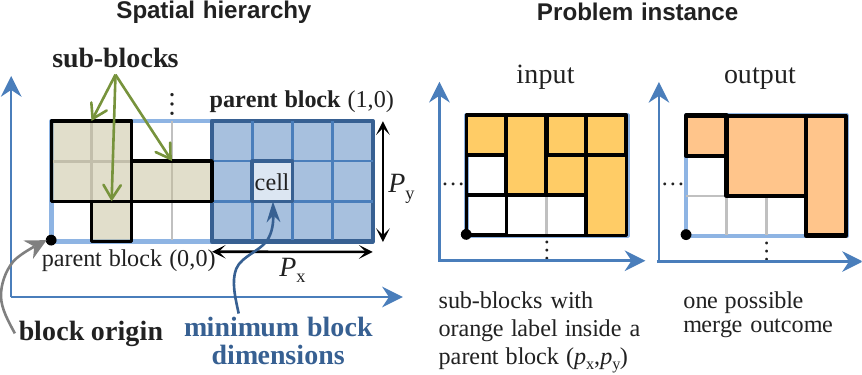}
\caption{Block merging spatial hierarchy}
\label{fig:bsu-spatial-hierarchy}
\end{figure}

Before describing the constraints, it is instructive to first explain the spatial hierarchy and understand the assumptions. Fig.~\ref{fig:bsu-spatial-hierarchy} illustrates the relationship between parent block, cells and input blocks (sub-blocks) of intermediate scale. Conceptually, the whole 3D space is spanned by \textit{parent blocks} which represent uniform, non-overlapping tiles positioned with respect to the anchor point, \textit{block origin}. Each parent block may be identified by an index $\mathbf{p}=(p_x,p_y,p_z)$ obtained via uniform quantisation given the origin $\mathbf{o}=(o_x,o_y,o_z)$ and parent block size, $(P_x,P_y,P_z)$. Each parent has internal structure --- each is divided by the \textit{minimum block size} into $(n_x\times n_y\times n_z)$ \textit{cells} in the same manner. A cell is the smallest spatial unit. The cell ``walls'' dictate what type of merges are possible within a parent block. All input blocks and merged blocks must adhere to this structure, i.e., each consisting of one or more whole cells.

The assumptions are: 1) all input sub-blocks have dimensions which are integer-multiples of the minimum block; 2) all blocks must be rectangular prisms; 3) no sub-block straddles the boundary of any parent block; 4) edges of input and merged blocks must align perfectly with the internal grid lines of the parent block to which they belong; 5) only sub-blocks from the same class and parent may be merged.

\subsection{Broad strategy}\label{sec:appendix-bsu-broad-strategy}
Beside some changes to the cell-expansion feasibility test, the block consolidation strategy based on coordinate-ascent merging is almost directly applicable to this problem. At a high level, the strategy comprises the following steps.
\begin{enumerate}
\item Establish an input sub-block to parent block mapping.
\item Divide and conquer (compartmental processing)
\begin{itemize}
\item Each problem instance is restricted to a set of input blocks associated with (indexed by) a parent block. This is highly amendable to parallel processing.
\end{itemize}
\item Within each parent block, process each category (collection of input blocks with the same class label) in turn.
  \begin{itemize}
  \item The position\,/\,extent of sub-blocks undergoing consolidation are maintained by a 3D cell occupancy map and stateful objects.
  \end{itemize}
\item A modified coordinate-ascent merging algorithm is used to merge blocks from the same parent and class.
  \begin{itemize}
  \item Feasibility of cell expansion is governed by specific rules which depend on the merging convention. However, the general goal remains the same, it still cycles through the x, y and z-coordinate one-by-one to consider if incremental expansion is possible.
  \end{itemize}
\end{enumerate}

\subsection{Feasibility of cell expansion}\label{sec:appendix-cell-expansion-feasibility}
For a parent block with cell dimensions $(K_\text{x},K_\text{y},K_\text{z})$, a 3D cell occupancy map $\theta$ with identical dimensions is used to manage merging states. To initialise this object, the cells occupied by each input block with the same label are set to 1 (active). A default value of 0 is set for the remaining (inactive) cells to signify a different domain classification. To advance this discussion, it is helpful to define a pooling function,
\begin{align}\label{eq:bsu-pooling-fn}
\zeta_v(\mathbf{n},\mathbf{k})=\sum_{d_\text{z}=0}^{k_\text{z}-1}\sum_{d_\text{y}=0}^{k_\text{y}-1}\sum_{d_\text{x}=0}^{k_\text{x}-1}\mathcal{I}(\theta(n_\text{x}+d_\text{x},n_\text{y}+d_\text{y},n_\text{z}+d_\text{z})=v)
\end{align}
which counts the number of cells with label value $v$ over a support interval that extends from $\mathbf{n}=(n_\text{x},n_\text{y},n_\text{z})\in\mathbf{Z}^3$ (the minimum cell coordinates) to $\mathbf{n}+\mathbf{k}-\mathbf{1}=(n_\text{x}+k_\text{x}-1,n_\text{y}+k_\text{y}-1,n_\text{z}+k_\text{z}-1)$ (the maximum cell coordinates) where $\mathbf{k}$ represents the provisional size of a block undergoing expansion. At any point during the coordinate-ascent algorithm, an incremental expansion $\boldsymbol{\delta}\in\mathbb{Z}^3$ --- typically $\boldsymbol{\delta}\in\{(1,0,0)$, $(0,1,0)$, $(0,0,1)\}$ --- is feasible if $\zeta_1(\mathbf{n},\mathbf{k}+\boldsymbol{\delta})=(k_\text{x}+\delta_\text{x})\cdot(k_\text{y}+\delta_\text{y})\cdot(k_\text{z}+\delta_\text{z})$ for a block with current cell dimensions $\mathbf{k}$.

Using this definition, the coordinate-ascent merging procedure from Section~\ref{sec:bsu-subblocks-consolidation} as used in the SRUS (spatial restructuring using surfaces) framework is formally described in \textbf{Algorithm~\ref{algo:bsu-coordinate-ascent}} on page~\pageref{algo:bsu-coordinate-ascent}.

\subsection{Modifications}\label{sec:appendix-bsu-bm-modifications}
There are two key differences in the BlockMerge (BM) case. First, the boolean occupancy map $\theta\in\{0,1\}^{ K_\text{x}\times K_\text{y}\times K_\text{z}}$ now holds sub-block indices and becomes multi-valued, viz., $\theta\in\mathbb{Z}^{ K_\text{x}\times K_\text{y}\times K_\text{z}}$. Second, when a block expansion step is feasible in one of the coordinate directions, the increment takes on the dimension of the block (or blocks) along the axis of expansion; this being typically larger than 1. Merging states are managed using an ordered\footnote{Objects of type $\mathcal{M}$ are sorted in ascending order by the number of cells within each block, then by the minimum vertex coordinates to break ties. This priority gives smaller blocks the earliest opportunity to grow.} list of structure similar to $\mathcal{M}$ in Algorithm~\ref{algo:bsu-coordinate-ascent}, where each structure initially contains the minimum vertex of an input block $\mathbf{v}_{\min}^{(b)}$, its cell dimensions $\mathbf{s}=(s_\text{x},s_\text{y},s_\text{z})\in\mathbb{Z}^3$ which can grow, the block label $\lambda^{(b)}$ and a boolean flag, \textit{subsumed}, which is set to false. The idea is to revise $\mathbf{s}$, the block dimensions expressed in terms of cells, as a block grows; blocks which have been swallowed are invalidated by setting \textit{subsumed} to true and will be ignored in subsequent iterations. This effectively results in a shrinking set, the coalesced blocks are the surviving entries when the algorithm terminates. The algorithm continues as long as the cell count changes for any block between iterations. Details are given in \textbf{Algorithm~\ref{algo:bsu-coordinate-ascent2}} on page~\pageref{algo:bsu-coordinate-ascent2}.

\subsection{Cell expansion feasibility test}\label{sec:appendix-bsu-cell-expansion-feasibility-test}
\begin{figure}[!th]
\centering
\includegraphics[width=87.5mm]{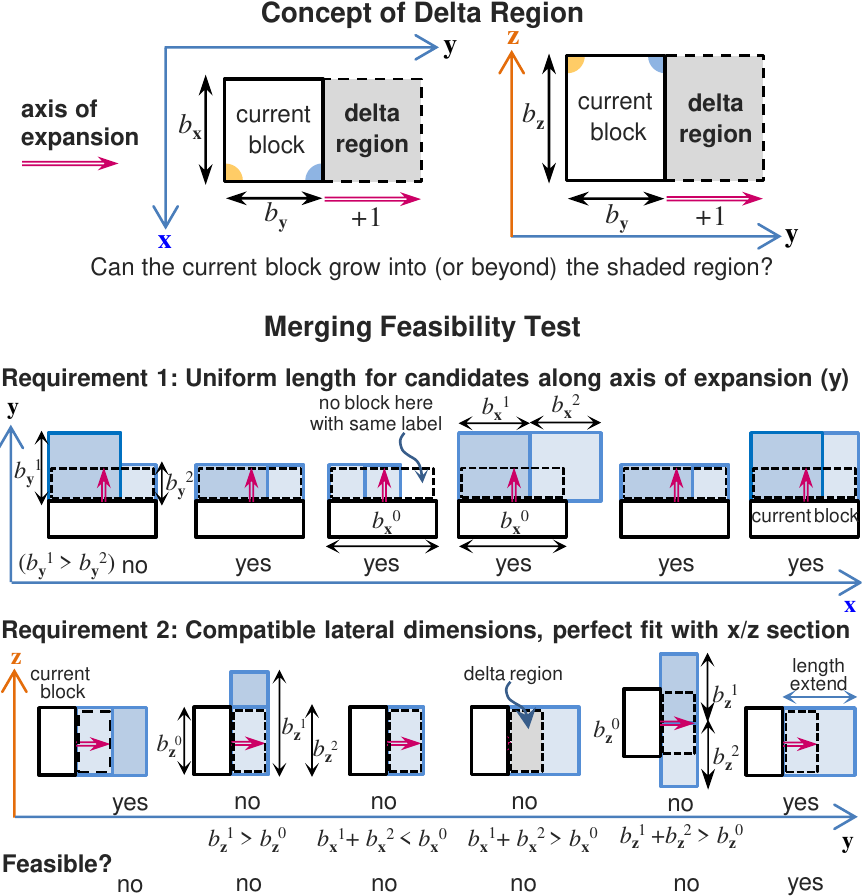}
\caption{Delta region and sub-block expansion feasibility tests}
\label{fig:bsu-delta-region-feasibility}
\end{figure}

For block merging, Algorithm~\ref{algo:bsu-coordinate-ascent2} has essentially the same blueprint as Algorithm~\ref{algo:bsu-coordinate-ascent}. The main difference is the acceptance criteria for each expansion step, see \textit{FeasibleCellExpansion} in lines \ref{algo:feasible-expansion1}, \ref{algo:feasible-expansion2} and \ref{algo:feasible-expansion3} in Algorithm~\ref{algo:bsu-coordinate-ascent2}. This is explained with the aid of Fig.~\ref{fig:bsu-delta-region-feasibility}. When block merging is attempted, the expansion step proposes an elongation of the current block along one of the axes of expansion. The volumetric difference, before and after the proposed expansion, is referred as the \textit{delta region}. Fig.~\ref{fig:bsu-delta-region-feasibility} further illustrates 5 situations where a merge with adjacent block(s) are infeasible. A proposed expansion step is feasible when two conditions are satisfied: 1) the dimension along the axis of expansion is the same for all adjoining blocks in the delta region; 2) the lateral dimensions of these adjoining blocks are compatible with the current block; in other words, their cross-sections must join perfectly. The computation inside \textit{FeasibleCellExpansion} is described in \textbf{Subroutine~\ref{algo:bsu-feasible-cell-expansion}}.

\subsection{Merging conventions}\label{sec:appendix-merging-convention}
In Algorithm~\ref{algo:bsu-coordinate-ascent2}, we have a block merging procedure that preserves the boundary of the input blocks, in the sense that it does not introduce new partitions (sub-divisions) that are not already present in a parent block. This is because when a sub-block is subsumed, it is swallowed whole by another block. This merging convention is referred as \textbf{persistent block memory} for future reference. A key property is that each input block is mapped uniquely to a single block in the merged model.

\begin{figure}[!th]
\centering
\includegraphics[width=82.5mm]{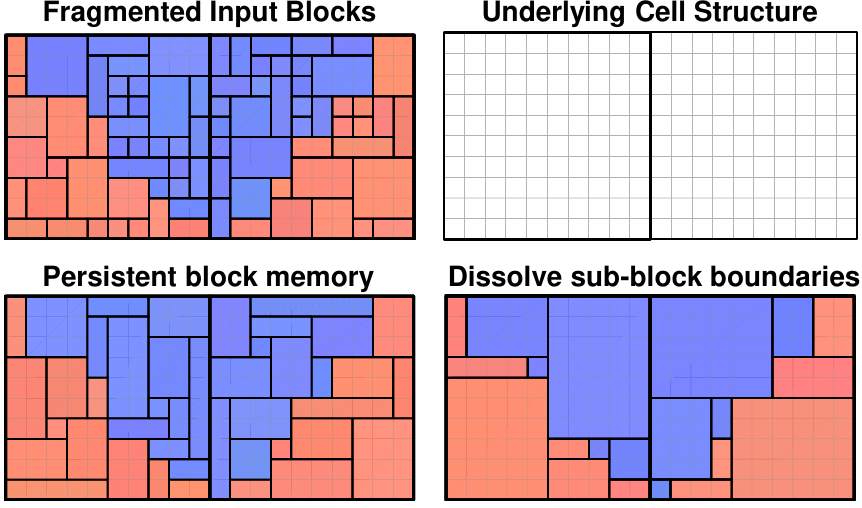}
\caption{Example of differences under the `persistent block memory' and `dissolve sub-block boundaries' block merging conventions}
\label{fig:bsu-merging-conventions}
\vspace{-3mm}
\end{figure}

In contrast, Algorithm~\ref{algo:bsu-coordinate-ascent} implicitly erases sub-block boundaries before block consolidation begins. This merging convention is referred as \textbf{dissolve sub-block boundaries}, it generally achieves higher compaction because it makes no distinction between input blocks from the same class and parent. It is able to grow blocks more freely and produce fewer merged blocks since the size compatibility constraints between individual blocks no longer apply when they are treated as one. This can be useful for healing a fractured block model. It can consolidate sub-blocks introduced by a false boundary from a previous surface update. Under the ``dissolve sub-block boundary'' convention, coordinate-ascent can start from a clean slate. Sub-blocks in a fragmented area may grow back to the largest possible extent even if individual sub-block dimensions or internal boundary alignments are otherwise incompatible. It does not suffer the negative consequences of block structure decomposition from previous iterations. Some of these differences are shown in Fig.~\ref{fig:bsu-merging-conventions}.

\subsection{Fairness and regulating parameters}\label{sec:appendix-merge-fairness}
Both algorithms include optional parameters. The token life span, $T$, limits the number of uninterrupted sequential merging steps a block can take during coordinate-ascent, to moderate aggressive merging behaviour. This token value is decremented by 1 after each x-y-z cycle. When it reaches zero, the current block must cease expansion and give other blocks the opportunity to grow. When every block in the  queue has had its turn, this block may resume expansion. The token value is reset to $T$ each time a block takes possession. By default, $T$ is set to infinity so no progress is ever halted. An upper bound on merged block dimensions is given by $(M_\text{x},M_\text{y},M_\text{z})\in\mathbb{R}^3$. By default, this is set to the parent block size to remove any restriction.

\subsection{Scan sequences to improve block aspect ratio}\label{sec:appendix-scan-sequences}
\begin{figure}[!th]
\centering
\includegraphics[width=87.5mm]{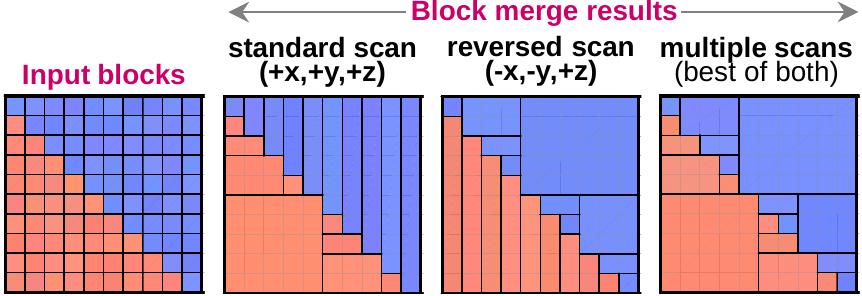}
\caption{Block merging results from different scan sequences}
\label{fig:bsu-scan-patterns}
\end{figure}

The final design consideration relates to the order in which input blocks are processed during coordinate-ascent. The main observation from Fig.~\ref{fig:bsu-scan-patterns} is that depending on the shape and direction of the class boundary, a sequential algorithm may generate a stair-case artefact, producing long narrow blocks which certain applications may find objectionable. The incremental block expansion may be obstructed by the boundary if it approaches from a certain direction as it cycles through each coordinate axis; this can lead to excessive growth in an unimpeded direction. In general, no single deterministic scanning sequence (e.g., increasing x, increasing y and increasing z as in the ``standard'' case) can be optimal in all situations. One way to overcome this is by introducing multiple scan patterns. For instance, instead of scanning (processing blocks) top-down, left-to-right, one can scan from bottom-up, from right-to-left. This is equivalent to flipping the x and y axes.

Accordingly, there are 8 distinct possibilities given we have 3 axes, these scan sequences may be abbreviated as $\pi_0=(+x,+y,+z)$, $\pi_1=(-x,+y,+z)$, $\pi_2=(+x,-y,+z)$ and so forth, where a negative sign indicates reversal of the relevant axis. The algorithm will try all 8 scan patterns and select the result which minimises an objective function. In this work, the preferred solution $\text{argmin}_{\pi} f_{\pi}(\{\boldsymbol{\Delta}^{(b,\pi)}\}_{b\in\mathcal{S}_{\mathbf{p},\lambda}})$ minimises the volume-weighted \textit{block aspect ratio}, the objective function may be expressed as
\begin{align}
f_{\pi}(\{\boldsymbol{\Delta}^{(b,\pi)}\}_{b\in\mathcal{S}_{\mathbf{p},\lambda}})=\dfrac{\sum_{b\in \mathcal{S}_{\mathbf{p},\lambda}} v^{(b,\pi)} \cdot \dfrac{\max\{\Delta_\text{x}^{(b,\pi)},\Delta_\text{y}^{(b,\pi)},\Delta_\text{z}^{(b,\pi)}\}}{\min\{\Delta_\text{x}^{(b,\pi)},\Delta_\text{y}^{(b,\pi)},\Delta_\text{z}^{(b,\pi)}\}}}{\sum_{b\in \mathcal{S}_{\mathbf{p},\lambda}} v^{(b,\pi)}}\label{eq:min-aspect-ratio-objective}
\end{align}
where merged block $b$ belongs to class $\lambda$ in parent block $\mathbf{p}$, $(\Delta_\text{x}^{(b,\pi)},\Delta_\text{y}^{(b,\pi)},\Delta_\text{z}^{(b,\pi)})\in\mathbb{R}^3$ and $v^{(b,\pi)}$ represent the dimensions and volume of the merged block, obtained from scan sequence $\pi$.

\subsection{Scan sequence implementation}\label{sec:appendix-scan-implementation}
In practice, the eight individual scan patterns are not programmed explicitly. Instead, sub-blocks are rearranged within a parent block before coordinate-ascent, in such a way that a specific scan sequence is attained when the permuted data is subject to the standard scan. This is done to avoid code duplication and preserve the existing logic.\footnote{An explicit  implementation for each scan pattern would involve $2^3$ nested \textsc{for} loops, this includes the standard\,/\,existing scan pattern --- 
for ($z=z_{\min}$; $z<z_{\max}$; $z$++) for ($y=y_{\min}$; $y<y_{\max}$; $y$++) for ($x=x_{\min}$; $x<x_{\max}$; $x$++) --- and seven other combinations including, for instance, the (-x,-y,+z) scan pattern --- for ($z=z_{\min}$; $z<z_{\max}$; $z$++) for ($y=y_{\max}-1$; $y\ge y_{\min}$; $y$-\,-) for ($x=x_{\max}-1$; $x\ge x_{\min}$; $x$-\,-). This is not easy to maintain.}

\begin{figure}[!ht]
\centering \includegraphics[width=87.5mm]{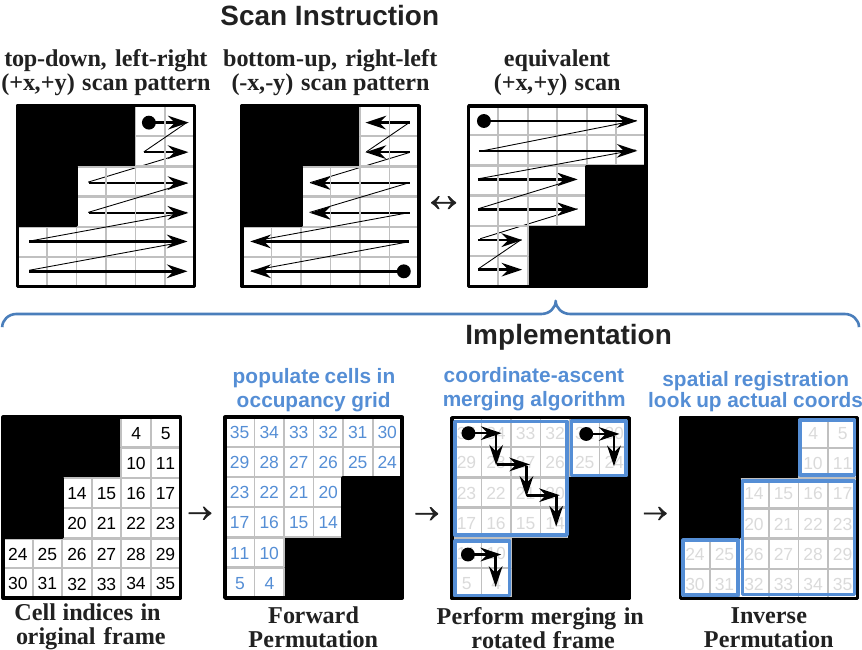}
\caption{Block merging scan sequence implementation}
\label{fig:bsu-scan-sequence-permutation}
\end{figure}

The approach is explained in Fig.~\ref{fig:bsu-scan-sequence-permutation}. The key observation is that only the \textsc{standard} scan is necessary (we do not need to implement 8 different scans directly) provided the cells occupied by the input blocks are permuted to reflect a reversal of the relevant axes. For instance, a bottom--up, right--left scan sequence on the original block data may be implemented by mapping the white cells from the south-east corner to north-west corner (see Fig.~\ref{fig:bsu-scan-sequence-permutation}\,(top)), then applying the ``standard'' top-down, left-right scan. The two are equivalent. Fig.~\ref{fig:bsu-scan-sequence-permutation}\,(bottom) outlines the steps involved.
\begin{itemize}
\item[a)] (\textbf{Forward permutation}) For each input block labelled \textit{white} in parent block $\mathbf{p}$, populate the occupancy map by sampling cells according to the direction of each axis specified in the scan instruction.\footnote{In reality, the occupancy map is a 3D array, but for simplicity, we only draw it in 2D.}
\item[b)] (\textbf{Perform merging in rotated frame}) Apply coordinate-ascent merging algorithm to permuted data using the standard scan pattern.
\item[c)] (\textbf{Inverse permutation}) Register the location of merged blocks in the original frame using table-lookup.
\end{itemize}

Synthesizing all the ideas, \textbf{Algorithm~\ref{algo:bsu-final-block-merging}} (page~\pageref{algo:bsu-final-block-merging}) describes the final block merging strategy which supports different merging conventions, multiple scan patterns and block aspect ratio optimisation. To elaborate on the the multi-threading aspect of the code, interleaved parent blocks are processed by individual threads within a region of interest. This choice, see interleaved parent indices in line \ref{algo:interleaved-parent-indices} of Algorithm~\ref{algo:bsu-final-block-merging}, is motivated by load balancing consideration. The intention is to spread the computation load evenly amongst the threads by decoupling spatial correlation, to avoid situations where too few (or too many) of the blocks processed by a thread actually intersect a surface.
\newpage
\section{Pseudocode}\label{sec:appendix-pseudocode}
This pseudocode comprises the following:
\paragraph{Algorithm 1:}\textbf{Coordinate-ascent merging algorithm v1}\newline (as used for spatial restructuring in Section~\ref{sec:bsu-subblocks-consolidation} of \cite{leung-19a})
\paragraph{Subroutine 1:}\textbf{Compute sub-block properties}
\paragraph{Algorithm 2:}\textbf{Coordinate-ascent merging algorithm v2}\newline (as used for model de-fragmentation in \ref{sec:appendix-bsu-extension})
\paragraph{Subroutine 2:}\textbf{Feasibility tests and state updates during block expansion}
\paragraph{Algorithm 3:}\textbf{Block merging with multiple scans and optimised block aspect ratios}

\begin{algorithm*}[!th]
\small
\caption{\quad \textbf{Coordinate-ascent merging algorithm} {\small(as used in the spatial restructuring SRUS framework in Section~\ref{sec:bsu-subblocks-consolidation})}}\label{algo:bsu-coordinate-ascent}
\begin{multicols}{2}
\begin{algorithmic}[1]
\renewcommand{\algorithmicrequire}{\textbf{Pre-requisite:}}
\REQUIRE Occupancy map, $\theta$, is populated s.t. all active cells that belong to sub-blocks of class $\lambda$ are set to 1.
\renewcommand{\algorithmicrequire}{\textbf{Assumption:}}
\REQUIRE Cells in occupancy map are enumerated in raster-scan order, thus index $i(n_\text{x},n_\text{y},n_\text{z})=(n_\text{z} K_\text{y} + n_\text{y}) K_\text{x} + n_\text{x}$. Parent block index is denoted $p$.
\renewcommand{\algorithmicrequire}{\textbf{Input:}}
\renewcommand{\algorithmicensure}{\textbf{Parameters:}}
\REQUIRE $\theta\in \{0,1\}^{K_\text{x}\times K_\text{y}\times K_\text{z}}$ 
\ENSURE Parent block cell dimensions: $K_\text{x}$, $K_\text{y}$, $K_\text{z}\in\mathbb{Z}$\\
\hspace{13mm} Min. block dimensions: $\boldsymbol{\Delta}_\text{min}^\text{block}\in\mathbb{R}^3$\\
\hspace{13mm} Max. merge cell dimensions: $M_\text{x}$, $M_\text{y}$, $M_\text{z}\in\mathbb{Z}$\\
\hspace{13mm} Token life span: $T\in\mathbb{Z}^{+}$\\
\renewcommand{\algorithmicensure}{\textbf{Variables:}}
\ENSURE Active cells: $\mathbf{a}=[]$ (initially an empty list)\\
\hspace{10mm} Merged blocks: $\mathcal{M}=\O$ (initially an empty set)\\
\hspace{10mm} Stride length: $\mathbf{s}=(s_\text{x}, s_\text{y}, s_\text{z})\in\mathbb{Z}^{3}$\\
\hspace{10mm} Provisional block dims: $\mathbf{d}=(d_\text{x}, d_\text{y}, d_\text{z})\in\mathbb{Z}^{3}$\\
\hspace{10mm} Min. coordinates of current block: $\mathbf{v}_{\min}^{(b)}\in\mathbb{R}^3$\\
\hspace{10mm} Obstacles count: $barriers$\\
\hspace{10mm} Iterations remaining: $i\in\mathbb{Z}$\\

\STATE \textbf{Find all active cells}: $\mathbf{a}\leftarrow \textit{IndexOfOccupants}(\theta)$
\STATE Set \textit{count} = 0, $n_\text{occupant}=|\mathbf{a}|$
\WHILE{number of active cells $|\mathbf{a}|\ge 1$}
  \STATE Set $i=T$ and $s_\text{x}=s_\text{y}=s_\text{z}=1$
  \IF{$|\mathbf{a}|=1$}
    \STATE  $\mathcal{M}$.append(\,\textit{SubBlockProperties}($\mathbf{v}_{\min}^{(b)}$, $\mathbf{s}$, $\boldsymbol{\Delta}_\text{min}^\text{(block)}$, $\lambda$)\,)$^\dag$\\
    \quad $^{\textsc{Note}\,\dag}$ see description in Subroutine \ref{algo:bsu-subblock-attributes}
    \STATE \textbf{break}
  \ENDIF
  \STATE Set $\mathbf{n}=(n_\text{x},n_\text{y},n_\text{z})=\textit{Subscript}(\,\text{cell }\mathbf{a}[0]\,)$\\
  where $n_\text{x},n_\text{y},n_\text{z}\ge 0$
  \WHILE{true}
    \STATE $\textit{barriers}=0$
    \STATE $(d_\text{x}, d_\text{y}, d_\text{z})\leftarrow (\min\{s_\text{x}+1, K_\text{x}-n_\text{x}\},s_\text{y},s_\text{z})$
    \IF{$(d_\text{x}\le M_\text{x}$ and $d_\text{y}\le M_\text{y}$ and $d_\text{z}\le M_\text{z})$\\
    and $(d_\text{x}>s_\text{x})$ and $\zeta_1(\mathbf{n},\mathbf{d})=d_\text{x}\cdot d_\text{y}\cdot d_\text{z}$}
    \STATE $s_\text{x}=d_\text{x}$
    \ELSE
    \STATE \textit{barrier} += 1
    \ENDIF
    \STATE $(d_\text{x}, d_\text{y}) = (\min\{s_\text{x}, K_\text{x}-n_\text{x}\}, \min\{s_\text{y}+1, K_\text{y}-n_\text{y}\})$
    \IF{$(d_\text{x}\le M_\text{x}$ and $d_\text{y}\le M_\text{y}$ and $d_\text{z}\le M_\text{z})$\\
    and $(d_\text{y}>s_\text{y})$ and $\zeta_1(\mathbf{n},\mathbf{d})=d_\text{x}\cdot d_\text{y}\cdot d_\text{z}$}
    \STATE $s_\text{y}=d_\text{y}$
    \ELSE
    \STATE \textit{barrier} += 1
    \ENDIF
    \STATE $(d_\text{y}, d_\text{z}) = (\min\{s_\text{y}, K_\text{y}-n_\text{y}\}, \min\{s_\text{z}+1, K_\text{z}-n_\text{z}\})$
    \IF{$(d_\text{x}\le M_\text{x}$ and $d_\text{y}\le M_\text{y}$ and $d_\text{z}\le M_\text{z})$\\
    and $(d_\text{z}>s_\text{z})$ and $\zeta_1(\mathbf{n},\mathbf{d})=d_\text{x}\cdot d_\text{y}\cdot d_\text{z}$}
    \STATE $s_\text{z}=d_\text{z}$
    \ELSE
    \STATE \textit{barrier} += 1
    \ENDIF
    \STATE $i$ --= 1
    \IF{$(\textit{count}+s_\text{x} s_\text{y} s_\text{z}\!=\!n_\text{occupant})$\\
     or $(\textit{barriers}\!=\!3)$ or $(i\!=\!0)$}
    \STATE \textbf{break} (no further expansion is possible)
    \ENDIF
  \ENDWHILE
\STATE Compute sub-block anchor point: $\mathbf{x}_\text{min}=\mathbf{v}_\text{min}^{(b)}+\mathbf{n}\circ\boldsymbol{\Delta}_\text{min}^\text{(block)}$
\STATE $\mathcal{M}$.append$(\, \textit{SubBlockProperties}(\mathbf{x}_\text{min},\mathbf{s},\boldsymbol{\Delta}_\text{min}^\text{(block)},\lambda)\,)$
\STATE \textbf{Update occupancy map}: set $\theta[c_\text{x},c_\text{y},c_\text{z}]$ to 0 (inactive)\\
for all cells bounded by $\mathbf{x}_\text{min}$ and $\mathbf{x}_\text{max}=\mathbf{x}_\text{min}+\mathbf{s}$.
\STATE \textit{count} += $s_\text{x} s_\text{y} s_\text{z}$
\STATE Find remaining active cells: $\mathbf{a}\leftarrow \textit{IndexOfOccupants}(\theta)$
\ENDWHILE
\renewcommand{\algorithmicensure}{\textbf{Output:}}
\ENSURE consolidated sub-blocks $\mathcal{M}$
\renewcommand{\algorithmicensure}{\textbf{Ensure:}}
\end{algorithmic}
\end{multicols}
\end{algorithm*}

\begin{algorithm}[!th]
\small
\setcounter{algorithm}{0}
\floatname{algorithm}{Subroutine}
\caption{\quad \textbf{Compute sub-block properties}}\label{algo:bsu-subblock-attributes}
\begin{algorithmic}[1]
\STATE \textit{SubBlockProperties}($\mathbf{v}_{\min}^{(b)}$, $\mathbf{s}$, $\boldsymbol{\Delta}_\text{min}^\text{(block)}$, $\lambda$)
\STATE Compute:\\
\hspace{5mm} sub-block dimensions: $\boldsymbol{\Delta}_\text{sub-block}^{(b)}=\mathbf{s}\circ\boldsymbol{\Delta}_\text{min}^\text{(block)}$,\\
\hspace{5mm} sub-block max coordinates: $\mathbf{v}_{\max}^{(b)}=\mathbf{v}_{\min}^{(b)}+\boldsymbol{\Delta}_\text{sub-block}^{(b)}$\\
\hspace{5mm} sub-block centroid: $\mathbf{c}_\text{sub-block}^{(b)}=\frac{1}{2}(\mathbf{v}_\text{min}+\mathbf{v}_{\max}^{(b)})\in\mathbb{R}^3$\\
\hspace{5mm} sub-block label: $\lambda^{(b)}\leftarrow \lambda$
\STATE \textbf{return} $\langle \mathbf{c}_\text{sub-block}^{(b)}, \boldsymbol{\Delta}_\text{sub-block}^{(b)}, \lambda^{(b)}\rangle$\\
note: $\circ$ denotes the Hadamard (element-wise) product.\\
\end{algorithmic}
\end{algorithm}

\begin{algorithm*}[!th]
\small
\caption{\quad \textbf{Coordinate-ascent merging algorithm} (as used in block model de-fragmentation in \ref{sec:appendix-bsu-bm-modifications})}\label{algo:bsu-coordinate-ascent2}
\begin{multicols}{2}
\begin{algorithmic}[1]
\renewcommand{\algorithmicrequire}{\textbf{Pre-requisites:}}
\REQUIRE The list of merged blocks $\mathcal{M}$ is initialised with one tuple $\langle \mathbf{v}_\text{min}^{(b)}$, $\mathbf{s}^{(b)}$, $\lambda^{(b)}$, $n_\text{cells}^{\text{prev}(b)}$, $n_\text{cells}^{\text{curr}(b)}$, $\textit{subsumed}^{(b)}\!=\!0\rangle$ for each sub-block $b$ in class $\lambda$ within the parent block, where $\mathbf{v}_\text{min}^{(b)}\in\mathbb{R}^3$, $\mathbf{s}^{(b)}\in\mathbb{Z}^3$, $n_\text{cells}^{\text{prev}(b)}$ and $n_\text{cells}^{\text{curr}(b)}$ denote the sub-block minimum vertex, sub-block cell-dimensions, number of cells in the previous and current iteration, respectively.  The occupancy map $\theta$ is populated such that each active cell is assigned the relevant sub-block index, viz., $b$; all remaining cells are set to -1 (inactive).
\renewcommand{\algorithmicrequire}{\textbf{Input:}}
\renewcommand{\algorithmicensure}{\textbf{Parameters:}}
\REQUIRE $\mathcal{M}$ (with all $n_\text{cells}^{\text{prev}(b)}$ set to 0) and $\theta\in \mathbb{Z}^{K_\text{x}\times K_\text{y}\times K_\text{z}}$ 
\ENSURE same as Algorithm~\ref{algo:bsu-coordinate-ascent}
\renewcommand{\algorithmicensure}{\textbf{Variables:}}
\ENSURE Active sub-blocks: $\mathbf{a}=[]$ (initially an empty list)\\
\hspace{10mm} Otherwise, similar to Algorithm~\ref{algo:bsu-coordinate-ascent}
\STATE \textbf{do}
  \STATE\quad \textbf{Sort} list of block properties, $\mathcal{M}$, by cell count,\\
  \quad then minimum vertex, in ascending order.
  \STATE\quad \textbf{Find all active sub-blocks}:\\
  \quad $\mathbf{a}\leftarrow \textit{FindAllActiveSubBlocks}(\mathcal{M})$ where \textit{subsumed}=0
  \STATE\quad \textbf{if} $|\mathbf{a}|=1$ \textbf{then}
    \STATE\quad\quad \textbf{break}
  \STATE\quad \textbf{end if}
  \STATE\quad \textbf{for} each $b$ in ordered sub-blocks $\mathbf{a}$ \textbf{do}
    \STATE\quad\quad Set $n_\text{cells}^{\text{prev}(b)} = n_\text{cells}^{\text{curr}(b)}$
    \STATE\quad\quad \textbf{if} \textit{subsumed}$^{(b)}$ \textbf{then}
    \STATE\quad\quad\quad \textbf{continue}
    \STATE\quad\quad \textbf{end if}
    \STATE\quad\quad Set $i=T$ and $(s_\text{x},s_\text{y},s_\text{z})=\left(s_\text{x}^{(b)},s_\text{y}^{(b)},s_\text{z}^{(b)}\right)$
    \STATE\quad\quad Set $\mathbf{n}\!=\!(n_\text{x},n_\text{y},n_\text{z})\!=\!\textit{Subscript}(\text{lowest cell in block }b)$
    \STATE\quad\quad \textbf{while} true \textbf{do}
      \STATE\quad\quad\quad $\textit{barriers}=0$
      \STATE\quad\quad\quad $(d_\text{x}, d_\text{y}, d_\text{z})\leftarrow (\min\{s_\text{x}+1, K_\text{x}-n_\text{x}\},s_\text{y},s_\text{z})$
      \STATE\quad\quad\quad \textbf{if} $(d_\text{x}>s_\text{x})$ and $\textit{FeasibleCellExpansion}(\theta,\mathcal{M},b\mid$\\ \label{algo:feasible-expansion1}
      \STATE\quad\quad\quad $(n_\text{x}\!+\!s_\text{x},n_\text{y},n_\text{z}), (n_\text{x}\!+\!d_\text{x},n_\text{y}\!+\!s_\text{y},n_\text{z}\!+\!s_\text{z}),\text{``x''})$ \textbf{then}\\
      \STATE\quad\quad\quad\quad $s_\text{x}=s^{(b)}_\text{x}{}^\star$
      \STATE\quad\quad\quad \textbf{else}
      \STATE\quad\quad\quad\quad \textit{barrier} += 1
      \STATE\quad\quad\quad \textbf{end if}
      \STATE\quad\quad\quad $d_\text{x} = \min\{s_\text{x}, K_\text{x}-n_\text{x}\},$
      \STATE\quad\quad\quad $d_\text{y} = \min\{s_\text{y}+1, K_\text{y}-n_\text{y}\}$
      \STATE\quad\quad\quad \textbf{if} $(d_\text{y}>s_\text{y})$ and $\textit{FeasibleCellExpansion}(\theta,\mathcal{M},b\mid$\\ \label{algo:feasible-expansion2}
      \STATE\quad\quad\quad $(n_\text{x},n_\text{y}\!+\!s_\text{y},n_\text{z}), (n_\text{x}\!+\!s_\text{x},n_\text{y}\!+\!d_\text{y},n_\text{z}\!+\!s_\text{z}),\text{``y''})$ \textbf{then}\\
      \STATE\quad\quad\quad\quad $s_\text{y}=s^{(b)}_\text{y}{}^\star$
      \STATE\quad\quad\quad \textbf{else}
      \STATE\quad\quad\quad\quad \textit{barrier} += 1
      \STATE\quad\quad\quad \textbf{end if}
      \STATE\quad\quad\quad $d_\text{y} = \min\{s_\text{y}, K_\text{y}-n_\text{y}\}$
      \STATE\quad\quad\quad $d_\text{z} = \min\{s_\text{z}+1, K_\text{z}-n_\text{z}\}$
      \STATE\quad\quad\quad \textbf{if} $(d_\text{z}>s_\text{z})$ and $\textit{FeasibleCellExpansion}(\theta,\mathcal{M},b\mid$\\ \label{algo:feasible-expansion3}
      \STATE\quad\quad\quad $(n_\text{x},n_\text{y},n_\text{z}\!+\!s_\text{z}), (n_\text{x}\!+\!s_\text{x},n_\text{y}\!+\!s_\text{y},n_\text{z}\!+\!d_\text{z}), \text{``z''})$ \textbf{then}\\
      \STATE\quad\quad\quad\quad $s_\text{z}=s^{(b)}_\text{z}{}^\star$
      \STATE\quad\quad\quad \textbf{else}
      \STATE\quad\quad\quad\quad \textit{barrier} += 1
      \STATE\quad\quad\quad \textbf{end if}
      \STATE\quad\quad\quad $i$ --= 1
      \STATE\quad\quad\quad \textbf{if} $(s_\text{x}=K_\text{x}-n_\text{x}$ and $s_\text{y}=K_\text{y}-n_\text{y}$ and
      \STATE\quad\quad\quad $s_\text{z}=K_\text{z}-n_\text{z})$ or $(\textit{barriers}\!=\!3)$ or $(i\!=\!0)$ \textbf{then}\\
      \STATE\quad\quad\quad\quad \textbf{break} (no further expansion is possible)
      \STATE\quad\quad\quad \textbf{end if}
    \STATE\quad\quad \textbf{end while}
  \STATE\quad \textbf{end for}
\STATE \textbf{while} $n_\text{cells}^{\text{curr}(b)}\ne n_\text{cells}^{\text{prev}(b)}$ for any block in $\mathcal{M}$
\renewcommand{\algorithmicensure}{\textbf{Output:}}
\STATE Remove all subsumed sub-blocks from $\mathcal{M}$
\ENSURE consolidated sub-blocks $\mathcal{M}$\\
\textsc{note} $^{\star}$: The properties of $\mathcal{M}^{(b)}$ are updated implicitly by \textit{FeasibleCellExpansion} when the expansion is feasible.\\ Details are given in Subroutine~\ref{algo:bsu-feasible-cell-expansion}.
\renewcommand{\algorithmicensure}{\textbf{Ensure:}}
\end{algorithmic}
\end{multicols}
\end{algorithm*}

\begin{algorithm*}[!th]
\small
\setcounter{algorithm}{1}
\floatname{algorithm}{Subroutine}
\caption{\quad \textbf{Feasibility tests and state updates during block expansion}}\label{algo:bsu-feasible-cell-expansion}
\begin{multicols}{2}
\begin{algorithmic}[1]
\renewcommand{\algorithmicensure}{\textbf{Parameters:}}
\ENSURE Parent block cell dimensions: $(K_\text{x},K_\text{y},K_\text{z})\in\mathbb{Z}^3$\\
\hspace{10mm}Current block cell dimensions: $(s_\text{x},s_\text{y},s_\text{z})\in\mathbb{Z}^3$\\
\hspace{10mm}Max. merge cell dimensions: $(M_\text{x},M_\text{y},M_\text{z})\in\mathbb{Z}^3$\\
\renewcommand{\algorithmicensure}{\textbf{Mutable objects:}}
\ENSURE Occupancy map: $\theta\in\mathbb{Z}^{K_\text{x}\times K_\text{y}\times K_\text{z}}$\\
\hspace{21mm}List of block properties: $\mathcal{M}$
\renewcommand{\algorithmicensure}{\textbf{Notations:}}
\ENSURE $\circ$ Delta region: $\mathcal{R}$\\
\hspace{11mm}$\circ$ Length along axis of expansion for sub-blocks\\
\hspace{13mm} found in the delta region: $l_{b'\in\mathcal{R}}(direction)$\\
\hspace{11mm}$\circ$ Number of cells from sub-blocks found in the\\
\hspace{13mm} delta region: $n_{\mathcal{R}}^\text{(cells)}$\\
\hspace{11mm}$\circ$ Unique set of sub-blocks in delta region: $\mathcal{S}$\\
\STATE \textit{FeasibleCellExpansion}($\theta$, $\mathcal{M}$, $b\mid$\\
\qquad\qquad $(n_\text{x}^0, n_\text{y}^0, n_\text{z}^0)$, $(n_\text{x}^1, n_\text{y}^1, n_\text{z}^1)$, \textit{direction})
\IF{$n_\text{x}^0\ge K_\text{x}$ or $n_\text{y}^0\ge K_\text{y}$ or $n_\text{z}^0\ge K_\text{z}$}
  \STATE \textbf{return} false
\ENDIF
\FOR{each cell $(c_\text{x},c_\text{y},c_\text{z})$ in $\mathcal{R}$}
  \STATE Let sub-block index $b'=\theta(c_\text{x},c_\text{y},c_\text{z})$
  \IF{$b'\ne -1$}
  \STATE $\mathcal{S}$.insert(\,$b'$\,)\\
  \hspace{3mm}\textbf{else}\ \ $\mathcal{R}$ contains at least one foreign cell
  \STATE \textbf{return} false
  \ENDIF
\ENDFOR
\IF{$l_{b'\in\mathcal{R}}(direction)$ is identical for all blocks in $\mathcal{R}$}
  \STATE Set $n_\text{extend} = l_{b'\in\mathcal{R}}(direction)\in\mathbb{Z^{+}}$
\ELSE
  \STATE \textbf{return} false\ \ (failed uniform length requirement)
\ENDIF
\STATE Let $\tilde{\mathcal{S}}=\{b'\in\mathcal{S}\mid \textit{subsumed}^{\,(b')}\!=\!\text{false}\}\subseteq \mathcal{S}$
\IF{\textit{direction} is ``x''}
  \IF{$(s_\text{x}+n_\text{extend},s_\text{y},s_\text{z})$ exceeds $(M_\text{x},M_\text{y},M_\text{z})$}
    \STATE \textbf{return} false
  \ELSE
    \STATE Compute $n_{\mathcal{R}}^\text{(cells)}$ from blocks $b'\in\tilde{\mathcal{S}}$
    \STATE Set $\textit{compatible}=(n_{\mathcal{R}}^\text{(cells)}=n_\text{extend}\cdot s_\text{y}\cdot s_\text{z})?$ true : false
  \ENDIF
\ELSIF{\textit{direction} is ``y''}
  \IF{$(s_\text{x},s_\text{y}+n_\text{extend},s_\text{z})$ exceeds $(M_\text{x},M_\text{y},M_\text{z})$}
    \STATE \textbf{return} false
  \ELSE
    \STATE Compute $n_{\mathcal{R}}^\text{(cells)}$ from blocks $b'\in\tilde{\mathcal{S}}$
    \STATE Set $\textit{compatible}=(n_{\mathcal{R}}^\text{(cells)}=s_\text{x}\cdot n_\text{extend}\cdot s_\text{z})?$ true : false
  \ENDIF
\ELSE
  \IF{$(s_\text{x},s_\text{y},s_\text{z}+n_\text{extend})$ exceeds $(M_\text{x},M_\text{y},M_\text{z})$}
    \STATE \textbf{return} false
  \ELSE
    \STATE Compute $n_{\mathcal{R}}^\text{(cells)}$ from blocks $b'\in\tilde{\mathcal{S}}$
    \STATE Set $\textit{compatible}=(n_{\mathcal{R}}^\text{(cells)}=s_\text{x}\cdot s_\text{y}\cdot n_\text{extend})?$ true : false
  \ENDIF
\ENDIF
\IF{\textit{compatible}}
  \STATE \textbf{Update block properties list $\mathcal{M}$}
  \STATE Set $\textit{subsumed}^{\,(\beta)}=\text{true}\ \ \forall \beta\in\tilde{\mathcal{S}}$
  \STATE Set $n_\text{cells}^{\text{prev}(b)}=n_\text{cells}^{\text{curr}(b)}$
  \STATE Set $n_\text{cells}^{\text{curr}(b)}$\,+=\,$\sum_{\beta\in\tilde{\mathcal{S}}} n_\text{cells}^{\text{curr}(\beta)}$
  \STATE \textbf{Update occupancy map $\theta$}
  \STATE Set $\theta(c_\text{x},c_\text{y},c_\text{z})=b$ for all cells in blocks $\beta\in\tilde{\mathcal{S}}$.
\ENDIF
\STATE \textbf{return} \textit{compatible}
\end{algorithmic}
\end{multicols}
\end{algorithm*}

\begin{algorithm*}[!th]
\small
\caption{\quad \textbf{Block merging with multiple scans and optimised block aspect ratios}}\label{algo:bsu-final-block-merging}
\begin{multicols}{2}
\begin{algorithmic}[1]
\renewcommand{\algorithmicfor}{\textbf{parallel for}}
\FOR{thread $t$ from 0 to $n_\text{thread}\!-\!1$}
\renewcommand{\algorithmicfor}{\textbf{for}}
  \STATE Set $\textit{coalesced\_blocks}^{(t)} = \O$
  \FOR{parent block $p$ in $\{t+i\cdot n_\text{thread}\}_{i\in\mathbb{Z}}$ and $p\!<\!n_\text{parent}^{\text{(block)}}$}\label{algo:interleaved-parent-indices}
    \STATE Find input blocks $\mathcal{B}_p$ contained in $p$
    \FOR{each class $\lambda$ within $p$}
        \STATE Find blocks $\mathcal{B}_{p,\lambda}$ with label $\lambda$
        \STATE Let the cost for current best solution $f_{*}=\infty$
        \IF{convention is \textit{DissolveSubBlockBoundaries}}
          \FOR{each scan pattern $\pi$}
              \STATE Populate occupancy map $\theta\!\in\!\{0,1\}^{K_\text{x}\times K_\text{y}\times K_\text{z}}$ s.t. active cells in $\mathcal{B}_{p,\lambda}$ are set to 1; 0 otherwise.
              \STATE Invoke coordinate-ascent (Algorithm~\ref{algo:bsu-coordinate-ascent}) to\\obtain the consolidated blocks $\mathcal{M}_{\pi}$
              \STATE Compute the cost $f(\mathcal{M}_{\pi})^{\,\dag}$
              \IF{$f_{*} > f(\mathcal{M}_{\pi})$}
                \STATE Set $f_{*}=f(\mathcal{M}_{\pi})$ and $\mathcal{M}_{*}=\mathcal{M}_{\pi}$
              \ENDIF
          \ENDFOR
          \STATE $\textit{coalesced\_blocks}^{(t)}.\text{append}(\,\mathcal{M}_{*}\,)$
        \ELSIF{convention is \textit{PersistentBlockMemory}}
          \FOR{each scan pattern $\pi$}
            \STATE Populate occupancy map $\theta\in\mathbb{Z}^{K_\text{x}\times K_\text{y}\times K_\text{z}}$ s.t. all active cells in $\mathcal{B}_{p,\lambda}$ are set to the relevant sub-block index $b\in \mathcal{B}_{p,\lambda}$; -1 otherwise.
            \STATE Invoke coordinate-ascent (Algorithm~\ref{algo:bsu-coordinate-ascent2}) to\\obtain the consolidated blocks $\mathcal{M}_{\pi}$
            \STATE Compute the cost $f(\mathcal{M}_{\pi})^{\,\dag}$
            \IF{$f_{*} > f(\mathcal{M}_{\pi})$}
              \STATE Set $f_{*}=f(\mathcal{M}_{\pi})$ and $\mathcal{M}_{*}=\mathcal{M}_{\pi}$
            \ENDIF
          \ENDFOR
          \STATE $\textit{coalesced\_blocks}^{(t)}.\text{append}(\,\mathcal{M}_{*}\,)$
        \ENDIF
    \ENDFOR
  \ENDFOR
  \STATE Signal when thread $t$ completes its task
\ENDFOR
\STATE Aggregate results: $\textit{solution}\leftarrow \{\textit{coalesced\_blocks}^{(t)}\}$
\renewcommand{\algorithmicensure}{\textbf{Output:}}
\ENSURE $\textit{solution}$\\
$\textsc{note:}^{\dag}$ using the objective function based on\\volume-weighted block aspect ratio, for instance.
\renewcommand{\algorithmicensure}{\textbf{Ensure:}}
\end{algorithmic}
\end{multicols}
\end{algorithm*}

\clearpage
\newpage
\section{Octree decomposition and merging}\label{sec:appendix-octree-decomposition-merging}
The two octree schemes considered in the paper are the standard octree decomposition, and octree with intra-scale merging. Starting at full resolution ($d=0$), at each level of the spatial hierarchy, a rectangular block with dimensions $\boldsymbol{\Delta}^{(d)}\in\mathbb{R}^3$ may be split into eight sub-blocks (or cells) called an octant, where the dimensions of each sub-block are essentially halved along each axis, yielding $\boldsymbol{\Delta}^{(d+1)}=\frac{1}{2}\boldsymbol{\Delta}^{(d)}$. A split is performed when 2 or more of its sub-blocks at resolution  $\boldsymbol{\Delta}^{(d+1)}$ carry different labels. This decomposition is performed recursively and stops only when the 3D block region becomes homogeneous (all 8 cells have the same label) or when the maximum decomposition level $D$ is reached. Such hierarchical structures are well studied in the literature, see \cite{samet-88} and \cite{tamminen1984efficient} for instance. The purpose of this section is to clarify what intra-scale block merging means in this work, and how it relates to the standard octree.

\begin{figure}[!th]
\centering
\includegraphics[width=87.5mm]{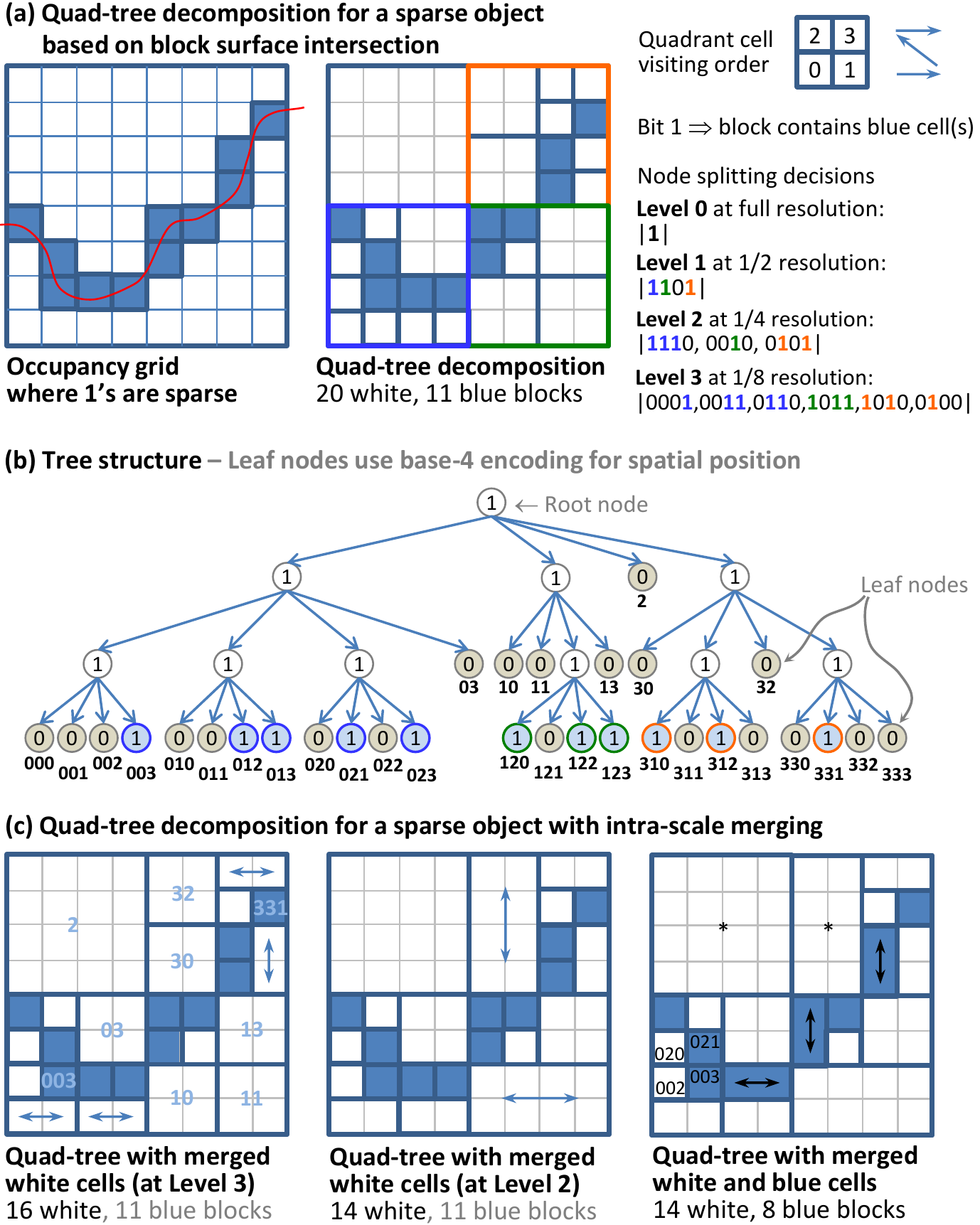}
\caption{Octree for encoding sparse object such as edges}
\label{fig:sm-octree-encoding-sparsity}
\end{figure}

Octree decomposition is a popular technique for encoding sparse data such as edge pixels in an image array. Fig.~\ref{fig:sm-octree-encoding-sparsity} provides an example whereby block surface intersections are localised by blue cells in (a-left). A complete quad-tree decomposition of this region into sub-blocks at $\frac{1}{2}$, $\frac{1}{4}$ and $\frac{1}{8}$ scale is shown in (a-middle). Following a particular quadrant cell scanning order, the 2D pattern may by represented by the tree-structure in (b). Intra-scale block merging has the specific meaning described in (c) where coalesced blocks are limited to adjacent cells within a quadrant; the arrows in (c-left) and (c-middle) show this happening at two spatial scales, $d=3$ and $d=2$. The final result after octree decomposition and intra-scale merging is shown in (c-right). This picture illustrates that further merging is in fact possible --- for instance between the white cells 002 and 020, or blue cells 003 and 021 --- if inter-scale merging is permitted. We opted not to challenge these rules for the octree approach since these merging opportunities have already been exploited by the proposed methods, and intra-scale merging has its place in our performance comparison. For simplicity, the region is treated as a 2D block, however all aspects generalise to three-dimensions (from quadrant to octant) and all processes involved in the actual experiments operate in 3D.

\begin{figure}[!th]
\centering
\includegraphics[width=87.5mm]{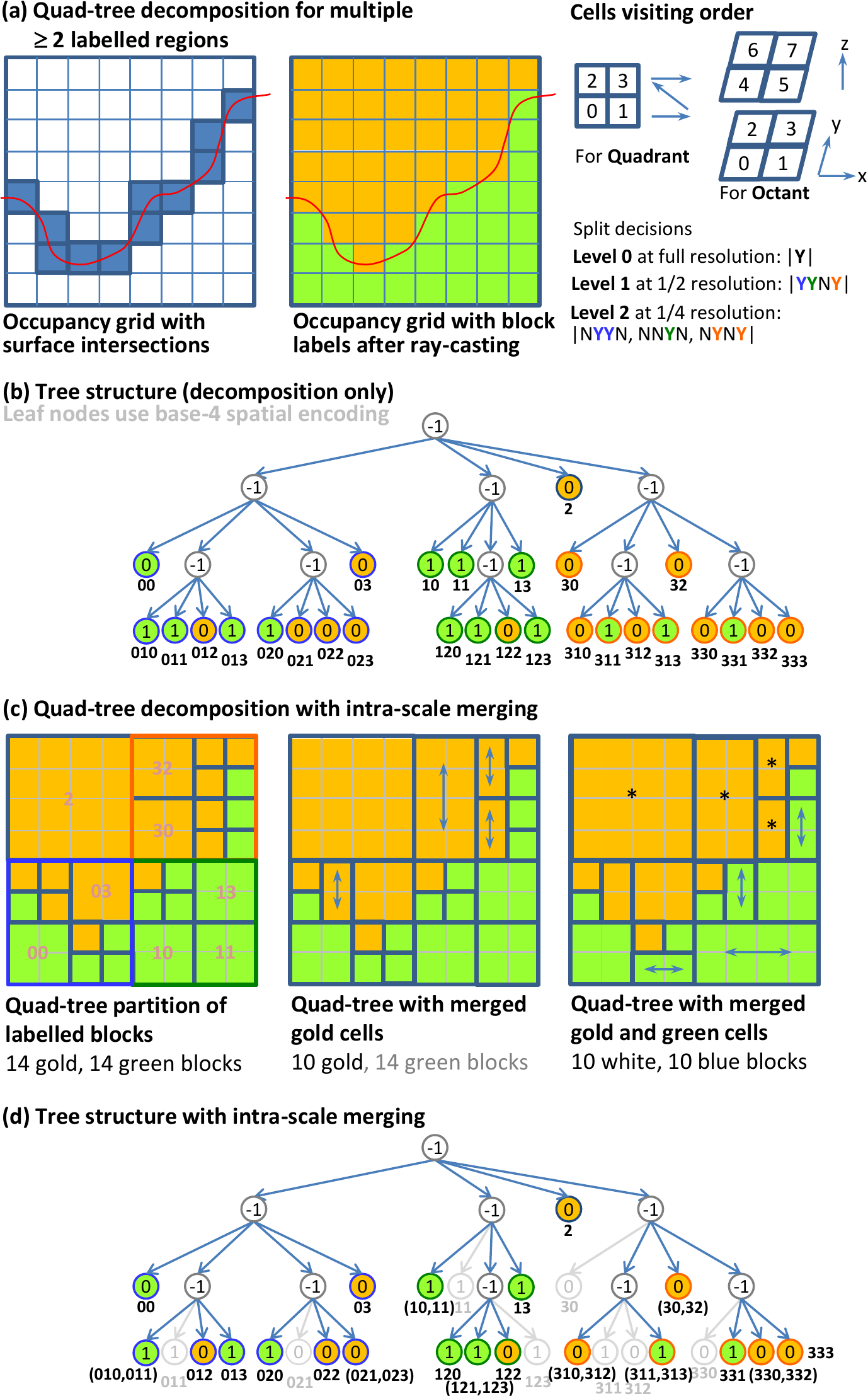}
\caption{Octree decomposition and intra-scale merging for multiple regions}
\label{fig:sm-octree-for-regions}
\end{figure}

Extending these ideas to encode non-sparse regions, we observe that standard octree decomposition works in a top-down manner and has no innate ability for labelling cells at the minimum block size. Therefore, ray-tracing is used (since it forms part of the block model spatial restructuring workflow) to label cells as 0 or 1; colouring cells in gold or green in Fig.~\ref{fig:sm-octree-for-regions}(a-middle) depending upon which side of the surface they are on.

Applying octree decomposition produces the tree-structure shown in Fig.~\ref{fig:sm-octree-for-regions}(b) where split nodes are labelled -1, leaf nodes are labelled 0 or 1 (when there are two regions) and coloured gold or green accordingly. Result obtained with further intra-scale merging is shown in (c). As before, arrows indicate the blocks which have been merged within a quadrant at a given scale. The resultant tree-structure after intra-scale merging is depicted in (d). Comparing with (b), branches connecting with blocks which have been subsumed are evidently pruned with the corresponding nodes removed. Our earlier remarks on further inter-scale merging opportunities also exist here, for instance, blocks marked with asterisk in (c-right) can all potentially be combined into a single block. Although we focused our attention on two regions in this example, all relevant aspects generalise to three or more regions when multiple surfaces are involved.

\subsection{Major difference between quadtree and octree}\label{sec:appendix-difference-quadtree-octree}
For an octree, the major difference with respect to quadtree are the candidates considered during intra-scale merging. Following the octant cell scanning order shown in Fig.~\ref{fig:sm-octree-for-regions}\,(a-right), prospective 2-cell merge candidates include basically 12 edges: viz., $\{(0,1), (0,2), (1,3), (2,3)\}$ and $\{(4,5), (4,6), (5,7), (6,7)\}$ from the top and bottom sides, and similarly $\{(2,6), (3,7)\} \cup \{(0,4), (1,5)\}$ from the north and south sides of an octant. Prospective 4-cell merge candidates include 6 square faces: \{(0,1,2,3), (4,5,6,7), (0,1,4,5), (2,3,6,7), (0,2,4,6), (1,3,5,7)\}.

\section{Detailed octree subblocking comparison}\label{sec:appendix-detailed-octree-subblocking-comparison}
This section provides a more detailed breakdown of the model block count results presented in Sec.~\ref{sec:block-count-vs-octree-analysis} of \cite{leung-19a}. Henceforth, we use the word `Octree' to denote standard octree decomposition. When the `+Merge' suffix is added, intra-scale merging is attempted between compatible cells within each octant. This means, edge-connected cells within the same octant may be combined in groups of two or four to form a rectangular or squared block as described above (in Appendix~\ref{sec:appendix-difference-quadtree-octree}). However, inter-scale merging across different decomposition levels is not permitted. `Proposed-P' refers to the proposed block merging algorithm performed under the \textit{persistent} block memory convention. `Proposed-D' refers to the same under the \textit{dissolved} subblock boundary convention. `Domain' means geological domain and `\%\,volume' means percentage of the total volume in the modelled region.

To promote spatial awareness, an animated sequence of the test site's domain structure is shown layer-by-layer in Table~\ref{tab:detailed-results-vs-octree-d3} where the domain colour palette matches the colour labels used in the tables. A geology background is not required to understand this data. However, domain labels annotated by M, N and H may be interpreted as `mineralised', `non-mineralised' and `hydrated' domains, respectively, by geologists.

\begin{table*}[!th]
\small
\setlength\tabcolsep{4pt}
\caption{Block model statistics: proposed methodology vs octree (with D=3 decomposition levels)}\label{tab:detailed-results-vs-octree-d3}
\resizebox{\textwidth}{!}{
\begin{tabular}{|l|c|cccc|cccc|}\hline
Domain & \%\,volume &\multicolumn{4}{c|}{block count} & \multicolumn{4}{c|}{volume-weighted block aspect ratio}\\ \hline
& & Octree & Octree\,+\,Merge & Proposed-P & Proposed-D & Octree & Octree\,+\,Merge & Proposed-P & Proposed-D\\ \hline
{\color{fire0}$\bsquare$} $N_{0}$ & 0.012340 & 2644 & 1154 & 772 & 704 & 2.5 & 3.954896 & 9.251789 & 3.685130\\ \hline
{\color{fire1}$\bsquare$} $M_{0}$ & 0.012216 & 2422 & 1057 & 791 & 594 & 2.5 & 3.923701 & 6.925476 & 3.811883\\ \hline
{\color{fire2}$\bsquare$} $N_{1}$ & 2.071587 & 326463 & 123708 & 59908 & 52925 & 2.5 & 2.435113 & 3.252310 & 2.386576\\ \hline
{\color{fire3}$\bsquare$} $N_{2}$ & 0.571025 & 27523 & 11878 & 7256 & 7679 & 2.5 & 3.301445 & 4.297926 & 2.631313\\ \hline
{\color{fire4}$\bsquare$} $M_{1}$ & 0.000045 & 17 & 12 & 12 & 12 & 2.5 & 3.529412 & 3.088235 & 3.088235\\ \hline
{\color{fire5}$\bsquare$} $N_{3}$ & 0.247183 & 27769 & 12558 & 8132 & 8728 & 2.5 & 3.954645 & 6.093452 & 2.532378\\ \hline
{\color{fire6}$\bsquare$} $N_{4}$ & 0.318811 & 34279 & 15510 & 10080 & 10732 & 2.5 & 3.954323 & 6.095554 & 2.586614\\ \hline
{\color{fire7}$\bsquare$} $N_{5}$ & 1.036601 & 80587 & 35968 & 23032 & 23655 & 2.5 & 3.552698 & 5.208733 & 2.565268\\ \hline
{\color{fire8}$\bsquare$} $M_{2}$ & 0.074106 & 13096 & 5958 & 3710 & 3788 & 2.5 & 3.723641 & 5.582627 & 3.214119\\ \hline
{\color{fire9}$\bsquare$} $N_{6}$ & 1.996944 & 158170 & 70237 & 44619 & 45183 & 2.5 & 3.610081 & 5.044647 & 2.649020\\ \hline
{\color{fire10}$\bsquare$} $M_{3}$ & 0.426214 & 53367 & 24250 & 15604 & 15490 & 2.5 & 3.638962 & 5.590088 & 2.862343\\ \hline
{\color{fire11}$\bsquare$} $N_{7}$ & 1.058833 & 166835 & 75399 & 46713 & 47876 & 2.5 & 3.871672 & 5.621888 & 2.996686\\ \hline
{\color{fire12}$\bsquare$} $M_{4}$ & 0.112265 & 23454 & 11036 & 7113 & 7121 & 2.5 & 3.780942 & 5.277800 & 3.288215\\ \hline
{\color{fire13}$\bsquare$} $H_{0}$ & 0.332034 & 64151 & 29953 & 21128 & 19535 & 2.5 & 3.784879 & 7.178617 & 3.074582\\ \hline
{\color{fire14}$\bsquare$} $N_{8}$ & 2.363637 & 241316 & 106972 & 67694 & 69211 & 2.5 & 3.942829 & 5.903363 & 2.878910\\ \hline
{\color{fire15}$\bsquare$} $M_{5}$ & 0.035386 & 8595 & 4190 & 2824 & 2828 & 2.5 & 3.720205 & 6.229485 & 3.344340\\ \hline
{\color{fire16}$\bsquare$} $N_{9}$ & 1.500652 & 249267 & 111807 & 67270 & 68843 & 2.5 & 3.928390 & 5.649683 & 3.162396\\ \hline
{\color{fire17}$\bsquare$} $M_{6}$ & 0.005508 & 1579 & 811 & 576 & 568 & 2.5 & 3.527364 & 4.199520 & 3.340855\\ \hline
{\color{fire18}$\bsquare$} $H_{1}$ & 0.052937 & 12489 & 6035 & 4357 & 4132 & 2.5 & 3.818116 & 6.895195 & 3.289241\\ \hline
{\color{fire19}$\bsquare$} $N_{10}$ & 5.062708 & 394979 & 173877 & 108587 & 110817 & 2.5 & 3.756453 & 5.372227 & 2.753414\\ \hline
{\color{fire20}$\bsquare$} $M_{7}$ & 0.230237 & 31048 & 13958 & 8972 & 8687 & 2.5 & 3.753732 & 4.868516 & 2.861059\\ \hline
{\color{fire21}$\bsquare$} $N_{11}$ & 4.327004 & 450788 & 198725 & 123289 & 126389 & 2.5 & 3.966091 & 5.891937 & 2.966619\\ \hline
{\color{fire22}$\bsquare$} $M_{8}$ & 0.119425 & 21038 & 9726 & 6387 & 6238 & 2.5 & 3.602132 & 5.310738 & 3.053729\\ \hline
{\color{fire23}$\bsquare$} $N_{12}$ & 6.059302 & 517338 & 225301 & 138735 & 141283 & 2.5 & 3.893502 & 5.591457 & 2.858506\\ \hline
{\color{fire24}$\bsquare$} $M_{9}$ & 0.092165 & 15999 & 7252 & 4671 & 4601 & 2.5 & 3.480721 & 5.185372 & 2.839088\\ \hline
{\color{fire25}$\bsquare$} $H_{2}$ & 0.193627 & 42416 & 20254 & 14320 & 13380 & 2.5 & 3.691417 & 6.955868 & 3.105813\\ \hline
{\color{fire26}$\bsquare$} $N_{13}$ & 3.049206 & 474133 & 208258 & 121712 & 124319 & 2.5 & 3.935326 & 5.502560 & 3.186988\\ \hline
{\color{fire27}$\bsquare$} $M_{10}$ & 0.005196 & 1692 & 822 & 563 & 526 & 2.5 & 3.379135 & 5.267176 & 3.388906\\ \hline
{\color{fire28}$\bsquare$} $N_{14}$ & 0.001007 & 339 & 180 & 138 & 133 & 2.5 & 3.366142 & 5.218110 & 3.547769\\ \hline
{\color{fire29}$\bsquare$} $N_{15}$ & 68.631674 & 797448 & 323952 & 175932 & 174585 & 2.5 & 2.599801 & 2.867128 & 2.506040\\ \hline
{\color{fire30}$\bsquare$} $M_{11}$ & 0.000127 & 48 & 28 & 20 & 23 & 2.5 & 3.645833 & 5.781250 & 3.564583\\ \hline
\hline
\multicolumn{2}{|l|}{Total (avg.\,by volume)} & 4241289 & 1830826 & 1094917 & 1100585 & 2.5 & 2.958828 & 3.668188 & 2.615373\\ \hline
\multicolumn{2}{|l|}{Total (avg.\,by block count)} & same & same & same & same & 2.5 & 3.535403 & 5.068056 & 2.839873\\ \hline
\hline
\multicolumn{2}{|l|}{Ratio} & 100.000 & 43.167 & 25.816 & 25.949 & \multicolumn{4}{c|}{}\\ \hline
\multicolumn{10}{c}{}\\
\multicolumn{10}{c}{\animategraphics[autoplay,loop,trim=0 0 0 4cm,width=105mm]{4}{fig30-}{00}{30}}\\
\multicolumn{10}{c}{\textbf{Animated sequence --- birds eye view of Site 8's spatial structure}\label{fig:appendix-animated-site8-geological-domains}}\\
\multicolumn{10}{c}{Geological domains are peeled back layer by layer in this animation}\\
\end{tabular}
}
\end{table*}

\begin{table*}[!th]
\small
\setlength\tabcolsep{4pt}
\caption{Block model statistics: proposed methodology vs octree (with D=4 decomposition levels)}\label{tab:detailed-results-vs-octree-d4}
\resizebox{\textwidth}{!}{
\begin{tabular}{|l|c|cccc|cccc|}\hline
Domain & \%\,volume &\multicolumn{4}{c|}{block count} & \multicolumn{4}{c|}{volume-weighted block aspect ratio}\\ \hline
& & Octree & Octree\,+\,Merge & Proposed-P & Proposed-D & Octree & Octree\,+\,Merge & Proposed-P & Proposed-D\\ \hline
{\color{fire0}$\bsquare$} $N_{0}$ & 0.012367 & 11742 & 5065 & 3317 & 2746 & 2.5 & 3.900329 & 15.549850 & 4.059046\\ \hline
{\color{fire1}$\bsquare$} $M_{0}$ & 0.012224 & 10946 & 4650 & 3420 & 2201 & 2.5 & 3.909655 & 10.466585 & 4.674493\\ \hline
{\color{fire2}$\bsquare$} $N_{1}$ & 2.071568 & 1429469 & 533237 & 223538 & 185677 & 2.5 & 2.438087 & 4.852193 & 3.363062\\ \hline
{\color{fire3}$\bsquare$} $N_{2}$ & 0.571042 & 108496 & 47062 & 26560 & 28328 & 2.5 & 3.330618 & 6.133705 & 2.987043\\ \hline
{\color{fire4}$\bsquare$} $M_{1}$ & 0.000039 & 112 & 62 & 45 & 44 & 2.5 & 3.592437 & 4.118487 & 3.170588\\ \hline
{\color{fire5}$\bsquare$} $N_{3}$ & 0.247097 & 116150 & 52202 & 31105 & 33811 & 2.5 & 3.882805 & 9.738552 & 3.590038\\ \hline
{\color{fire6}$\bsquare$} $N_{4}$ & 0.318835 & 143282 & 64352 & 38541 & 41093 & 2.5 & 3.928394 & 9.994956 & 3.620897\\ \hline
{\color{fire7}$\bsquare$} $N_{5}$ & 1.036702 & 342263 & 150890 & 87578 & 89680 & 2.5 & 3.623427 & 8.200493 & 3.193486\\ \hline
{\color{fire8}$\bsquare$} $M_{2}$ & 0.074196 & 60521 & 26754 & 14949 & 15198 & 2.5 & 3.768124 & 8.326254 & 3.820477\\ \hline
{\color{fire9}$\bsquare$} $N_{6}$ & 1.996801 & 667620 & 292632 & 169023 & 170337 & 2.5 & 3.644469 & 7.759336 & 3.312802\\ \hline
{\color{fire10}$\bsquare$} $M_{3}$ & 0.426296 & 235590 & 104610 & 60747 & 59749 & 2.5 & 3.610870 & 8.889136 & 3.776128\\ \hline
{\color{fire11}$\bsquare$} $N_{7}$ & 1.058725 & 728816 & 324163 & 182289 & 186255 & 2.5 & 3.892632 & 8.553657 & 3.970692\\ \hline
{\color{fire12}$\bsquare$} $M_{4}$ & 0.112501 & 110475 & 49990 & 28969 & 28582 & 2.5 & 3.779309 & 7.580900 & 4.003741\\ \hline
{\color{fire13}$\bsquare$} $H_{0}$ & 0.331797 & 298282 & 135474 & 88804 & 78141 & 2.5 & 3.779701 & 11.899441 & 3.762755\\ \hline
{\color{fire14}$\bsquare$} $N_{8}$ & 2.362666 & 1022988 & 450290 & 259668 & 261922 & 2.5 & 3.936937 & 9.418536 & 3.851701\\ \hline
{\color{fire15}$\bsquare$} $M_{5}$ & 0.035805 & 45500 & 21242 & 12706 & 12578 & 2.5 & 3.677538 & 9.116177 & 3.699823\\ \hline
{\color{fire16}$\bsquare$} $N_{9}$ & 1.501456 & 1098197 & 485024 & 264369 & 268390 & 2.5 & 3.963615 & 8.493589 & 4.080947\\ \hline
{\color{fire17}$\bsquare$} $M_{6}$ & 0.005730 & 9258 & 4734 & 2985 & 2929 & 2.5 & 3.564548 & 5.640211 & 3.770489\\ \hline
{\color{fire18}$\bsquare$} $H_{1}$ & 0.053013 & 62734 & 29214 & 19290 & 17460 & 2.5 & 3.701885 & 10.991420 & 3.706879\\ \hline
{\color{fire19}$\bsquare$} $N_{10}$ & 5.064220 & 1663577 & 727924 & 410459 & 414897 & 2.5 & 3.794050 & 8.463884 & 3.543067\\ \hline
{\color{fire20}$\bsquare$} $M_{7}$ & 0.228323 & 142484 & 61726 & 35310 & 33610 & 2.5 & 3.726269 & 7.352394 & 3.805164\\ \hline
{\color{fire21}$\bsquare$} $N_{11}$ & 4.325235 & 1895420 & 828568 & 466319 & 473527 & 2.5 & 3.942704 & 9.354241 & 4.004534\\ \hline
{\color{fire22}$\bsquare$} $M_{8}$ & 0.119550 & 99113 & 44273 & 25985 & 24900 & 2.5 & 3.579750 & 7.963333 & 3.817275\\ \hline
{\color{fire23}$\bsquare$} $N_{12}$ & 6.059267 & 2143315 & 928956 & 519128 & 522118 & 2.5 & 3.921219 & 8.812658 & 3.766239\\ \hline
{\color{fire24}$\bsquare$} $M_{9}$ & 0.092176 & 73872 & 32446 & 18975 & 18255 & 2.5 & 3.478816 & 8.269290 & 3.744605\\ \hline
{\color{fire25}$\bsquare$} $H_{2}$ & 0.193962 & 201251 & 92622 & 60682 & 54519 & 2.5 & 3.722921 & 11.130710 & 3.710936\\ \hline
{\color{fire26}$\bsquare$} $N_{13}$ & 3.050039 & 2038137 & 885891 & 469694 & 475684 & 2.5 & 3.970251 & 8.126027 & 4.180359\\ \hline
{\color{fire27}$\bsquare$} $M_{10}$ & 0.005201 & 9332 & 4259 & 2611 & 2405 & 2.5 & 3.427114 & 7.226263 & 3.727820\\ \hline
{\color{fire28}$\bsquare$} $N_{14}$ & 0.001046 & 2086 & 1004 & 647 & 610 & 2.5 & 3.463970 & 7.412168 & 3.983059\\ \hline
{\color{fire29}$\bsquare$} $N_{15}$ & 68.632008 & 2643443 & 1065825 & 525508 & 501270 & 2.5 & 2.605537 & 3.283658 & 2.576202\\ \hline
{\color{fire30}$\bsquare$} $M_{11}$ & 0.000115 & 284 & 143 & 89 & 102 & 2.5 & 3.832853 & 8.900865 & 2.621326\\ \hline
\hline
\multicolumn{2}{|l|}{Total (avg.\,by volume)} & 17414755 & 7455284 & 4053310 & 4007018 & 2.5 & 2.968178 & 4.901418 & 2.941797\\ \hline
\multicolumn{2}{|l|}{Total (avg.\,by block count)} & same & same & same & same & 2.5 & 3.585874 & 7.861600 & 3.650470\\ \hline
\hline
\multicolumn{2}{|l|}{Ratio} & 100.000 & 42.810 & 23.275 & 23.009 & \multicolumn{4}{c|}{}\\ \hline
\multicolumn{10}{c}{}\\
\multicolumn{10}{c}{\includegraphics[clip,trim=0 0 0 4cm,width=105mm]{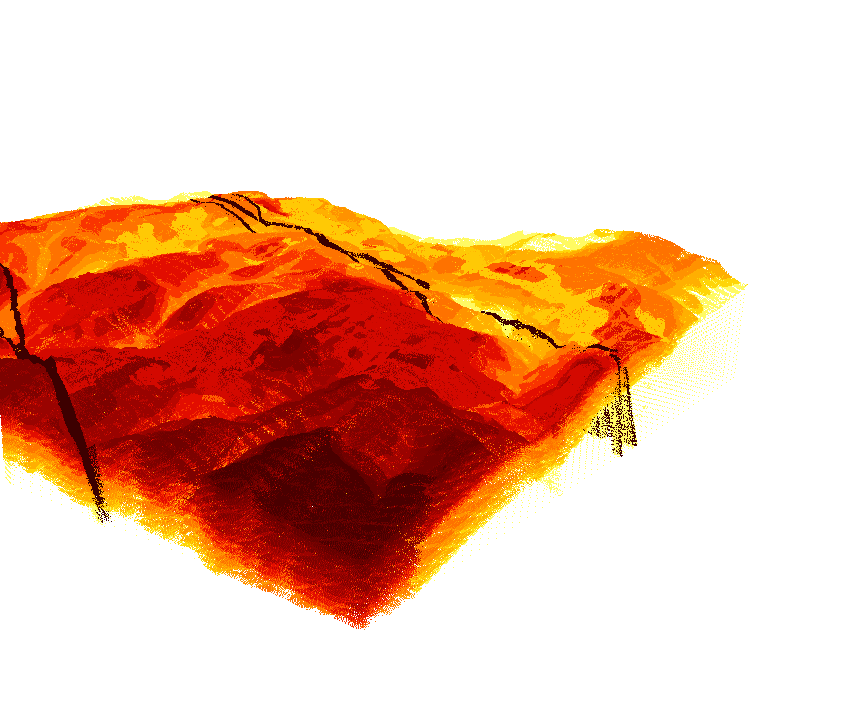}}\\
\multicolumn{10}{c}{X cross-sections of the same geological domains\label{fig:appendix-animated-site8-geological-domains-dx}}\\
\end{tabular}
}
\end{table*}

\begin{table*}[!th]
\small
\setlength\tabcolsep{4pt}
\caption{Block model statistics: proposed methodology vs octree (with D=5 decomposition levels)}\label{tab:detailed-results-vs-octree-d5}
\resizebox{\textwidth}{!}{
\begin{tabular}{|l|c|cccc|cccc|}\hline
Domain & \%\,volume &\multicolumn{4}{c|}{block count} & \multicolumn{4}{c|}{volume-weighted block aspect ratio}\\ \hline
& & Octree & Octree\,+\,Merge & Proposed-P & Proposed-D & Octree & Octree\,+\,Merge & Proposed-P & Proposed-D\\ \hline
{\color{fire0}$\bsquare$} $N_{0}$ & 0.012187 & 54650 & 22512 & 13500 & 10598 & 2.5 & 3.937759 & 28.207846 & 5.272790\\ \hline
{\color{fire1}$\bsquare$} $M_{0}$ & 0.012225 & 47666 & 19995 & 14373 & 8581 & 2.5 & 3.880556 & 17.203926 & 6.303496\\ \hline
{\color{fire2}$\bsquare$} $N_{1}$ & 2.069887 & 5966122 & 2206483 & 856985 & 691068 & 2.5 & 2.448336 & 8.103377 & 5.186617\\ \hline
{\color{fire3}$\bsquare$} $N_{2}$ & 0.569810 & 465501 & 196686 & 101427 & 106403 & 2.5 & 3.341298 & 9.875999 & 3.403828\\ \hline
{\color{fire4}$\bsquare$} $M_{1}$ & 0.000040 & 646 & 333 & 225 & 211 & 2.5 & 3.638946 & 7.322417 & 2.796074\\ \hline
{\color{fire5}$\bsquare$} $N_{3}$ & 0.246775 & 482523 & 214397 & 120786 & 128908 & 2.5 & 3.872228 & 17.465091 & 5.425162\\ \hline
{\color{fire6}$\bsquare$} $N_{4}$ & 0.318412 & 593948 & 264225 & 149991 & 157914 & 2.5 & 3.893686 & 18.004315 & 5.075332\\ \hline
{\color{fire7}$\bsquare$} $N_{5}$ & 1.035730 & 1441073 & 626411 & 340497 & 345133 & 2.5 & 3.648167 & 14.088016 & 3.945783\\ \hline
{\color{fire8}$\bsquare$} $M_{2}$ & 0.074129 & 266189 & 114990 & 59476 & 59692 & 2.5 & 3.725392 & 14.391746 & 5.381985\\ \hline
{\color{fire9}$\bsquare$} $N_{6}$ & 1.994655 & 2809366 & 1215883 & 656664 & 653464 & 2.5 & 3.665340 & 13.196620 & 4.206202\\ \hline
{\color{fire10}$\bsquare$} $M_{3}$ & 0.426116 & 994488 & 437691 & 238458 & 232800 & 2.5 & 3.634994 & 15.740771 & 5.344710\\ \hline
{\color{fire11}$\bsquare$} $N_{7}$ & 1.057534 & 3076723 & 1352508 & 715135 & 728884 & 2.5 & 3.909520 & 14.793386 & 6.200759\\ \hline
{\color{fire12}$\bsquare$} $M_{4}$ & 0.112500 & 484807 & 215842 & 116435 & 114125 & 2.5 & 3.801134 & 12.542677 & 5.895113\\ \hline
{\color{fire13}$\bsquare$} $H_{0}$ & 0.331752 & 1295410 & 578825 & 360946 & 308160 & 2.5 & 3.779019 & 21.440046 & 5.554469\\ \hline
{\color{fire14}$\bsquare$} $N_{8}$ & 2.361292 & 4261395 & 1856015 & 1009360 & 997440 & 2.5 & 3.932490 & 16.538731 & 5.283464\\ \hline
{\color{fire15}$\bsquare$} $M_{5}$ & 0.035589 & 209842 & 94582 & 51685 & 50865 & 2.5 & 3.681174 & 15.741827 & 5.199280\\ \hline
{\color{fire16}$\bsquare$} $N_{9}$ & 1.500113 & 4615285 & 2017890 & 1041881 & 1049804 & 2.5 & 3.979482 & 14.656347 & 6.338921\\ \hline
{\color{fire17}$\bsquare$} $M_{6}$ & 0.005824 & 53198 & 25599 & 14448 & 14215 & 2.5 & 3.544744 & 8.629379 & 5.305402\\ \hline
{\color{fire18}$\bsquare$} $H_{1}$ & 0.052644 & 302851 & 133967 & 79652 & 70437 & 2.5 & 3.725562 & 19.455017 & 5.425445\\ \hline
{\color{fire19}$\bsquare$} $N_{10}$ & 5.059155 & 6903632 & 2989708 & 1582574 & 1573708 & 2.5 & 3.815540 & 14.647774 & 4.527278\\ \hline
{\color{fire20}$\bsquare$} $M_{7}$ & 0.229129 & 615542 & 261940 & 139458 & 129839 & 2.5 & 3.719217 & 12.449978 & 5.553382\\ \hline
{\color{fire21}$\bsquare$} $N_{11}$ & 4.322323 & 7841613 & 3398938 & 1800920 & 1801115 & 2.5 & 3.935243 & 16.360033 & 5.554397\\ \hline
{\color{fire22}$\bsquare$} $M_{8}$ & 0.119576 & 434074 & 190030 & 104029 & 98436 & 2.5 & 3.584765 & 13.754793 & 5.630620\\ \hline
{\color{fire23}$\bsquare$} $N_{12}$ & 6.055483 & 8847559 & 3800130 & 1995917 & 1975291 & 2.5 & 3.923066 & 15.285411 & 4.941555\\ \hline
{\color{fire24}$\bsquare$} $M_{9}$ & 0.092180 & 322413 & 139347 & 76104 & 71667 & 2.5 & 3.464583 & 14.724576 & 5.738100\\ \hline
{\color{fire25}$\bsquare$} $H_{2}$ & 0.193918 & 892552 & 402913 & 249416 & 217807 & 2.5 & 3.716055 & 19.805877 & 5.544980\\ \hline
{\color{fire26}$\bsquare$} $N_{13}$ & 3.048503 & 8456955 & 3650182 & 1837766 & 1853527 & 2.5 & 3.977186 & 13.698514 & 6.438980\\ \hline
{\color{fire27}$\bsquare$} $M_{10}$ & 0.005207 & 45786 & 20074 & 11478 & 10381 & 2.5 & 3.470138 & 11.670511 & 4.750513\\ \hline
{\color{fire28}$\bsquare$} $N_{14}$ & 0.001024 & 10736 & 4770 & 2806 & 2567 & 2.5 & 3.561028 & 13.049818 & 4.983007\\ \hline
{\color{fire29}$\bsquare$} $N_{15}$ & 68.656172 & 12271192 & 4599110 & 1865318 & 1741347 & 2.5 & 2.608728 & 4.109259 & 2.667184\\ \hline
{\color{fire30}$\bsquare$} $M_{11}$ & 0.000117 & 1564 & 760 & 452 & 457 & 2.5 & 3.635483 & 12.809443 & 2.827696\\ \hline
\hline
\multicolumn{2}{|l|}{Total (avg.\,by volume)} & 74065301 & 31052736 & 15608162 & 15204844 & 2.5 & 2.972378 & 7.395642 & 3.456381\\ \hline
\multicolumn{2}{|l|}{Total (avg.\,by block count)} & same & same & same & same & 2.5 & 3.585541 & 13.523983 & 5.078064\\ \hline
\hline
\multicolumn{2}{|l|}{Ratio} & 100.000 & 41.926 & 21.074 & 20.529 & \multicolumn{4}{c|}{}\\ \hline
\multicolumn{10}{c}{}\\
\end{tabular}
}
\end{table*}

\end{document}